\documentclass[prd,preprint,superscriptaddress,tightenlines,nofootinbib,
  eqsecnum,showpacs]{revtex4}

\usepackage{amsmath}
\usepackage{amsfonts}
\usepackage{amssymb}
\usepackage{bm}
\usepackage{hyperref}
\usepackage{mathrsfs}
\usepackage{graphicx}

\usepackage{ulem}
\usepackage[usenames]{color}

\definecolor{darkgreen}{rgb}{0,0.5,0}

\hypersetup{
    bookmarks=true,         
    unicode=false,          
    pdftoolbar=true,        
    pdfmenubar=true,        
    pdffitwindow=false,     
    pdfstartview={FitH},    
    pdftitle={My title},    
    pdfauthor={Author},     
    pdfsubject={Subject},   
    pdfcreator={Creator},   
    pdfproducer={Producer}, 
    pdfkeywords={keyword1} {key2} {key3}, 
    pdfnewwindow=true,      
    colorlinks=true,       
    linkcolor=red,          
    citecolor=cyan,        
    filecolor=magenta,      
    urlcolor=darkgreen,           
    linktocpage=true
}

\allowdisplaybreaks

\DeclareSymbolFontAlphabet{\mathrsfs}{rsfs}
\DeclareMathAlphabet{\mathcal}{OMS}{cmsy}{m}{n}

\newcommand{\ud}{\mathrm{d}}

\newcommand{\beq}{\begin{equation}}
\newcommand{\eeq}{\end{equation}}
\newcommand\calO{{\mathcal{O}}}

\newcounter{theorem} 
\newenvironment{theorem}
           {\refstepcounter{theorem}
   \vspace{1em}
   \noindent\textbf{Lemma~\thetheorem}:
   \begin{em}}
  {\end{em}
   \newline}

\begin{document}

\title{Fokker action of non-spinning compact binaries\\at the fourth
  post-Newtonian approximation}

\author{Laura Bernard}\email{bernard@iap.fr}
\affiliation{$\mathcal{G}\mathbb{R}\varepsilon{\mathbb{C}}\mathcal{O}$,
  Institut d'Astrophysique de Paris,\\ UMR 7095, CNRS, Sorbonne
  Universit{\'e}s \& UPMC Univ Paris 6,\\ 98\textsuperscript{bis}
  boulevard Arago, 75014 Paris, France}

\author{Luc Blanchet}\email{blanchet@iap.fr}
\affiliation{$\mathcal{G}\mathbb{R}\varepsilon{\mathbb{C}}\mathcal{O}$,
  Institut d'Astrophysique de Paris,\\ UMR 7095, CNRS, Sorbonne
  Universit{\'e}s \& UPMC Univ Paris 6,\\ 98\textsuperscript{bis}
  boulevard Arago, 75014 Paris, France}

\author{Alejandro Boh\'e}\email{alejandro.bohe@aei.mpg.de}
\affiliation{Albert Einstein Institut, Am Muehlenberg 1, 14476
  Potsdam-Golm, Germany}

\author{Guillaume Faye}\email{faye@iap.fr}
\affiliation{$\mathcal{G}\mathbb{R}\varepsilon{\mathbb{C}}\mathcal{O}$,
  Institut d'Astrophysique de Paris,\\ UMR 7095, CNRS, Sorbonne
  Universit{\'e}s \& UPMC Univ Paris 6,\\ 98\textsuperscript{bis}
  boulevard Arago, 75014 Paris, France}

\author{Sylvain Marsat}\email{sylvain.marsat@aei.mpg.de}
\affiliation{Maryland Center for Fundamental Physics \& Joint
  Space-Science Center, Department of Physics, University of Maryland,
  College Park, MD 20742, USA}\affiliation{Gravitational Astrophysics
  Laboratory, NASA Goddard Space \\Flight Center, Greenbelt, MD 20771,
  USA}

\date{\today}

\begin{abstract}
  The Fokker action governing the motion of compact binary systems without
  spins is derived in harmonic coordinates at the fourth post-Newtonian
  approximation (4PN) of general relativity. Dimensional regularization is
  used for treating the local ultraviolet (UV) divergences associated with
  point particles, followed by a renormalization of the poles into a
  redefinition of the trajectories of the point masses. Effects at the 4PN
  order associated with wave tails propagating at infinity are included
  consistently at the level of the action. A finite part procedure based on
  analytic continuation deals with the infrared (IR) divergencies at spatial
  infinity, which are shown to be fully consistent with the presence of near
  zone tails. Our end result at 4PN order is Lorentz invariant and has the
  correct self-force limit for the energy of circular orbits. However, we find
  that it differs from the recently published result derived within the ADM
  Hamiltonian formulation of general relativity~[T. Damour, P. Jaranowski, and
  G. Sch\"afer, Phys. Rev. D \textbf{89}, 064058 (2014)]. More work is needed
  to understand this discrepancy.
\end{abstract}


\maketitle

\section{Introduction} 
\label{sec:intro}

Gravitational waves emitted by inspiraling and merging compact
(neutron stars and/or black holes) binary systems are likely to be
routinely detected by the ground-based network of advanced laser
interferometric detectors~\cite{BuonSathya}. Banks of extremely
accurate replica of theoretical predictions (templates) are a
compulsory ingredient of a successful data analysis for these
detectors --- both on-line and off-line. In the early inspiral phase
the post-Newtonian (PN) approximation of general relativity should be
pushed to extremely high order~\cite{Bliving14}. Furthermore high
accuracy comparison and matching of PN results are performed with
numerical relativity computations appropriate for the final merger and
ringdown phases~\cite{BBHS15}.

With these motivations in mind we tackle the problem of the equations
of motion of compact binaries (without spin) at the fourth
post-Newtonian (4PN) order.\footnote{As usual the $n$PN order refers
  to the terms of order $1/c^{2n}$ in the equations of motion beyond
  the Newtonian acceleration.}  Solving this problem is also important
for various applications (numerical/analytical self-force comparisons,
last stable circular orbit, effective-one-body
calculations~\cite{MNT81,BuonD99}) and paves the way to the problem of
radiation and orbital phase evolution at the 4PN order beyond the
Einstein quadrupole formalism --- whose solution is needed for
building 4PN accurate templates.

Historical works on the PN equations of motion of compact binaries
include Lorentz \& Droste~\cite{LD17}, Einstein, Infeld \&
Hoffmann~\cite{EIH}, Fock~\cite{Fock39,Fock}, Chandrasekhar and
collaborators~\cite{C65,CN69,CE70}, as well as Otha \textit{et
  al.}~\cite{OO73,OO74a,OO74b}. These works culminated in the 1980s
with the derivation of the equations of motion up to 2.5PN order,
where radiation reaction effects appear~\cite{BeDD81,DD81b,D82}
(see also~\cite{S85,S86,DS85,Kop85,GKop86,BFP98} for alternative
derivations), and led to the successful analysis of the time of
arrival of the radio pulses from the Hulse-Taylor binary
pulsar~\cite{TW82,D83}.

In the early 2000s the equations of motion were derived at the 3PN
order using different methods: the ADM Hamiltonian formalism of
general
relativity~\cite{JaraS98,JaraS99,JaraS00,DJSpoinc,DJSequiv,DJSdim},
the PN iteration of the equations of motion in harmonic
coordinates~\cite{BF00,BFeom,BFreg,ABF01,BI03CM,BDE04}, some
surface-integral method~\cite{IFA00,IFA01,itoh1,itoh2,ItohLR}, and
effective field theory schemes~\cite{FS3PN}. Furthermore, radiation
reaction effects at 3.5PN order were
added~\cite{JaraS97,PW02,KFS03,NB05,itoh3}, and spin contributions
have been extensively
investigated~\cite{HaS11so,HaS11ss,HSS10,MBFB13,BMFB13,
BFMP15,PoR08a,PoR08b,LS15a,LS15b,LS15c}.

Works in the early 2010s partially obtained the equations of motion at
the 4PN order using the ADM Hamiltonian
formalism~\cite{JaraS12,JaraS13,JaraS15} and the effective field
theory~\cite{FS4PN}. More recently, the important effect of
gravitational wave tails at 4PN order~\cite{BD88,B93} was included in
the ADM Hamiltonian. This permitted to understand the IR divergencies
in this calculation and to complete the 4PN dynamics~\cite{DJS14} (see
also~\cite{DJS15}). Notice however that the latter work~\cite{DJS14}
did not perform a full consistent PN analysis but resorted to an
auxiliary self-force calculation~\cite{BDLW10b,LBW12,BiniD13} to fix a
last coefficient.

In the present paper we derive the Fokker Lagrangian~\cite{Fokker} at the 4PN
order in harmonic coordinates. We combine a dimensional regularization of the
UV divergencies associated with point particles with a finite part
regularization based on analytic continuation dealing with IR divergencies. We
show that the IR divergencies are perfectly consistent with the presence of
the tail effect at 4PN order, which is incorporated consistently into the
Fokker action. However, like in~\cite{DJS14}, we are obliged to introduce an
arbitrary coefficient relating the IR cut-off scale to the \textit{a priori}
different scale present in the tail integral. This coefficient is determined
by using a self-force calculation (both numerical~\cite{BDLW10b,LBW12} and
analytical~\cite{BiniD13}), so that our end result for the energy of circular
orbits at 4PN order has the correct self-force limit. We also checked that it
is manifestly Lorentz-Poincar\'e invariant. In a companion
paper~\cite{BBBFM15b} we shall study the conserved integrals of the motion,
the reduction to the center-of-mass frame and the dynamics of
quasi-circular orbits.

Up to quadratic order in Newton's constant, our Lagrangian is
equivalent to the Lagrangian derived by means of effective field
theory techniques~\cite{FS4PN}. However, trying to relate our result to the
result obtained from the ADM Hamiltonian
approach~\cite{JaraS12,JaraS13,DJS14,JaraS15}, we find a difference
with the latter works, occuring at orders $G^4$ and $G^5$ in the
Hamiltonian. Part of the difference is due to the fact that we
disagree with the treatment of the tail part of the Hamiltonian for
circular orbits in Ref.~\cite{DJS14}. However, even when using our own
treatment of tails in their results, there still remains a
discrepancy with the works~\cite{JaraS12,JaraS13,DJS14,JaraS15} that
we cannot resolve. More work is needed to understand the origin of
this remaining difference and resolve it.

In Sec.~\ref{sec:fokker} we show how to use the Fokker action in the
context of PN approximations. In particular we split the action into a
term depending on the PN field in the near zone and a term depending
on the field in the far zone. The latter term is crucial to control
the tails which are then computed consistently in the action at 4PN
order in Sec.~\ref{sec:tail}. We explain our method for iterating the
PN approximation of the Fokker action in Sec.~\ref{sec:PNiterate}. In
Sec.~\ref{sec:lagrangian}, we present the full-fledged Lagrangian of
compact binaries at the 4PN order, both in harmonic and ADM like
coordinates, and we compare with the results~\cite{DJS14} obtained for
the 4PN ADM Hamiltonian in Sec.~\ref{sec:hamiltonian}. Finally we
explain in Sec.~\ref{sec:Ehamiltonian} our disagreement with
Ref.~\cite{DJS14} regarding the treatment of the tail term. The paper
ends with several technical Appendices.

\section{The Fokker action}
\label{sec:fokker}

\subsection{General statements}
\label{sec:gen}

We consider the complete Einstein-Hilbert gravitation-plus-matter
action $S=S_g+S_m$, where the gravitational piece $S_g$ takes the
Landau-Lifshitz form with the usual harmonic gauge-fixing
term,\footnote{We also denote $S_g=\int \ud t\,L_g$ and
  $L_g=\int\ud^3\mathbf{x}\,\mathcal{L}_g$. The Lagrangian $L_g$ is
  defined modulo a total time derivative and the Lagrangian density
  $\mathcal{L}_g$ modulo a space-time derivative.}
\begin{equation}\label{Sg}
S_{g} = \frac{c^{3}}{16\pi G} \int \ud^{4}x \, \sqrt{-g} \left[
  g^{\mu\nu} \left( \Gamma^{\rho}_{\mu\lambda}
  \Gamma^{\lambda}_{\nu\rho} - \Gamma^{\rho}_{\mu\nu}
  \Gamma^{\lambda}_{\rho\lambda} \right) -\frac{1}{2}
  g_{\mu\nu}\Gamma^{\mu}\Gamma^{\nu} \right] \,,
\end{equation}
with $\Gamma^{\mu}\equiv g^{\rho\sigma}\Gamma^{\mu}_{\rho\sigma}$, and where
$S_{m}$ denotes the matter piece appropriate for two point particles
($A=1,2$) without spin nor internal structure,
\begin{equation}\label{Sm}
S_{m} = - \sum_A m_A c^{2} \int \ud t \sqrt{-(g_{\mu\nu})_A
  \,v_A^{\mu}v_A^{\nu}/c^{2}}\,.
\end{equation}
Here $m_A$ is the mass of the particles, $v_A^\mu=\ud y_A^\mu/\ud
t=(c,\bm{v}_A)$ is the usual coordinate velocity, $y_A^\mu=(c
t,\bm{y}_A)$ the usual trajectory, and $(g_{\mu\nu})_A$ stands for the
metric evaluated at the location of the particle $A$ following the
dimensional regularization scheme.

A closed-form expression for the gravitational action can be written
with the help of the gothic metric
$\mathfrak{g}^{\mu\nu}=\sqrt{-g}g^{\mu\nu}$ and its inverse
$\mathfrak{g}_{\mu\nu}=g_{\mu\nu}/\sqrt{-g}$ as
\begin{align}\label{gothic}
S_{g} &= \frac{c^{3}}{32\pi G} \int \ud^{4} x
\biggl[-\frac{1}{2}\left(\mathfrak{g}_{\mu\rho}\mathfrak{g}_{\nu\sigma}
  - \frac{1}{2}\mathfrak{g}_{\mu\nu}\mathfrak{g}_{\rho\sigma}\right)
  \mathfrak{g}^{\lambda\tau}\partial_{\lambda}
  \mathfrak{g}^{\mu\nu}\partial_{\tau}
  \mathfrak{g}^{\rho\sigma}\nonumber\\&~\qquad\qquad\qquad\quad
  +\mathfrak{g}_{\mu\nu}\Bigl(\partial_{\rho}\mathfrak{g}^{\mu\sigma}
  \partial_{\sigma}\mathfrak{g}^{\nu\rho} -
  \partial_{\rho}\mathfrak{g}^{\mu\rho}\partial_{\sigma}
  \mathfrak{g}^{\nu\sigma}\Bigr)\biggr]\,.
\end{align}
Expanding around Minkowski space-time we pose $\mathfrak{g}^{\mu\nu} =
\eta^{\mu\nu} + h^{\mu\nu}$ which defines the metric perturbation
variable $h^{\mu\nu}$. The action appears then as an infinite
non-linear power series in $h$, where indices on $h$ and on partial
derivatives $\partial$ are lowered and raised with the Minkowski
metric $\eta^{\mu\nu}=\eta_{\mu\nu}=\text{diag}(-1,1,1,1)$. The
Lagrangian density $\mathcal{L}_g$ can take various forms obtained
from each other by integrations by parts. For our purpose the best
form starts at quadratic order by terms like $\sim h\Box h$, where
$\Box=\eta^{\rho\sigma}\partial^2_{\rho\sigma}$ is the flat
d'Alembertian operator. So the general structure of the Lagrangian we
shall use is $\mathcal{L}_g\sim h\Box h+h \partial h\partial h +
h\,h\partial h\partial h + \cdots$. See~\eqref{Lg23} for the explicit
expressions of the quadratic and cubic terms.

The Einstein field equations derived from the harmonic gauge fixed
action read
\begin{subequations}\label{EFE}
\begin{align}
\Box h^{\mu\nu} &= \frac{16\pi G}{c^4}
\tau^{\mu\nu}\,,\\ \tau^{\mu\nu} &\equiv \vert g\vert T^{\mu\nu} +
\frac{c^4}{16\pi G}\Sigma^{\mu\nu}[h, \partial h, \partial^2
  h]\,.\label{tau}
\end{align}
\end{subequations}
The above quantity $\tau^{\mu\nu}$ denotes the pseudo stress-energy
tensor of the matter and gravitational fields, with
$T^{\mu\nu}=\frac{2}{\sqrt{-g}}\delta S_m/\delta g_{\mu\nu}$ and with
the gravitational source term $\Sigma^{\mu\nu}$, at least quadratic in
$h$ and its first and second derivatives, being given by
\begin{equation}\label{Lambda}
\Sigma^{\mu\nu} = \Lambda^{\mu\nu} - H^\mu H^\nu - H^\rho
\partial_\rho h^{\mu\nu} -
\frac{1}{2}\mathfrak{g}^{\mu\nu}\mathfrak{g}_{\rho\sigma} H^\rho
H^\sigma +
2\mathfrak{g}_{\rho\sigma}\mathfrak{g}^{\lambda(\mu}\partial_\lambda
h^{\nu)\rho}H^\sigma\,,
\end{equation}
where $\Lambda^{\mu\nu}$ takes the standard expression valid in
harmonic coordinates while the ``harmonicity'' is defined by $H^\mu
\equiv \partial_\nu h^{\mu\nu} = -
\sqrt{-g}\,\Gamma^\mu$.\footnote{The expression of $\Lambda^{\mu\nu}$
  is given by Eq.~(24) in~\cite{Bliving14}. Later we shall also need
  its generalization to $d$ space dimensions as given by~(175)
  in~\cite{Bliving14}. The harmonicity terms shown in~\eqref{Lambda}
  are the same in $d$ dimensions.} As we see, the gravitational source
term contains all required harmonicities $H^\mu$, which will not be
assumed to be zero in the PN iteration of the field
equations~\eqref{EFE}.

The Fokker action is obtained by inserting back
into~\eqref{Sg}--\eqref{Sm} an explicit PN iterated solution of the
field equations~\eqref{EFE} given as a functional of the particle's
trajectories, \textit{i.e.}, an explicit PN metric
$g_{\mu\nu}(\mathbf{x}; \bm{y}_B(t), \bm{v}_B(t), \cdots)$. Here the
ellipsis indicate extra variables coming from the fact that we solve
Eqs.~\eqref{EFE} including all harmonicity terms and without
replacement of accelerations, so that the equations of motion are
off-shell at this stage and the solution for the metric depends also
on accelerations $\bm{a}_B(t)$, derivative of accelerations
$\bm{b}_B(t)$, and so on. In particular, the metric in the matter
action evaluated at the location of the particle $A$ will be some
$(g_{\mu\nu})_A=g_{\mu\nu}(\bm{y}_A(t); \bm{y}_B(t), \bm{v}_B(t),
\cdots)$. Thus, the Fokker generalized PN action, depending not only
on positions and velocities but also on accelerations and their
derivatives, is given by
\begin{align}\label{SF}
S_\text{F}\left[\bm{y}_B(t), \bm{v}_B(t), \cdots\right] &=
\int\ud t\!\int \ud^{3}\mathbf{x} \,\mathcal{L}_g\left[\mathbf{x};
  \bm{y}_B(t), \bm{v}_B(t), \cdots\right] \nonumber\\ &-
\sum_A m_A c^{2} \int \ud t \sqrt{-g_{\mu\nu}\left(\bm{y}_A(t);
  \bm{y}_B(t), \bm{v}_B(t), \cdots\right)
  \,v_A^{\mu}v_A^{\nu}/c^{2}}\,,
\end{align}
where $\mathcal{L}_g$ is the Lagrangian density of the gravitational
action~\eqref{gothic}. As is well known, it is always possible to
eliminate from a generalized PN action a contribution that is
quadratic in the accelerations by absorbing it into a ``double-zero''
term which does not contribute to the dynamics~\cite{DS85}. The
argument can be extended to any term polynomial in the accelerations
(and their derivatives). The PN equations of motion of the particles
are obtained as the generalized Lagrange equations
\begin{align}\label{EOM}
\frac{\delta S_\text{F}}{\delta \bm{y}_B} \equiv \frac{\partial
  L_\text{F}}{\partial \bm{y}_B} - \frac{\ud}{\ud
  t}\left(\frac{\partial L_\text{F}}{\partial \bm{v}_B}\right) +
\cdots = 0\,,
\end{align}
where $L_\text{F}$ is the corresponding Lagrangian
($S_\text{F}=\int\ud t\,L_\text{F}$). Once they have been constructed,
the equations~\eqref{EOM} can be order reduced by replacing
iteratively all the higher-order accelerations with their expressions
coming from the PN equations of motion themselves. The classical
Fokker action should be equivalent, in the ``tree-level''
approximation, to the effective action used by the effective field
theory~\cite{FS3PN,FS4PN,FStail,GLPR15}.

\subsection{Fokker action in the PN approximation}
\label{sec:fokkerPN}

In the Fokker action~\eqref{SF} the gravitational term integrates over
the whole space a solution of the Einstein field equations obtained by
PN iteration. The problem is that the PN solution is valid only in the
near zone of the source --- made here of a system of particles. Let us
denote by $\overline{h}$ the PN expansion of the full-fledged metric
perturbation $h$.\footnote{We are here dealing with an explicit
  solution of the Einstein field equations~\eqref{EFE} for insertion
  into the Fokker action [see~\eqref{SF}]. Hence the metric
  perturbation depends on the particles, $h(\mathbf{x}; \bm{y}_A(t),
  \bm{v}_A(t), \cdots)$, as does its PN expansion
  $\overline{h}(\mathbf{x}; \bm{y}_A(t), \bm{v}_A(t), \cdots)$ and
  multipole expansion $\mathcal{M}(h)(\mathbf{x}; \bm{y}_A(t),
  \bm{v}_A(t), \cdots)$ considered below. For simplicity we shall not
  indicate the dependence on the particles.} Outside the near zone of
the source, $\overline{h}$ is not expected to agree with $h$ and, in
fact, will typically diverge at infinity.\footnote{For instance, we
  know that $\overline{h}$ cannot be ``asymptotically flat'' starting
  at the 2PN or 3PN order, depending on the adopted coordinate
  system~\cite{Rend92}.} On the other hand, the \textit{multipole
  expansion} of the metric perturbation, that we denote by
$\mathcal{M}(h)$, will agree with $h$ in all the exterior region of
the source, but will blow up when formally extended inside the near
zone, and diverge when $r\to 0$. Indeed $\mathcal{M}(h)$ is a vacuum
solution of the field equation differing from the true solution inside
the matter source. The PN expansion $\overline{h}$ and the multipole
expansion $\mathcal{M}(h)$ are matched together in their overlapping
domain of validity, namely the exterior part of the near zone. Note
that such overlapping regions always exist for PN sources. The equation
that realizes this match states that the near zone expansion ($r/c\to
0$) of the multipole expansion is identical, in a sense of formal
series, to the multipole expansion ($a/r\to 0$, with $a$ being the
size of the source) of the PN expansion. It reads
(see~\cite{Bliving14} for more details)
\begin{equation}\label{matching}
\overline{\mathcal{M}(h)} = \mathcal{M}\bigl(\overline{h}\bigr)\,.
\end{equation}
The question we want to answer now is: How to transform the Fokker
action~\eqref{SF} into an expression involving integrals over PN
expansions that are obtained by formal PN iteration of the field
equations in the near zone and can be computed in practice\,?
Obviously, the problem concerns only the gravitational part of the
action $S_g=\int\ud t\!\int\ud^3\mathbf{x}\,\mathcal{L}_g$. Note that
the PN expansion of the Lagrangian density has the structure
$\overline{\mathcal{L}}_g\sim \overline{h}\Box \overline{h} +
\overline{h}\partial \overline{h}\partial
\overline{h}+\cdots$. Similarly, we can define the multipole expansion
of the integrand, which takes the form $\mathcal{M}(\mathcal{L}_g)\sim
\mathcal{M}(h)\Box \mathcal{M}(h)
+\cdots$. We are now in a position to state the following lemma.

\begin{theorem} \label{th1}
The gravitational part of the Fokker Lagrangian can be written as a
space integral over the looked-for {\rm PN} Lagrangian density plus an
extra contribution involving the multipole expansion,
\begin{equation}\label{lemma1}
L_g = \mathop{\text{{\rm FP}}}_{B=0}\int \ud^{3}\mathbf{x}
\,\Bigl(\frac{r}{r_0}\Bigr)^B\,\overline{\mathcal{L}}_g +
\mathop{\text{{\rm FP}}}_{B=0}\int \ud^{3}\mathbf{x}
\,\Bigl(\frac{r}{r_0}\Bigr)^B\mathcal{M}(\mathcal{L}_g)\,.
\end{equation}
A regulator $(r/r_0)^B$ and a finite part (FP) at $B=0$, with $r_0$
being an arbitrary constant and $B$ a complex number, cure the
divergencies of the {\rm PN} expansion when
$r\equiv\vert\bm{x}\vert\to+\infty$ in the first term while dealing
with the singular behaviour of the multipole expansion when $r\to 0$
in the second term. The constant $r_0$ represents an IR scale in the
first term and a UV scale in the second; it cancels out between the
two terms.
\end{theorem}

\noindent
This lemma relies on the common general structure of the two sides of
the matching equation~\eqref{matching}, which implies a similar
structure for the gravitational part of the Lagrangian density, namely
(see \textit{e.g.}~\cite{Bliving14})
\begin{equation}\label{structure}
\overline{\mathcal{M}(\mathcal{L}_g)} =
\mathcal{M}(\overline{\mathcal{L}}_g) \sim \sum \hat{n}_L \,r^a(\ln
r)^b F(t)\,,
\end{equation}
where $\hat{n}_L=\text{STF}(n_L)$ denotes an angular factor made of the
symmetric-trace-free (STF) product of unit vectors $n^i = x^i/r$, with
$L=i_1\cdots i_\ell$ and $n_L=n_{i_1}\cdots n_{i_\ell}$. The powers of $r$ can
take any positive or negative integer values $a\in\mathbb{Z}$, while the
powers of the logarithm are positive integers $b\in\mathbb{N}$. The functions
$F(t)$ denote very complicated multi-linear functionals of the multipole
moments describing the source. The formal structure~\eqref{structure} can be
either seen as a near-zone expansion when $r\to 0$ or as a far-zone expansion
when $r\to +\infty$.

To prove~\eqref{lemma1}, we consider the difference between $L_g$ and the
second term, namely
\begin{equation}\label{Delta1}
\Delta_g \equiv L_g - \mathop{\text{{\rm FP}}}_{B=0}\int
\ud^{3}\mathbf{x}
\,\Bigl(\frac{r}{r_0}\Bigr)^B\mathcal{M}(\mathcal{L}_g)\,.
\end{equation}
Since $L_g$ is perfectly convergent, it does not need any
regularization; the regulator $(r/r_0)^B$ and the finite part at
$B=0$ can be inserted into it without altering the result. Hence
\begin{equation}\label{Delta2}
\Delta_g = \mathop{\text{FP}}_{B=0}\int \ud^{3}\mathbf{x}
\,\Bigl(\frac{r}{r_0}\Bigr)^B \Bigl[\mathcal{L}_g -
  \mathcal{M}(\mathcal{L}_g)\Bigr]\,.
\end{equation}
Now we remark that the difference between $\mathcal{L}_g$ and its
multipole expansion $\mathcal{M}(\mathcal{L}_g)$ is zero in the
exterior region and is therefore of compact support, limited to the PN
source, which is always smaller than the near zone size. Thus, we can
replace it with the near-zone or PN expansion, so that
\begin{equation}\label{Delta3}
\Delta_g = \mathop{\text{FP}}_{B=0}\int \ud^{3}\mathbf{x}
\,\Bigl(\frac{r}{r_0}\Bigr)^B \Bigl[\overline{\mathcal{L}_g -
    \mathcal{M}(\mathcal{L}_g)}\Bigr]\,.
\end{equation}
Finally the integral over a formal near-zone expansion of a multipolar
expansion, \textit{i.e.}, an object like
$\overline{\mathcal{M}(\mathcal{L}_g)}$, multiplied by a regulator
$(r/r_0)^B$, is always zero by analytic continuation in $B$. To see why it is
so, we evaluate the integral by inserting the formal
structure~\eqref{structure}. After angular integration there remains a series
of radial integrals of the type $\int_0^{+\infty}\ud r\,r^{B+a+2}(\ln r)^b$
which are all separately zero by analytic continuation in $B$. Indeed, one may
split the previous integral into near-zone $\int_0^\mathcal{R}$ and far-zone
$\int_\mathcal{R}^{+\infty}$ contributions. The near-zone integral is computed
for $\Re(B)>-a-3$ and analytically continued for any $B\in\mathbb{C}$, except
for a multiple pole at $B=-a-3$. Likewise, the far-zone integral is computed
for $\Re(B)<-a-3$ and analytically continued for any $B\in\mathbb{C}$, except
$-a-3$. The two analytic continuations cancel each other and the result is
exactly zero for any $B\in\mathbb{C}$, without poles (see~\cite{Bliving14} for
more details). Finally, our lemma is proved,
\begin{equation}\label{Delta4}
\Delta_g = \mathop{\text{FP}}_{B=0}\int \ud^{3}\mathbf{x}
\,\Bigl(\frac{r}{r_0}\Bigr)^B \,\overline{\mathcal{L}}_g\,.
\end{equation}

We now examine the fate of the second, multipolar term
in~\eqref{lemma1} and show that it is actually negligible at the 4PN
order. For this purpose, we shall prove that this term is non-zero
only for ``hereditary'' terms depending on the whole past history of
the source. Recall indeed that the multipolar expansion
$\mathcal{M}(h)$ is constructed from a post-Minkowskian (PM) algorithm
starting from the most general solution of the linear Einstein field
equations in the vacuum region outside the source
(see~\cite{Bliving14} for a review, as well as Appendix~\ref{app:np2}
below). This linear solution is a functional of the multipole moments
of the source, \textit{i.e.}, the two series of mass-type and
current-type moments $I_L(u)$ and $J_L(u)$ that describe the source
($L=i_1\cdots i_\ell$, with $\ell$ being the multipole order),
evaluated at the retarded time of harmonic coordinates
$u=t-r/c$.\footnote{The retarded cone in harmonic coordinates differs
  from a null coordinate cone by the famous logarithmic deviation, say
  $U = u - \frac{2G M}{c^3}\ln(\frac{c^2r}{2GM}) +
  \mathcal{O}(\frac{1}{r})$. Such logarithmic deviation is taken into
  account in the formalism but in the form of a PM expansion,
  \textit{i.e.}, it is formally expanded when $G\to 0$; it is then
  responsible for the appearance of powers of logarithms.} It is
``instantaneous'' in the sense that it depends on the state of the
source, characterized by the moments $I_L$ and $J_L$, only at time
$u$. The PM iteration of this solution generates many terms that are
likewise instantaneous and many hereditary terms that involve an
integration over the past of the source, say $\int_{-\infty}^u\ud v
\,Q(1+\frac{u-v}{r})[I_L(v)~\text{or}~J_L(v)]$, where $Q$ is typically
a Legendre function of the second kind~\cite{B98tail, BFW14a,
  BFW14b}. One feature of the instantaneous terms is that, for them,
the dependence on $u$ can be factorized out through some function
$G(u)$ which is a multi-linear product of the multipole moments
$I_L(u)$ or $J_L(u)$ and their derivatives. By contrast, for
hereditary terms, such a factorization is in general impossible.

This motivates our definition of instantaneous terms in the multipole
expansion $\mathcal{M}(\mathcal{L}_g)$ (supposed to be generated by a
PM algorithm) as being those with general structure of type
\begin{equation}\label{inst}
\mathcal{M}(\mathcal{L}_g){\big|}_{\text{inst}} = \sum
\frac{\hat{n}_L}{r^k}(\ln r)^q \,G(u)\,,
\end{equation}
where $G(u)$ is any functional of the moments $I_L(u)$ or $J_L(u)$ and
their time-derivatives (or anti time-derivatives), while $k$, $q$ are
positive integers with $k\geqslant 2$. By contrast the hereditary
terms will have a more complicated structure. For instance, recalling
that $\mathcal{M}(\mathcal{L}_g)$ is highly non-linear in
$\mathcal{M}(h)$, the hereditary terms could consist of the
interactions between instantaneous terms and tail terms producing the
so-called ``tails-of-tails''. The corresponding structure would
be\footnote{In our computation we consider only the conservative part
  of the dynamics and neglect the usual radiation reaction
  terms. Then, in the instantaneous terms~\eqref{inst}, we should
  replace the retarded argument with the advanced one, while in
  hereditary terms of type~\eqref{hered} we should consider an
  appropriate symmetrization between retarded and advanced integrals.}
\begin{equation}\label{hered}
\mathcal{M}(\mathcal{L}_g){\big|}_{\text{hered}} = \sum
\frac{\hat{n}_L}{r^k}(\ln r)^q \,H(u)\int_{-\infty}^u\ud v
\,Q\Bigl(1+\frac{u-v}{r}\Bigr)K(v)\,,
\end{equation}
where $H(u)$ and $K(u)$ are multi-linear functionals of $I_L(u)$ and
$J_L(u)$. Obviously, more complicated structures are possible. For
hereditary terms in the previous sense, the dependence over $u$ cannot
be factorized out independently from $r$. Now, we have the following
lemma.

\begin{theorem}
The second term in the gravitational part of the
Lagrangian~\eqref{lemma1} gives no contribution to the action for any
instantaneous contribution of type~\eqref{inst},
\begin{equation}\label{lemma2}
\int\ud t\!\int \ud^{3}\mathbf{x} \,\Bigl(\frac{r}{r_0}\Bigr)^B
\mathcal{M}(\mathcal{L}_g){\big|}_{\text{{\rm inst}}} = 0\,.
\end{equation}
Thus, only hereditary contributions of type~\eqref{hered} or more
complicated will contribute.
\end{theorem}

\noindent
The proof goes on in one line. Plugging~\eqref{inst} into the action,
changing variable from $t$ to $u$ and using the factorization of the
function $G(u)$, we get after angular integration a series of radial
integrals of the type $\int_0^{+\infty}\ud r\,r^{B+2-k}(\ln r)^q$,
which are zero by analytic continuation in $B$ as before.

We emphasize as a caveat that the object $\mathcal{M}(\mathcal{L}_g)$,
made of instantaneous and hereditary pieces~\eqref{inst}
and~\eqref{hered}, should be carefully distinguished from
$\overline{\mathcal{M}(\mathcal{L}_g)}$ whose general structure was
given in~\eqref{structure}. The multipole expansion
$\mathcal{M}(\mathcal{L}_g)$ is defined all over the exterior zone and
can be constructed by means of a PM algorithm. At any PM order and for
a given set of multipole moments $I_L$ and $J_L$,
$\mathcal{M}(\mathcal{L}_g)$ is always made of a finite number of
terms like~\eqref{inst} or~\eqref{hered}. On the contrary,
$\overline{\mathcal{M}(\mathcal{L}_g)}$ represents a formal
\textit{infinite} Taylor series when $r\to 0$ which, as we have seen
from the matching equation~\eqref{matching}, can also be interpreted
as a formal series $\mathcal{M}(\overline{\mathcal{L}_g})$ when $r\to
+\infty$. In such a formal sense,
$\overline{\mathcal{M}(\mathcal{L}_g)}$ is in fact valid
``everywhere''.

Finally we are in a position to show that the multipolar contribution
to the action --- \textit{i.e.}, the second term in~\eqref{lemma1} ---
is negligible at the 4PN order. Indeed, with the choice we have made
to write the original action by starting at quadratic order with terms
$\sim h \Box h$ (after suitable integration by parts), we see that the
multipole expansion of the Lagrangian density, which is at least
quadratic in $\mathcal{M}(h)$, takes the form
\begin{equation}\label{structMult1}
\mathcal{M}(\mathcal{L}_g) \sim \mathcal{M}(h)\Box\mathcal{M}(h) +
\mathcal{M}(h)\partial\mathcal{M}(h)\partial\mathcal{M}(h) + \cdots\,.
\end{equation}
Furthermore, $\mathcal{M}(h)$ is a \textit{vacuum} solution of the
field equations~\eqref{EFE}, physically valid only in the exterior of
the source. Hence $\Box\mathcal{M}(h) = \mathcal{M}(\Sigma)$ with no
matter source terms, and this quantity is therefore of the type
\begin{equation}\label{structMult2}
\Box\mathcal{M}(h) \sim \mathcal{M}(h)\partial^2\mathcal{M}(h) +
\partial\mathcal{M}(h)\partial\mathcal{M}(h) + \cdots\,.
\end{equation}
Combining~\eqref{structMult1} and~\eqref{structMult2} we see that
$\mathcal{M}(\mathcal{L}_g)$ is at least \textit{cubic} in $\mathcal{M}(h)$.
In a PM expansion of $\mathcal{M}(h)$ [see~\eqref{multPM} in the
Appendix~\ref{app:np2}], this term is at least of order $\mathcal{O}(G^3)$.
Now, from~\eqref{lemma2}, we know that $\mathcal{M}(\mathcal{L}_g)$ must
necessarily be made of some multipole interaction involving hereditary terms,
as displayed in~\eqref{hered}, and these must be cubic. But we know that at
dominant order such terms are the so-called ``tails-of-tails'', made of
multipole interactions $M\times M\times I_L(u)$ or $M\times M\times J_L(u)$
($M$ is the ADM mass), which arise at least at the 5.5PN order~\cite{BFW14a,
  BFW14b}. Therefore, in our calculation limited to 4PN, we are able to
completely neglect the multipolar contribution in the
Lagrangian~\eqref{lemma1}, which becomes a pure functional of the PN expansion
$\overline{h}$ of the metric perturbation up to the 4PN order,
\begin{equation}\label{Lg4PN}
L_g = \mathop{\text{{\rm FP}}}_{B=0}\int \ud^{3}\mathbf{x}
\,\Bigl(\frac{r}{r_0}\Bigr)^B\,\overline{\mathcal{L}}_g \,.
\end{equation}
Note that since the constant $r_0$ cancels out from the two terms
of~\eqref{lemma1}, the term~\eqref{Lg4PN} at 4PN order must \textit{in
  fine} be independent of that constant. We shall explicitly verify
the independence of our final Lagrangian over the IR cut-off scale
$r_0$.

\section{The tail effect at 4PN order}
\label{sec:tail}

Let us recall that there is an imprint of tails in the local PN
dynamics of the source at the 4PN order. The effect appears as a
tail-induced modification of the dissipative radiation reaction force
at the relative 1.5PN order beyond the leading 2.5PN
contribution~\cite{BD88,B93}. Associated with it there exists a non
dissipative piece that contributes to the conservative dynamics at the
4PN order. Here we shall show how to consistently include this piece
into the Fokker action, starting from the result~\eqref{Lg4PN}. To
this end we first need the explicit expressions for the parts
of~\eqref{Lg4PN} that are quadratic and cubic in $\overline{h}$,
namely
\begin{subequations}\label{Lg23}
\begin{align}
L^{(2)}_{g} &= \frac{c^{4}}{32\pi G} \,\mathop{\text{{\rm
      FP}}}_{B=0}\int \ud^{3}\mathbf{x}\,\Bigl(\frac{r}{r_0}\Bigr)^B
\biggl[ \frac{1}{2}\overline{h}_{\mu\nu}\Box \overline{h}^{\mu\nu} -
  \frac{1}{4}\overline{h}\Box \overline{h} \biggr]
\,,\label{Lg2}\\ L^{(3)}_{g} &= \frac{c^{4}}{32\pi G}
\,\mathop{\text{{\rm FP}}}_{B=0}\int
\ud^{3}\mathbf{x}\,\Bigl(\frac{r}{r_0}\Bigr)^B \biggl[
  \overline{h}^{\rho\sigma}\Bigl(-\frac{1}{2}\partial_\rho
  \overline{h}_{\mu\nu}\partial_\sigma \overline{h}^{\mu\nu} +
  \frac{1}{4}\partial_\rho \overline{h}\partial_\sigma \overline{h}
  \Bigr) - \overline{h}_{\mu\nu}\Bigl( \partial_\rho
  \overline{h}^{\mu\sigma}\partial_\sigma \overline{h}^{\nu\rho} -
  \overline{H}^\mu \overline{H}^\nu \Bigr)
  \nonumber\\ &\qquad\qquad\qquad\qquad\quad +
  \overline{h}_{\mu\nu}\Bigl( \partial_\rho
  \overline{h}^\mu_\sigma\partial^\rho \overline{h}^{\nu\sigma} -
  \frac{1}{2}\partial_\rho \overline{h}\partial^\rho
  \overline{h}^{\mu\nu} \Bigr) \biggr] \,.\label{Lg3}
\end{align}
\end{subequations}

We shall insert in~\eqref{Lg23} the general expression for the PN
expansion of the field in the near zone obtained by solving the
matching equation~\eqref{matching} to any PN
order~\cite{PB02,BFN05}. This solution incorporates all tails and
related effects (both dissipative and conservative). It is built from
a particular $B$-dependent solution of the wave equation, defined from
the PN expansion of the pseudo stress-energy tensor~\eqref{tau},
$\overline{\tau}^{\mu\nu}$, by
\begin{equation}\label{instpotentials}
\overline{h}_\text{part}^{\mu\nu} \equiv \frac{16\pi G}{c^4}
\,\mathop{\text{{\rm
      FP}}}_{B=0}\mathcal{I}^{-1}\Bigl[\Bigl(\frac{r}{s'_0}\Bigr)^B
  \overline{\tau}^{\mu\nu}\Bigr]\,,
\end{equation}
where the action of the operator $\mathcal{I}^{-1}$ of the
``instantaneous'' potentials (in the terminology of~\cite{B93}) is
given by
\begin{equation}\label{operatorI}
\mathcal{I}^{-1}\Bigl[\Bigl(\frac{r}{s'_0}\Bigr)^B
  \overline{\tau}^{\mu\nu}\Bigr] =
\sum_{k=0}^{+\infty}\left(\frac{\partial}{c\partial t}\right)^{\!\!2k}
\Delta^{-k-1}
\Bigl[\Bigl(\frac{r}{s'_0}\Bigr)^B\overline{\tau}^{\mu\nu}\Bigr]\,,
\end{equation}
in terms of the $k$-th iterated Poisson integral operator,
\begin{equation}\label{genpoisson}
\Delta^{-k-1}
\Bigl[\Bigl(\frac{r}{s'_0}\Bigr)^B\overline{\tau}^{\mu\nu}\Bigr] =
-\frac{1}{4\pi}\int \ud^3\mathbf{x}'\,\Bigl(\frac{r'}{s'_0}\Bigr)^B\,
\frac{\vert\mathbf{x}-\mathbf{x}'\vert^{2k-1}}{(2k)!}
\,\overline{\tau}^{\mu\nu}(\mathbf{x}',t) \,.
\end{equation} 
The general PN solution that matches an exterior solution with
retarded boundary conditions at infinity is then the sum of the
particular solution~\eqref{instpotentials} and of a homogeneous
multipolar solution regular inside the source, \textit{i.e.}, of the
type retarded minus advanced, and expanded in the near
zone,\footnote{Here and below, the presence of an overbar denoting the
  near-zone expansion $r\to 0$ is explicitly understood on the regular
  retarded-minus-advanced homogeneous solutions like the second term
  in~\eqref{hgen}.}
\begin{equation}\label{hgen}
\overline{h}^{\mu\nu} = \overline{h}_\text{part}^{\mu\nu} -
\frac{2G}{c^4}\sum^{+\infty}_{\ell=0}
\frac{(-)^\ell}{\ell!}\,\partial_L \!\left\{
\frac{\mathcal{A}^{\mu\nu}_L (t-r/c)-\mathcal{A}^{\mu\nu}_L
  (t+r/c)}{r} \right\}\,.
\end{equation} 
Note that the particular solution~\eqref{instpotentials} involves the
scale $s'_0$. Similarly, as we shall see, the homogeneous solution
in~\eqref{hgen} will also depend on the scale $s'_0$ (through
$s_0=s'_0\,e^{-11/12}$ introduced below).

The multipole moments $\mathcal{A}_L(t)$ in~\eqref{hgen} are STF in
$L=i_1\cdots i_\ell$ and can be called radiation-reaction
moments. They are composed of two parts,
\begin{equation} \label{AFR}
\mathcal{A}^{\mu\nu}_L(t) = \mathcal{F}^{\mu\nu}_L(t)
+\mathcal{R}^{\mu\nu}_L(t)\,.
\end{equation}
The first one, $\mathcal{F}_L$, corresponds essentially to linear
radiation reaction effects and yields the usual radiation damping
terms at half integral 2.5PN and 3.5PN orders. These terms will not
contribute to the conservative dynamics (they yield
total time derivatives in the action) and we ignore them.

Important for our purpose is the second part, $\mathcal{R}_L$, which
depends on boundary conditions imposed at infinity. It is a functional
of the multipole expansion of the gravitational source term in the
Einstein field equations, \textit{i.e.}, $\mathcal{M}(\Sigma)$, and is
given by Eq.~(4.11) in~\cite{PB02}. The function $\mathcal{R}_L$ is
responsible for the tail effects in the near zone metric and its
contribution to~\eqref{hgen} starts precisely at 4PN order. We
evaluate it for quadrupolar tails, corresponding to the interaction
between the total ADM mass $M$ of the source and its STF quadrupole
moment $I_{ij}$. Denoting the corresponding homogeneous solution by
\begin{equation}\label{Hhom}
\overline{\mathcal{H}}^{\mu\nu} = -
\frac{2G}{c^4}\sum^{+\infty}_{\ell=0}
\frac{(-)^\ell}{\ell!}\,\partial_L \left\{
\frac{\mathcal{R}^{\mu\nu}_L (t-r/c)-\mathcal{R}^{\mu\nu}_L
  (t+r/c)}{r} \right\}\,,
  \end{equation}
the relevant calculation at the quadrupole level was done in
Eq.~(3.64) of~\cite{B93}:
\begin{subequations}\label{Hhomcomp}
\begin{align}
\overline{\mathcal{H}}^{00} &= - \frac{4G^2
  M}{c^{5}}\int_0^{+\infty} \ud\tau \ln\left(\frac{c\tau}{2
  s_0}\right)\partial_{ij} \biggl\{
\frac{I_{ij}^{(2)}(t-\tau-r/c)-I_{ij}^{(2)}(t-\tau+r/c)}{r}
\biggr\}\,,\\ \overline{\mathcal{H}}^{0i} &= \frac{4G^2
  M}{c^{6}}\int_0^{+\infty} \ud\tau \ln\left(\frac{c\tau}{2
  s_0}\right)\partial_{j} \biggl\{
\frac{I_{ij}^{(3)}(t-\tau-r/c)-I_{ij}^{(3)}(t-\tau+r/c)}{r}
\biggr\}\,,\\ \overline{\mathcal{H}}^{ij} &= - \frac{4G^2
  M}{c^{7}}\int_0^{+\infty} \ud\tau \ln\left(\frac{c\tau}{2
  s_0}\right) \biggl\{
\frac{I_{ij}^{(4)}(t-\tau-r/c)-I_{ij}^{(4)}(t-\tau+r/c)}{r}
\biggr\}\,,
\end{align}
\end{subequations}
with the shorthand $s_0=s'_0\,e^{-11/12}$. For systems of
particles the quadrupole moment reads $I_{ij}=\sum_A m_A y_A^{\langle
  i}y_A^{j\rangle}$, where $\langle\rangle$ denotes the STF
projection. Time-derivatives are indicated as $I_{ij}^{(n)}$. Here,
note that the constant $s_0$ in the logarithms \textit{a priori}
differs from $r_0$ [the IR cut-off in~\eqref{Lg4PN}] and we pose
\begin{equation}\label{s0}
s_0 = r_0 \,e^{-\alpha}\,.
\end{equation}
In this work we shall view the parameter $\alpha$ in~\eqref{s0} as an
``ambiguity'' parameter reflecting some incompleteness of the present
formalism. We do not seem to be able to control this
ambiguity,\footnote{Notice that we do not control the ``bulk'' PN
  near-zone metric outside the particles; the present formalism is
  incomplete in this sense.} which we therefore leave as
arbitrary. The parameter $\alpha$ is the equivalent of the parameter
$C$ in~\cite{DJS14} and we shall later fix it, like in~\cite{DJS14},
by requiring that the conserved energy for circular orbits contains
the correct self-force limit already known by
numerical~\cite{BDLW10b,LBW12} and analytical~\cite{BiniD13}
calculations [see~\eqref{valeuralpha}]. We have checked that if we
integrate by part the quadratic contributions in the PN
Lagrangian~\eqref{Lg2}, so that we start with terms $\sim
r^B\partial\overline{h}\partial\overline{h}$, the surface terms that
are generated by the presence of the regulator factor $r^B$ do not
contribute to the dynamics modulo a mere redefinition of $\alpha$.

At the leading 4PN order the expressions~\eqref{Hhomcomp} reduce to
\begin{subequations}\label{Hhom4PN}
\begin{align}
\overline{\mathcal{H}}^{00} &= \frac{8G^2
  M}{15c^{10}}x^ix^j\int_0^{+\infty} \ud\tau \ln\left(\frac{c\tau}{2
  s_0}\right)I_{ij}^{(7)}(t-\tau) +
\mathcal{O}\left(12\right)\,,\\ \overline{\mathcal{H}}^{0i} &= -
\frac{8G^2 M}{3c^{9}}x^j\int_0^{+\infty} \ud\tau
\ln\left(\frac{c\tau}{2 s_0}\right)I_{ij}^{(6)}(t-\tau) +
\mathcal{O}\left(11\right)\,,\\ \overline{\mathcal{H}}^{ij} &=
\frac{8G^2 M}{c^{8}}\int_0^{+\infty} \ud\tau \ln\left(\frac{c\tau}{2
  s_0}\right)I_{ij}^{(5)}(t-\tau) + \mathcal{O}\left(10\right)\,,
\end{align}
\end{subequations}
where we denote the PN remainder by $\calO(n)\equiv\calO(c^{-n})$. We
insert the PN solution
${\overline{h}=\overline{h}_\text{part}+\overline{\mathcal{H}}}$ into
the action~\eqref{Lg4PN} and compute the contributions of the tail
part $\overline{\mathcal{H}}$ (the instantaneous parts are discussed
later). The quadratic terms in the action [see~\eqref{Lg2}] will yield
some $\sim\overline{\mathcal{H}}\Box \overline{h}_\text{part}$ that
are very simple to compute since at 4PN order we can use the leading
expressions for $\overline{h}_\text{part}$ [see
\textit{e.g.}~\eqref{metricpot} below]. Furthermore we find that some
contributions $\sim\overline{\mathcal{H}}\,\partial
\overline{h}_\text{part}\,\partial\overline{h}_\text{part}$ coming
from the cubic part of the action must also be included at 4PN
order.\footnote{From~\eqref{Lg3}, these cubic terms are easily
  identified as $\propto\overline{\mathcal{H}}^{ij}\,\partial_i
  \overline{h}^{00}_\text{part}\,\partial_j\overline{h}^{00}_\text{part}
  - \frac{1}{2}\overline{\mathcal{H}}^{ij}\,\partial_i
  \overline{h}_\text{part}\,\partial_j\overline{h}_\text{part}$.}
Finally, inserting $\overline{\mathcal{H}}$ into the matter part $S_m$
of the action makes obviously further contributions. We thus obtain
the following 4PN tail effect in the total Fokker action as (skipping
the PN remainder)
\begin{equation}\label{Stail0}
S_\text{F}^\text{tail} = \sum_A m_A c^2 \int\!\ud t
\left[-\frac{1}{8}\overline{\mathcal{H}}^{00}_A +
  \frac{1}{2c}\overline{\mathcal{H}}^{0i}_A
  v^i_A-\frac{1}{4c^2}\overline{\mathcal{H}}^{ij}_A
  v^i_Av^j_A\right] - \frac{1}{16\pi G}\int\!\ud t\!\int
\ud^{3}\mathbf{x} \,\overline{\mathcal{H}}^{ij}\partial_i U\partial_j
U \,.\\
\end{equation}
Most terms have a compact support and have been straightforwardly
evaluated for particles with mass $m_A$ and ordinary coordinate
velocity $v_A^i$ ($A=1,2$). However the last term in~\eqref{Stail0} is
non compact and contains the Newtonian potential $U=\sum_A G
m_A/\vert\mathbf{x}-\bm{y}_A\vert$. Next we
substitute~\eqref{Hhom4PN} into~\eqref{Stail0} and obtain after
some integrations by parts,
\begin{equation}\label{Stail}
S_\text{F}^\text{tail} = -\frac{2G^2M}{5c^8} \int_{-\infty}^{+\infty}
\ud t \,I_{ij}(t) \int_0^{+\infty} \ud\tau
\ln\left(\frac{c\tau}{2s_0}\right)I_{ij}^{(7)}(t-\tau) + S_\eta\,.
\end{equation}
We observe that the non-compact support term in~\eqref{Stail0} has
nicely combined with the other terms to give a bilinear expression in
the time derivatives of the quadrupole moment $I_{ij}$. The last term
$S_\eta$ denotes an irrelevant gauge term associated with a harmonic
gauge transformation with vector $\eta^i$ at the 4PN order. Such gauge
term is due to a replacement of accelerations in the action that we
did in order to arrive at the form~\eqref{Stail}. For completeness we
give the ``zero-on-shell'' form of this gauge term as
\begin{subequations}\label{Seta}
\begin{align}
S_\eta &= - \sum_A m_A \int_{-\infty}^{+\infty} \ud t \,\Bigl[ a_A^i -
  (\partial_i U)_A\Bigl] \eta_A^i\,,\\ \text{where}\quad\eta^i &=
-\frac{2G^2M}{c^8} x^j \int_0^{+\infty} \ud\tau
\ln\left(\frac{c\tau}{2s_0}\right)I_{ij}^{(5)}(t-\tau)\,.
\end{align}
\end{subequations}
An important point to notice is that the result~\eqref{Stail} can be
rewritten in a manifestly time-symmetric way. Thus the procedure
automatically selects some ``conservative'' part of the tail at 4PN
order --- the dissipative part giving no contribution to the
action. Indeed we can alternatively write (ignoring from now on the
gauge term)
\begin{equation}\label{Stailsym}
S_\text{F}^\text{tail} = -\frac{G^2M}{5c^8} \int_{-\infty}^{+\infty}
\ud t \,I_{ij}(t) \int_0^{+\infty} \ud\tau
\ln\left(\frac{c\tau}{2s_0}\right)\left[I_{ij}^{(7)}(t-\tau) -
  I_{ij}^{(7)}(t+\tau)\right] \,,
\end{equation}
which can also be transformed to a simpler form (after
  integrations by parts) with the help of the Hadamard \textit{partie
  finie} (Pf)~\cite{Hadamard, Schwartz}, as
\begin{equation}\label{StailHad}
S_\text{F}^\text{tail} = \frac{G^2M}{5c^8}
\mathop{\text{Pf}}_{2s_0/c}\int\!\!\!\int \frac{\ud t\ud t'}{\vert
  t-t'\vert} \,I_{ij}^{(3)}(t) \,I_{ij}^{(3)}(t') \,.
\end{equation}
The dependence on the scale $s_0$ [see~\eqref{s0}] enters here
\textit{via} the arbitrary constant present in the definition of the
Hadamard partie finie.\footnote{For any regular function $f(t)$
  tending to zero sufficiently rapidly when $t\to\pm\infty$ we have
$$\mathop{\text{Pf}}_{\tau_0} \int_{-\infty}^{+\infty} \ud
  t'\,\frac{f(t')}{\vert t-t'\vert} \equiv \int_0^{+\infty} \ud\tau
  \ln\left(\frac{\tau}{\tau_0}\right)\left[f^{(1)}(t-\tau) -
    f^{(1)}(t+\tau)\right]\,.$$
} The result~\eqref{StailHad} agrees with the non-local action for the
4PN tail term which has been considered in~\cite{DJS14} [see Eq.~(4.4)
there] and investigated in the
effective field theory approach~\cite{FStail,GLPR15}. Note however 
that while this contribution was added by hand to the 4PN local action
in~\cite{DJS14}, we have shown here how to derive it from
scratch. Varying the action~\eqref{StailHad} with respect to the
particle world-lines we obtain
\begin{equation}\label{verStail}
\frac{\delta S_\text{F}^\text{tail}}{\delta y_A^i(t)} =
-\frac{4G^2M}{5c^8} m_A y_A^j(t)\!\mathop{\text{Pf}}_{2s_{0}/c}
\int_{-\infty}^{+\infty} \frac{\ud t'}{\vert t-t'\vert}
I_{ij}^{(6)}(t') \,,
\end{equation}
which coincides with the conservative part of the known 4PN tail
contribution in the equations of motion~\cite{BD88,B93}.

\section{PN iteration of the Fokker action}
\label{sec:PNiterate}

\subsection{The method ``$n+2$''}
\label{sec:np2}

In the previous section we inserted the explicit PN solution
$\overline{h} = \overline{h}_\text{part} + \overline{\mathcal{H}}$
given by~\eqref{hgen} into the Fokker action, and showed that the
regular homogeneous solution $\overline{\mathcal{H}}$ produces the
expected tails at 4PN order [see~\eqref{StailHad}]. We now deal with
the terms generated by the particular solution in that
decomposition. For simplicity, since the tails have now been
determined, we shall just call that particular solution $\overline{h}
= \overline{h}_\text{part}$.

We first check that the variation of the Fokker action with PN
gravitational term~\eqref{Lg4PN} yields back the PN expansion of the
Einstein field equations. Indeed, because of the factor $r^B$ we have
to worry about the surface term that is generated when performing the
variation with respect to $\overline{h}$. Schematically we have the
structure $\mathcal{L}_g\sim r^B(\overline{h}\Box\overline{h} +
\overline{h}\partial\overline{h}\partial\overline{h} + \cdots)$. When
varying for instance the first term we get a contribution $\sim
r^B\overline{h}\Box\delta\overline{h}$ on which we must shift the box
operator to the left side modulo a surface term. However the surface
term will contain the regulator $r^B$, so we see that it is actually
rigourously zero by analytic continuation in $B$, since it is zero
when starting from the case where $\Re(B)$ is a large negative
number. Computing then the functional derivative of the Fokker action
with respect to the PN expansion of the field, we still have some
factors $r^B$ but in some local (non integrated) expression, on which
the FP prescription reduces to taking the value at $B=0$. Thus we
obtain the PN field equations as expected, say
\begin{equation}\label{deltaSFokker} 
\frac{\delta S_\text{F}}{\delta\overline{h}} \sim c^4\Bigl[\Box
  \overline{h} - \overline{\Sigma} - c^{-4}
  \overline{\mathcal{T}}\Bigr]\,,
\end{equation}
where $\overline{\Sigma}$ denotes the non-linear gravitational source
term and $\mathcal{T} \sim \vert g\vert T$ symbolizes the matter
tensor [see~\eqref{EFE}]. In anticipation of the PN counting we address
below, we have inserted into~\eqref{deltaSFokker} the appropriate PN
factor $c^4/16\pi G\sim c^{4}$.

We now discuss our practical method by which we control the PN
expansion of the components of the metric perturbation $\overline{h}$
in order to obtain the Fokker action accurate to order $n$PN. As we
shall see, thanks to the properties of the Fokker
action~\cite{Fokker}, we essentially need to insert the metric
perturbation at half the PN order which would have been naively
expected.\footnote{This point has been suggested to us by T. Damour
  (private communication).} To this end we decompose the PN metric
perturbation according to
\begin{equation}\label{decomp}
\overline{h}^{\mu\nu} \longrightarrow \left\{
 \begin{array}{l}
	\overline{h}^{00ii} \equiv \overline{h}^{00} +
        \overline{h}^{ii} \,,
        \\ \overline{h}^{0i} \,,
        \\ \overline{h}^{ij} \,.
\end{array}\right.
\end{equation} 
 Written in terms of these
variables the (gauge fixed) gravitational action takes the
form\footnote{We present here the expression in 3 dimensions. Later we
  shall use dimensional regularization, so we shall need the easily
  generalized $d$-dimensional expression.}
\begin{align}\label{SgPN}
S_{g} = \frac{c^{4}}{64\pi G} \mathop{\text{{\rm FP}}}_{B=0} \int
\!\ud t \!\int \ud^{3}\mathbf{x} \,\Bigl(\frac{r}{r_0}\Bigr)^B \biggl[
  \frac{1}{2}\overline{h}^{00ii}\Box \overline{h}^{00ii} - 2
  \overline{h}^{0i}\Box \overline{h}^{0i} + \overline{h}^{ij}\Box
  \overline{h}^{ij} - \overline{h}^{ii} \Box \overline{h}^{jj} +
  \calO\bigl(\overline{h}^{3}\bigr) \biggr] \,.
\end{align}
Similarly the matter action reads at dominant order as
\begin{equation}\label{SmPN} 
S_{m} = \sum_A m_A c^{2} \int\!\ud t \biggl[ -1 +
  \frac{v_A^{2}}{2c^{2}} -\frac{1}{4}\overline{h}_A^{00ii} +
  \frac{v_A^{i}}{c}\,\overline{h}_A^{0i} -
  \frac{v_A^{i}v_A^{j}}{2c^{2}}\,\overline{h}_A^{ij} +
  \frac{v_A^{2}}{2c^{2}}\,\overline{h}_A^{ii} +
  \calO\bigl(\overline{h}_A^{2},c^{-2}\overline{h}_A\bigr) \biggr] \,,
\end{equation}
where the remainder term includes both higher-order terms in $\overline{h}$ as
well as sub-dominant PN corrections. Varying independently with respect to
these components of $\overline{h}$, we recover the fact that to lowest order
$(\overline{h}^{00ii}, \overline{h}^{0i}, \overline{h}^{ij}) = \calO(2,3,4)$,
where we recall that $\calO(n)=\calO(c^{-n})$.

Consider first the usual PN iteration scheme, in which one solves the
field equations up to order $n$, \textit{i.e.}, up to order $c^{-n}$
\textit{included}, where $n$ is an \textit{even} integer. This means
that $(\overline{h}^{00ii}, \overline{h}^{0i}, \overline{h}^{ij})$ are
known up to order $\calO(n+2,n+1,n)$ included, corresponding for $n$
even to the usual \textit{conservative} expansion --- neglecting the
radiation reaction dissipative terms.\footnote{For this discussion we
  can neglect conservative half-integral PN approximations, which
  arise to higher orders~\cite{BFW14a, BFW14b}.}  We collectively
denote by $\overline{h}_{n}[\bm{y}_A]$ the PN solution of the field
equation up to that order, functional of the trajectories of the
particles $\bm{y}_A(t)$ together with their velocities, accelerations
and derivatives of accelerations, not indicated
here. From~\eqref{deltaSFokker}, we see that the PN order of the
functional derivative of the Fokker action evaluated for the
approximate solution $\overline{h}_{n}[\bm{y}_A]$ will be given by the
committed error in that solution. Hence we have for $n$ \textit{even}
(and ignoring the non-conservative odd PN orders)
\begin{subequations}\label{estimates}
\begin{align}
\frac{\delta S_\text{F}}{\delta \overline{h}^{00ii}}\bigl[
  \overline{h}_{n}[\bm{y}_B],\bm{y}_A \bigr] &= \calO\big(n\big)\,,
\\ \frac{\delta S_\text{F}}{\delta \overline{h}^{0i}}\bigl[
  \overline{h}_{n}[\bm{y}_B],\bm{y}_A \bigr] &= \calO\big(n-1\big)\,,
\\ \frac{\delta S_\text{F}}{\delta \overline{h}^{ij}}\bigl[
  \overline{h}_{n}[\bm{y}_B], \bm{y}_A \bigr] &= \calO\big(n-2\big)\,.
\end{align}
\end{subequations}
If now we write the complete solution as $\overline{h}[\bm{y}_B] =
\overline{h}_{n}[\bm{y}_B] + \overline{r}_{n+2}$, introducing some
un-controlled PN remainder term
\begin{equation}\label{remainderestimates}
\overline{r}_{n+2} =
(\overline{r}^{00ii}_{n+4},\overline{r}^{0i}_{n+3},
\overline{r}^{ij}_{n+2}) = \calO(n+4,n+3,n+2)\,,
\end{equation}
the Fokker action expanded around the known approximate solution
reads\footnote{The complete justification of this expansion is
  actually not trivial because of the presence of the regulator
  $(r/r_0)^B$ coming from the PN gravitational term~\eqref{Lg4PN} and
  the integrations by parts that are necessary in order to arrive
  at~\eqref{SFexpand}. We deal with this point in
  Appendix~\ref{app:np2}.}
\begin{align}\label{SFexpand}
    S_\text{F}\bigl[\overline{h}[\bm{y}_B], \bm{y}_A\bigr] &=
    S_\text{F}\bigl[ \overline{h}_{n}[\bm{y}_B],\bm{y}_A\bigr]
    \nonumber\\ & + \mathop{\text{{\rm FP}}}_{B=0} \int \!\ud t \!\int
    \ud^{3}\mathbf{x} \,\Bigl(\frac{r}{r_0}\Bigr)^B
    \biggl[\frac{\delta S_\text{F}}{\delta \overline{h}^{00ii}}\bigl[
        \overline{h}_{n}[\bm{y}_B],\bm{y}_A\bigr]
      \overline{r}^{00ii}_{n+4} \nonumber\\ & \qquad\qquad
      +\frac{\delta S_\text{F}}{\delta \overline{h}^{0i}}\bigl[
        \overline{h}_{n}[\bm{y}_B],\bm{y}_A\bigr]\overline{r}^{0i}_{n+3}
      + \frac{\delta S_\text{F}}{\delta \overline{h}^{ij}} \bigl[
        \overline{h}_{n}[\bm{y}_B],\bm{y}_A\bigr]\overline{r}^{ij}_{n+2}
      + \cdots \biggr]\,.
\end{align}
The ellipsis stand for the quadratic and higher-order terms in the
remainders $\overline{r}_{n+2}$. Inserting both the orders of
magnitude estimates~\eqref{estimates} as well as the orders of the
remainders~\eqref{remainderestimates} we readily obtain
\begin{equation}\label{n1PN}
S_\text{F}\bigl[\overline{h}[\bm{y}_B], \bm{y}_A\bigr] =
S_\text{F}\bigl[ \overline{h}_{n}[\bm{y}_B],\bm{y}_A\bigr] +
\mathcal{O}\left(2n\right) \,,
\end{equation}
which means that the Fokker action has been determined at the
$(n-1)$PN order. This is not yet the $n$PN accuracy we were aiming
for.

However, we notice that in this scheme the term $\overline{h}^{ij}$ is
responsible for the dominant error $\calO(2n)$, together with a term
of the same order, associated with the second variation
$(\delta^{2}S_\text{F}/(\delta \overline{h}^{ij}\delta
\overline{h}^{kl}))\,\overline{r}^{ij}_{n+2}\overline{r}^{kl}_{n+2}$. Thus,
if one pushes by one order the precision of the component
$\overline{h}^{ij}$, denoting $\overline{h}'_{n}[\bm{y}_A]$ the
corresponding solution which is now accurate up to order
${\mathcal{O}(n+2,n+1,n+2)}$ included, we see that
\begin{subequations}\label{newestimates}
\begin{align}
\frac{\delta S_\text{F}}{\delta \overline{h}}\bigl[
  \overline{h}'_{n}[\bm{y}_B],\bm{y}_A \bigr] &=
\calO\big(n,n-1,n\big)\,,\\ \text{and}\quad\overline{r}'_{n+2} &=
\calO\big(n+4,n+3,n+4\big)\,.
\end{align}\end{subequations}
Here the remainders are such that $\overline{h}[\bm{y}_B] =
\overline{h}'_{n}[\bm{y}_B] + \overline{r}'_{n+2}$. With the
estimates~\eqref{newestimates} we now obtain our looked for $n$PN
precision, namely
\begin{equation}\label{nPN}
S_\text{F}\bigl[\overline{h}[\bm{y}_B], \bm{y}_A\bigr] =
S_\text{F}\bigl[ \overline{h}'_{n}[\bm{y}_B],\bm{y}_A\bigr] +
\mathcal{O}\left(2n+2\right) \,.
\end{equation}
Concerning the terms with higher order functional derivatives --- the
ellipsis in~\eqref{SFexpand} --- we can remark that the derivatives
are at most a factor $c^4$ multiplied by a remainder term that is
squared at least. In the scheme~\eqref{newestimates} the dominant
source of error is now the term $\overline{h}^{0i}$. 

Since we ignore non-conservative odd PN terms, solving for
$(\overline{h}^{00ii}, \overline{h}^{0i}, \overline{h}^{ij}) $ to
order ${\calO(n+1,n+2,n+2)}$ with $n$ an \textit{odd} integer gives
\begin{subequations}\label{newestimates2}
\begin{align}
\frac{\delta S_\text{F}}{\delta \overline{h}}\bigl[
  \overline{h}''_{n}[\bm{y}_B],\bm{y}_A \bigr] &=
\calO\big(n-1,n,n-1\big)\,,\\ \text{and}\quad\overline{r}''_{n+2} &=
\calO\big(n+3,n+4,n+3\big)\,,
\end{align}\end{subequations}
hence the error is still $\calO(2n+2)$. In conclusion, we find that in
order to control the Fokker action to the $n$PN order, it is necessary
and sufficient to insert the components of the metric perturbation
\begin{equation}\label{resultnp2}
\overline{h}=(\overline{h}^{00ii}, \overline{h}^{0i},
\overline{h}^{ij})~~\text{up to order}~~\left\{
 \begin{array}{l}
\displaystyle \calO\big(n+2,n+1,n+2\big)~~\text{when $n$ is even}\,,
\\[0.4cm] \displaystyle \calO\big(n+1,n+2,n+1\big)~~\text{when $n$ is
  odd}\,.
\end{array}\right.
\end{equation}
Since in both cases all the components of $\overline{h}$ (for the
conservative dynamics) are to be computed up to order $\calO(n+2)$ we
call the PN iteration up to that order the ``method $n+2$''.

\subsection{Metric potentials in $d$ dimensions}
\label{sec:metric}

From the previous result, we see that at the 4PN order we need the
components of the metric perturbation up to order $\calO(6,5,6)$
included. To that order we shall parametrize the metric by means of
usual PN potentials (see \textit{e.g.}~\cite{Bliving14}). But since we
use dimensional regularization for treating the local divergencies we
provide the requested expression of the metric in $d$ spatial
dimensions. To this end the appropriate generalization of the
variables~\eqref{decomp} is
\begin{equation}
\overline{h}^{00ii} =
2\frac{(d-2)\overline{h}^{00}+\overline{h}^{ii}}{d-1}\,,
\quad\overline{h}^{0i}\,, \quad\overline{h}^{ij}\,.
\end{equation}
We have (with $\hat{W}=\hat{W}_{ii}$ and $\hat{Z}=\hat{Z}_{ii}$)
\begin{subequations}\label{metricpot}
\begin{align} 
\overline{h}^{00ii} &= -\frac{4}{c^{2}}V - \frac{4}{c^{4}}
\left[\frac{d-1}{d-2}V^{2}-2\frac{d-3}{d-2}K\right] \\& \quad -
\frac{8}{c^{6}} \biggl[2\hat{X} + V \hat{W} +
  \frac{1}{3}\left(\frac{d-1}{d-2}\right)^2V^{3}
  -2\frac{d-3}{d-1}V_iV_i-2\frac{(d-1)(d-3)}{(d-2)^2}K V\biggr] +
\calO\left(8\right)\,,\nonumber\\
\overline{h}^{0i} &= - \frac{4}{c^{3}} V_{i} - \frac{4}{c^{5}}
\biggl(2\hat{R}_{i} + \frac{d-1}{d-2}V V_i\biggr) +
\calO\left(7\right)\,,\\
\overline{h}^{ij} &= - \frac{4}{c^{4}}\biggl(\hat{W}_{ij} -
\frac{1}{2} \delta_{ij} \hat{W}\biggr) - \frac{16}{c^{6}} \biggl(
\hat{Z}_{ij} - \frac{1}{2} \delta_{ij} \hat{Z} \biggr) +
\calO\left(8\right) \,.
\end{align}
\end{subequations}
Each of these potentials obeys a flat space-time wave equation (in $d$
dimensions) sourced by lower order potentials in the same family, and
by appropriate matter density components. The list of requested wave
equations is
{\allowdisplaybreaks
\begin{subequations}\label{defpotentials}
\begin{align}
\Box V &= - 4 \pi G\, \sigma \,,\label{V}\\
\Box K &= - 4 \pi G\, \sigma\,V\,,\\
\Box \hat{X} &= -4\pi G\biggl[\frac{V\sigma_{ii}}{d-2} +
  \frac{2(d-3)}{d-1}\sigma_i V_i +\left(\frac{d-3}{d-2}\right)^2
  \sigma\left(\frac{V^2}{2} +K\right)\biggr] +\hat
W_{ij}\, \partial^2_{ij}V \nonumber\\ & +2 V_i\,\partial_t\partial_i V +
\frac{d-1}{2(d-2)}V \partial^2_t V
+\frac{d(d-1)}{4(d-2)^2}\left(\partial_t V\right)^2 -2 \partial_i
V_j\, \partial_j V_i + \Box\delta\hat{X} \,,\\
\Box V_{i} &= - 4 \pi G\, \sigma_{i}\,,\\
\Box \hat{R}_{i} & = -\frac{4\pi G}{d-2}\biggl[\frac{5-d}{2}\, V
  \sigma_i -\frac{d-1}{2}\, V_i\, \sigma\biggr] -
\frac{d-1}{d-2}\,\partial_k V\partial_i V_k
-\frac{d(d-1)}{4(d-2)^2}\,\partial_t V \partial_i V\,, \\
\Box \hat{W}_{ij} & = -4\pi G\biggl(\sigma_{ij}
-\delta_{ij}\,\frac{\sigma_{kk}}{d-2}\biggr)
-\frac{d-1}{2(d-2)}\partial_i V \partial_j V\,,\\
\Box \hat{Z}_{ij} & = -\frac{4\pi G}{d-2}\, V\biggl(\sigma_{ij}
-\delta_{ij}\,\frac{\sigma_{kk}}{d-2}\biggr) -\frac{d-1}{d-2}\,
\partial_t V_{(i}\, \partial_{j)}V +\partial_i V_k\, \partial_j V_k
+\partial_k V_i\, \partial_k V_j \nonumber\\ & -2 \partial_k V_{(i}\,
\partial_{j)}V_k - \frac{\delta_{ij}}{d-2}\, \partial_k V_m
\left(\partial_k V_m -\partial_m V_k\right) -\frac{d(d-1)}{8(d-2)^3}\,
\delta_{ij}\left(\partial_t V\right)^2
\nonumber\\ &+\frac{(d-1)(d-3)}{2(d-2)^2}\, \partial_{(i}
V\partial_{j)} K\,.
\end{align}
\end{subequations}}\noindent
The presence of additional terms proportional to the harmonicity $H^{\mu}$ in
the gravitational source term~\eqref{Lambda} leads in principle to differences
in these wave equations with respect to previous works which used harmonic
coordinates~\cite{BDE04}. At the order we are considering, the only such
additional contribution enters the potential $\hat{X}$. It is denoted by
$\delta\hat{X}$ above. The corresponding source reads
\begin{equation}
	\Box\delta\hat{X} = \partial_{i}V \left[ \partial_{t}V_{i} + 
\partial_{j}\left( \hat{W}_{ij} - \frac{1}{2}\delta_{ij}\hat{W} \right) \right]\,.
\end{equation}

When the equations of motion are satisfied the above potentials
are linked by the following differential identities coming from the
harmonic gauge condition,
{\allowdisplaybreaks
\begin{subequations}
\label{diffident}
\begin{align}
&\partial_t\biggl\{ \frac{d-1}{2(d-2)} V +\frac{1}{2 c^2}\biggl[\hat W
    +\left(\frac{d-1}{d-2}\right)^2 V^2 -\frac{2(d-1)(d-3)}{(d-2)^2}\,
    K\biggr] \biggr\}\nonumber\\ &\qquad\qquad +\partial_i\biggl\{V_i
  +\frac{2}{c^2}\biggl[\hat R_i +\frac{d-1}{2(d-2)} V V_i\biggr]
  \biggr\} = \mathcal{O}\left(4\right)\,,\\ &\partial_t\biggl\{ V_i
  +\frac{2}{c^2}\biggl[ \hat R_i + \frac{d-1}{2(d-2)} V V_i
    \biggr]\biggr\}\nonumber\\ &\qquad\qquad +\partial_j\biggl\{ \hat
  W_{ij} -\frac{1}{2}\, \hat W \delta_{ij} +\frac{4}{c^2}\biggl[\hat
    Z_{ij} -\frac{1}{2}\, \hat Z \delta_{ij}\biggr] \biggr\} =
  \mathcal{O}\left(4\right)\,.
\end{align}
\end{subequations}}\noindent
Note that we generally do not use these relations, which are true only
``on-shell'', at the level of the Fokker action. The only relation we are
allowed to use for simplifications is
\begin{equation}
	\frac{d-1}{2(d-2)}\partial_{t}V + \partial_{i}V_{i} = \calO(2) \,,
\end{equation}
since it will hold for the Newtonian potentials $V$ and $V_{i}$ regardless of
the equations of motion. According to Eqs.~\eqref{instpotentials}
and~\eqref{hgen}, in order to recover the ``particular'' solution, one should
integrate the latter wave equations by means of the operator of symmetric
potentials $\mathcal{I}^{-1}$, and in principle, one should implement the
calculation by means of a factor $(r/r'_0)^B$ to cure possible IR
divergencies. But at this relatively low level $\mathcal{O}(6,5,6)$ we find
that the IR regulator is in fact not necessary, and we can use the usual
symmetric propagator $\Delta^{-1}+c^{-2}\partial_t^2\Delta^{-2}+\cdots$. The
matter source terms are defined by
\begin{equation}\label{sourcedensity}
\sigma = 2\frac{(d-2)T^{00} + T^{ii}}{(d-1)c^2}\,,\qquad\sigma_i =
\frac{T^{0i}}{c}\,,\qquad\sigma_{ij} = T^{ij}\,,
\end{equation} 
from the components of the stress-energy tensor of the point
particles,
\begin{equation}\label{Tmunu}
T^{\mu\nu} = \sum_A \frac{m_A \,v_A^\mu
  v_A^\nu}{\sqrt{-(g_{\rho\sigma})_A \,v_A^{\rho}v_A^{\sigma}/c^{2}}}
\frac{\delta^{(d)}(\mathbf{x}-\bm{y}_A)}{\sqrt{-g}}\,.
\end{equation}
Finally the constant $G$ is related to Newton's constant $G_\text{N}$
in three dimensions by 
\begin{equation}\label{G}
G = G_\text{N}\,\ell_0^{d-3}\,,
\end{equation}
where $\ell_0$ denotes the characteristic length scale associated with
dimensional regularization.

\subsection{Implementation of the calculation}
\label{sec:implem}

Having determined in~\eqref{metricpot}--\eqref{defpotentials} the
metric components for insertion into the Fokker
action~\eqref{Sg}--\eqref{Sm}, we tackle the difficult (and very
lengthy) calculation of all the spatial integrals in the gravitational
part $S_g$ of the action.\footnote{Extensive use is made of the
  algebraic software Mathematica together with the tensor package
  \textit{xAct}~\cite{xtensor}.} To reach the 4PN precision we must
include non-linear terms in the action up to the sixth non-linear
level, say
\begin{equation}\label{Lgsim}
\mathcal{L}_g \sim c^4\Bigl[\overline{h}\Box \overline{h} +
  \overline{h} \partial \overline{h}\partial \overline{h} + \cdots +
  \overline{h} \overline{h} \overline{h} \overline{h}\partial
  \overline{h}\partial \overline{h}\Bigr] +
\calO\left(10\right)\,.
\end{equation}
The matter part $S_m$ of the action is much simpler and will not be
discussed.

Following previous works on the 3PN equations of motion~\cite{BFeom,
  BDE04} we shall proceed in several steps. The
potentials~\eqref{defpotentials} are first computed for any point
in 3-dimensional space and then inserted into the gravitational
part of the action. The computation of potentials extensively uses the
famous function $g=\ln(r_1+r_2+r_{12})$, solution of the elementary
Poisson equation $\Delta g = r_1^{-1}r_2^{-1}$~\cite{Fock}, which
permits one to deal with quadratic source terms of type ${\sim\partial
V\partial V}$. One needs also to integrate a cubic source term $\sim
\hat{W}\partial^2 V$ and for that we use more complicated elementary
solutions given by Eqs.~(6.3)--(6.5) in~\cite{BFP98}.

The integration is then implemented by means of the Hadamard
regularization (HR)\footnote{Or, more precisely, the so-called
  pure-Hadamard-Schwartz regularization~\cite{BDE04}. See~\cite{BFreg}
  for precise definitions of various concepts of Hadamard's partie
  finie.} to treat the UV divergencies associated with point
particles. We thus compute with HR the spatial integral (with
non-compact support) of the terms in~\eqref{Lgsim}, say some
generic function $F(\mathbf{x})$ resulting from the PN iteration
performed in 3 dimensions,
\begin{equation}\label{intF}
I = \mathop{\text{Pf}}_{s_1, s_2} \int
\ud^3\mathbf{x}\,F(\mathbf{x})\,.
\end{equation}
The function $F$ is singular at the two points $\bm{y}_1$ and
$\bm{y}_2$, and the Hadamard partie finie Pf depends on two UV scales
denoted $s_A$. We assume that the integral extends on some finite
volume surrounding the singularities so that we do not include
  the IR regulator $r^B$ at this stage (see below for discussion of
the IR divergencies). The HR is simple and very convenient for
practical calculations but is unfortunately plagued with ambiguities
starting at the 3PN order. Therefore, in a second step, we shall
correct for the possible ambiguities of HR by adding to the HR result
the \textit{difference} ``DR$-$HR'' between the corresponding result
of the more powerful \textit{dimensional} regularization
(DR)~\cite{tHooft, Bollini} and the one of HR. While at 3PN and 4PN
orders the HR result contains logarithmic divergences yielding
ambiguities, the DR result gives some simple poles, \textit{i.e.},
$\propto 1/\varepsilon$ where $\varepsilon=d-3$. The poles are
followed by a finite part $\propto\varepsilon^0$ which is free of
ambiguities, and all terms of order $\mathcal{O}(\varepsilon)$ are
neglected.

We perform the similar PN iteration in $d$ dimensions to obtain a
generic non-compact $d$ dimensional integral of some function
$F^{(d)}(\mathbf{x})$, say
\begin{equation}\label{intFd}
I^{(d)} = \int \ud^d\mathbf{x}\,F^{(d)}(\mathbf{x})\,.
\end{equation}
When $r_1\rightarrow 0$ the function $F^{(d)}$ admits a singular
expansion more complicated than in 3 dimensions, as it involves
complex powers of $r_1$ of the type $p+q\varepsilon$ (instead of
  merely $p$), 
\begin{equation}\label{Fexpd}
  F^{(d)}(\mathbf{x}) = \sum_{p, q} r_1^{p+q\varepsilon}
  \mathop{f}_1{}_{p,q}^{(\varepsilon)}(\mathbf{n}_1) + o(r_1^{N})\,,
\end{equation}
where $p$ and $q$ are relative integers whose values are limited by
some $p_0 \leqslant p \leqslant N$ and $q_0 \leqslant q\leqslant q_1$
(with $p_0, q_0, q_1\in\mathbb{Z}$). The coefficients
$\mathop{f}_1{}_{p,q}^{\!\!(\varepsilon)}$ depend on the direction of
approach to the singularity $\mathbf{n}_1=(\mathbf{x}-\bm{y}_1)/r_1$,
and are linked to their counterparts $\mathop{f}_1{}_{\!p}$ associated
with the function $F$ in 3 dimensions by
\begin{equation}\label{constr}
  \sum_{q=q_0}^{q_1}\mathop{f}_1{}_{p,q}^{(0)} (\mathbf{n}_1) =
  \mathop{f}_1{}_{\!p}(\mathbf{n}_1)\,.
\end{equation}
One can show that at the 4PN order the functions $F^{(d)}$ have no
poles as $\varepsilon \rightarrow 0$, so the limit $\varepsilon=0$
in~\eqref{constr} is well defined.

The point is that the difference DR$-$HR can be computed purely
\textit{locally}, \textit{i.e.}, in the vicinity of the two particles,
as it is entirely determined, in the limit $\varepsilon\to 0$, by the
coefficients $\mathop{f}_1{}_{p,q}^{\!\!(\varepsilon)}$ of the local
expansion of the function $F^{(d)}$. This is clear because the parts
of the integrals outside the singularities cancel out in the
difference when $\varepsilon\to 0$. Denoting such difference by
$\mathcal{D}I=I^{(d)}-I$, for any of the non-compact support
integrals composing the gravitational action, we have the basic
formula
\begin{equation}\label{Dres}
\mathcal{D}I = \frac{1}{\varepsilon}\sum_{q=q_0}^{q_1}
\left[\frac{1}{q+1}+\varepsilon \ln s_1\right]
\bigl<\mathop{f}_1{}_{-3,q}^{(\varepsilon)}\bigr>_{2+\varepsilon} +
1\leftrightarrow 2 + \mathcal{O}(\varepsilon)\,.
\end{equation}
Here $1\leftrightarrow 2$ is the particle label permutation, $s_1$ and
$s_2$ are the HR scales in~\eqref{intF}, the
$\mathcal{O}(\varepsilon)$ remainder is neglected, and the spherical
angular integrals read
\begin{equation}\label{averaged}
\bigl< f \bigr>_{d-1} = \int
\ud\Omega_{d-1}(\mathbf{n}_1)\,f(\mathbf{n}_1)\,,
\end{equation}
with $\ud\Omega_{d-1}$ being the usual differential surface element in
$d-1$ dimensions. Notice the sum ranging over the integer $q$
in~\eqref{Dres} and the problematic case $q=-1$. An important test of
the calculation (and more generally of the adequacy of DR to treat the
classical problem of point particles in GR), is that the spherical
integral~\eqref{averaged} is always zero in the case of the offending
value $q=-1$.

The potentials~\eqref{defpotentials} in $d$ dimensions are in
principle computed with $d$-dimensional generalizations of the
elementary solutions used in HR, notably the function $g^{(d)}$ which
generalizes the function $g=\ln(r_1+r_2+r_{12})$. This function is
known in explicit closed form for any $d$ (see the Appendix C
in~\cite{BDE04}). Here we need only its local expansion when $r_1\to
0$. In practice, the local expansion of a potential is obtained by
integrating term by term the local expansion of its source, and adding
the appropriate homogeneous solution. We obtain in
Appendix~\ref{app:g} the local expansion of the function $g^{(d)}$. We
have checked that at the 4PN order we do not need to consider the
$d$-dimensional generalizations of the elementary
solutions~(6.3)--(6.5) in~\cite{BFP98}. We also found that the final
4PN results are unchanged if we add to the potentials some arbitrary
homogeneous solutions at order $\varepsilon$, provided that the
harmonic coordinate conditions~\eqref{diffident} for the potentials
remain satisfied when the potentials are ``on-shell''.

Once the HR calculation has been completed and the difference ``DR$-$HR''
added,\footnote{Notice that the scales $s_A$ cancel out in the final DR
  result.} the next step consists in renormalizing the result by absorbing the
poles $\propto 1/\varepsilon$ into appropriate shifts of the trajectories of
the particles. There is a lot of freedom for such shifts. Here we adopt some
non-minimal prescription in order to recover the earlier shifts at 3PN
order~\cite{BDE04}, which yielded precisely the 3PN harmonic coordinate
equations of motion in~\cite{BFeom}. The latter 3PN equations of motion depend
on two gauge constants $r'_1$ and $r'_2$ that we therefore introduce into the
shifts in replacement of the characteristic DR length scale $\ell_0$. For
convenience we extend this prescription to 4PN order in the simplest way, so
that $\ell_0$ disappears from the Lagrangian and the logarithmic terms (in
harmonic coordinates) are only of the form $\ln(r_{12}/r'_1)$ or
$\ln(r_{12}/r'_2)$, where $r_{12}$ is the separation between particles, and
are symmetric under $1\leftrightarrow 2$ exchange. Our 4PN shifts read then
\begin{subequations}\label{xi1text}
\begin{align}
  \bm{\xi}_1 &= \frac{11}{3}\frac{G_\text{N}^2\,m_1^2}{c^6} \left[
    \frac{1}{\varepsilon}-2\ln\left(
    \frac{\overline{q}^{1/2}r'_1}{\ell_0}\right)
    -\frac{327}{1540}\right] \bm{a}^{(d)}_{1,\,\mathrm{N}} +
  \frac{1}{c^8}\bm{\xi}_{1,\,\mathrm{4PN}}\,,\\ \bm{\xi}_2 &= 1
  \leftrightarrow 2\,.
\end{align}\end{subequations}
At 3PN order we recognize the shift given by Eq.~(1.13)
in~\cite{BDE04}, where $\ell_0$ is defined by~\eqref{G}, $\overline{q}
= 4\pi e^{\gamma_\text{E}}$ depends on Euler's constant
$\gamma_\text{E}\simeq 0.577$, and $\bm{a}^{(d)}_{1,\,\mathrm{N}}$ is
the Newtonian acceleration of the particle 1 in $d$ dimensions. The
complete expression of the shift at 4PN order is given in
Appendix~\ref{app:shift}. After applying the shifts~\eqref{xi1text}
the poles $\propto 1/\varepsilon$ cancel out and the result is UV
finite.

We also found that our ``brute'' Lagrangian depends on the individual
positions $\bm{y}_A$ of the particles. Such dependence is pure gauge
and we removed it by including appropriate terms in the shift, so that the
shifted Lagrangian depends only on the relative position $\bm{y}_{12}$
and is manifestly translation invariant. More generally we found that our initial
Lagrangian is not manifestly Poincar\'e invariant, but that we can
adjust the shift~\eqref{xi1text} so that it becomes Poincar\'e
invariant \textit{in a manifest way} (modulo a total time
derivative). The (global) Lorentz-Poincar\'e invariance is a very
satisfying property of our final 4PN dynamics.

Finally we discuss the very important problem of IR divergencies,
which appear specifically at the 4PN order. As we see in the complete
formula for the 4PN shift [Eq.~\eqref{shift4PN} in
Appendix~\ref{app:shift}], besides the UV logarithms $\ln(r_{12}/r'_1)$
and $\ln(r_{12}/r'_2)$, there are also some logarithms
$\ln(r_{12}/r_0)$ at the 4PN order, where $r_0$ was introduced in the
gravitational part of the action [see Eq.~\eqref{Lg4PN}] as an IR
cut-off dealing with the divergences of three-dimensional volume
integrals such as~\eqref{intF}, caused by the PN expansion
$\overline{h}$ diverging at infinity. The fact that the constant $r_0$
can be completely removed from the calculation by applying the
shift~\eqref{xi1text} constitutes a very important test of the
calculation. This is made possible by the presence of the 4PN
non-local tails~[\eqref{Stailsym} or \eqref{StailHad}]. To see that, we
rewrite the logarithmic kernel in the tail integrals (containing the
constant $s_0$) as
\begin{equation}\label{logkernel}
\ln\left(\frac{c\tau}{2s_0}\right) =
\ln\left(\frac{c\tau}{2r_{12}}\right) +
\ln\left(\frac{r_{12}}{r_0}\right) + \alpha\,,
\end{equation}
where $\alpha$ links $s_0$ to $r_0$ and is defined by~\eqref{s0}. We
find that the second term of~\eqref{logkernel} combines with the IR
divergences of the 3-dimensional volume integrals to exactly produce a
term removable by a shift, hence the $\ln(r_{12}/r_0)$ contributions
in~\eqref{shift4PN}. With the first term in~\eqref{logkernel} we shall
rewrite the tail integrals using the separation
$r_{12}$,\footnote{Specifically, our choice is to insert $r_{12}$ into
  Eq.~\eqref{StailHad} of the tail term, \textit{i.e.}, after
  integrations by parts.} being careful that $r_{12}$ is no longer a
constant and will have to be varied and participate in the
dynamics. Finally our end result will not only be UV finite but also
IR finite.

The constant $\alpha$ which remains is the analogue of the constant
$C$ in~\cite{DJS14}. It does not seem possible to determine its value
within the present method. Like in~\cite{DJS14} we shall compute it by
comparison with self-force calculations, which have determined the 4PN
term in the conserved energy for circular
orbits~\cite{BDLW10b,LBW12,BiniD13}. Let us check that $\alpha$ is a
pure numerical constant, \textit{i.e.}, does not depend on the
masses. Since $\alpha$ is dimensionless and is necessarily a symmetric
function of the two masses $m_1$ and $m_2$, it can only depend on the
symmetric mass ratio $\nu=m_1m_2/(m_1+m_2)^2$. Thus, we can write very
generally (with a finite or infinite sum)
\begin{equation}
\alpha = \sum_{n} \alpha_n \,\nu^n\,.
\end{equation}
In the Lagrangian $\alpha$ is in factor of $\sim(I_{ij}^{(3)})^2$. We
derive the corresponding terms in the acceleration of the particles
and look at the mass dependence of these terms. Imposing that the
acceleration should be a polynomial in the two masses
separately,\footnote{This can be justified from a diagrammatic
  expansion of the $N$-body problem based on the post-Minkowskian
  approximation~\cite{Dgef96}.} we find that the only admissible case
is indeed a pure constant $\alpha=\alpha_0$. Finally we adjust
$\alpha$ so that the conserved energy for circular orbits (that we
shall compute in the sequel paper~\cite{BBBFM15b}, see also
Sec.~\ref{sec:Ehamiltonian}) agrees with self-force calculations in
the small mass ratio limit --- see the coefficient of $\nu$ at 4PN
order in Eq.~(5.5) of~\cite{DJS14}. Anticipating the result we find
\begin{equation}\label{valeuralpha}
\alpha = \frac{811}{672}\,.
\end{equation}

\section{Lagrangian of compact binaries at the 4PN order}
\label{sec:lagrangian}

\subsection{Result in harmonic coordinates}
\label{sec:res}

The Lagrangian in harmonic coordinates at the 4PN order will be a
generalized one, depending on the positions of particles $\bm{y}_A$
and velocities $\bm{v}_A=\ud \bm{y}_A/\ud t$, and also accelerations
$\bm{a}_A=\ud \bm{v}_A/\ud t$, derivatives of accelerations
$\bm{b}_A$, and so on. However, by adding suitable double-zero or
multi-zero terms~\cite{DS85} we have removed all terms that are
non-linear in accelerations and derivatives of
accelerations. Furthermore, by adding suitable total time derivatives
we have eliminated the dependence on derivatives of
accelerations. Note that the process is iterative, since the latter
time derivatives reintroduce some terms non-linear in accelerations,
that need to be removed by further double-zeros. Thus the generalized
4PN harmonic-coordinate Lagrangian depends on $\bm{y}_A$ and
$\bm{v}_A$, and is linear in accelerations $\bm{a}_A$.\footnote{An
  exception is the 4PN tail piece which will be left as a functional
  of $\bm{y}_A$, $\bm{v}_A$, $\bm{a}_A$ and $\bm{b}_A$.} As said in
Sec.~\ref{sec:implem}, after a suitable shift the Lagrangian depends
only on the relative separation $r_{12}=\vert\bm{y}_1-\bm{y}_2\vert$
between the particles in harmonic coordinates. We denote the
corresponding unit direction by
$\bm{n}_{12}=(\bm{y}_1-\bm{y}_2)/r_{12}$. We systematically use
parenthesis to denote ordinary scalar products, \textit{e.g.}
$(n_{12}v_1)=\bm{n}_{12}\cdot\bm{v}_1$ and
$(a_1v_2)=\bm{a}_1\cdot\bm{v}_2$.

We write the full 4PN Lagrangian in the form
\begin{equation}\label{Lstruct}
L = L_\text{N} + \frac{1}{c^2}L_\text{1PN} + \frac{1}{c^4}L_\text{2PN}
+ \frac{1}{c^6}L_\text{3PN} + \frac{1}{c^8}L_\text{4PN} +
\mathcal{O}\left(10\right)\,,
\end{equation}
where the pieces up to 3PN order are known from previous works (see
\textit{e.g.}~\cite{Bliving14}) as
{\allowdisplaybreaks
\begin{subequations}\label{L3PN}
\begin{align}
L_\text{N} &= \frac{G m_1 m_2}{2 r_{12}} + \frac{m_1 v_1^2}{2} + 1
\leftrightarrow 2\,,\\
L_\text{1PN} &= - \frac{G^2 m_1^2 m_2}{2 r_{12}^2} + \frac{m_1
  v_1^4}{8} \nonumber \\ & + \frac{G m_1 m_2}{r_{12}} \left( -
\frac{1}{4} (n_{12}v_1) (n_{12}v_2) + \frac{3}{2} v_1^2 - \frac{7}{4}
(v_1v_2) \right) + 1 \leftrightarrow 2\,,\\
L_\text{2PN} &= \frac{G^3 m_1^3 m_2}{2 r_{12}^3} + \frac{19 G^3 m_1^2
  m_2^2}{8 r_{12}^3} \nonumber \\ &  + \frac{G^2 m_1^2
  m_2}{r_{12}^2} \left( \frac{7}{2} (n_{12}v_1)^2 - \frac{7}{2}
(n_{12}v_1) (n_{12}v_2) + \frac{1}{2}(n_{12}v_2)^2 + \frac{1}{4} v_1^2
- \frac{7}{4} (v_1v_2) + \frac{7}{4} v_2^2 \right) \nonumber \\ &
 + \frac{G m_1 m_2}{r_{12}} \bigg( \frac{3}{16}
(n_{12}v_1)^2 (n_{12}v_2)^2 - \frac{7}{8} (n_{12}v_2)^2 v_1^2 +
\frac{7}{8} v_1^4 + \frac{3}{4} (n_{12}v_1) (n_{12}v_2) (v_1v_2)
\nonumber \\ &     - 2 v_1^2 (v_1v_2) +
\frac{1}{8} (v_1v_2)^2 + \frac{15}{16} v_1^2 v_2^2 \bigg) + \frac{m_1
  v_1^6}{16} \nonumber \\ &  + G m_1 m_2 \left( -
\frac{7}{4} (a_1 v_2) (n_{12}v_2) - \frac{1}{8} (n_{12} a_1)
(n_{12}v_2)^2 + \frac{7}{8} (n_{12} a_1) v_2^2 \right) + 1
\leftrightarrow 2\,,\\
L_\text{3PN} &= \frac{G^2 m_1^2 m_2}{r_{12}^2} \bigg( \frac{13}{18}
(n_{12}v_1)^4 + \frac{83}{18} (n_{12}v_1)^3 (n_{12}v_2) - \frac{35}{6}
(n_{12}v_1)^2 (n_{12}v_2)^2 - \frac{245}{24} (n_{12}v_1)^2 v_1^2
\nonumber \\ & + \frac{179}{12} (n_{12}v_1) (n_{12}v_2) v_1^2 -
\frac{235}{24} (n_{12}v_2)^2 v_1^2 + \frac{373}{48} v_1^4 +
\frac{529}{24} (n_{12}v_1)^2 (v_1v_2) \nonumber \\ & - \frac{97}{6}
(n_{12}v_1) (n_{12}v_2) (v_1v_2) - \frac{719}{24} v_1^2 (v_1v_2) +
\frac{463}{24} (v_1v_2)^2 - \frac{7}{24} (n_{12}v_1)^2 v_2^2 \nonumber
\\ & - \frac{1}{2} (n_{12}v_1) (n_{12}v_2) v_2^2 + \frac{1}{4}
(n_{12}v_2)^2 v_2^2 + \frac{463}{48} v_1^2 v_2^2 - \frac{19}{2}
(v_1v_2) v_2^2 + \frac{45}{16} v_2^4 \bigg) \nonumber \\ & + G m_1 m_2
\bigg(\frac{3}{8} (a_1 v_2) (n_{12}v_1) (n_{12}v_2)^2 + \frac{5}{12}
(a_1 v_2) (n_{12}v_2)^3 + \frac{1}{8} (n_{12} a_1) (n_{12}v_1)
(n_{12}v_2)^3 \nonumber \\ & + \frac{1}{16} (n_{12} a_1) (n_{12}v_2)^4
+ \frac{11}{4} (a_1 v_1) (n_{12}v_2) v_1^2 - (a_1 v_2) (n_{12}v_2)
v_1^2 \nonumber \\ & - 2 (a_1 v_1) (n_{12}v_2) (v_1v_2) + \frac{1}{4}
(a_1 v_2) (n_{12}v_2) (v_1v_2) \nonumber \\ & + \frac{3}{8} (n_{12}
a_1) (n_{12}v_2)^2 (v_1v_2) - \frac{5}{8} (n_{12} a_1) (n_{12}v_1)^2
v_2^2 + \frac{15}{8} (a_1 v_1) (n_{12}v_2) v_2^2 \nonumber \\ & -
\frac{15}{8} (a_1 v_2) (n_{12}v_2) v_2^2 - \frac{1}{2} (n_{12} a_1)
(n_{12}v_1) (n_{12}v_2) v_2^2 \nonumber \\ & - \frac{5}{16} (n_{12}
a_1) (n_{12}v_2)^2 v_2^2 \bigg) + \frac{5m_1 v_1^8}{128} \nonumber
\\ & + \frac{G^2 m_1^2 m_2}{r_{12}} \bigg( - \frac{235}{24} (a_2 v_1)
(n_{12}v_1) - \frac{29}{24} (n_{12} a_2) (n_{12}v_1)^2 -
\frac{235}{24} (a_1 v_2) (n_{12}v_2) \nonumber \\ & - \frac{17}{6}
(n_{12} a_1) (n_{12}v_2)^2 + \frac{185}{16} (n_{12} a_1) v_1^2 -
\frac{235}{48} (n_{12} a_2) v_1^2 \nonumber \\ & - \frac{185}{8}
(n_{12} a_1) (v_1v_2) + \frac{20}{3} (n_{12} a_1) v_2^2 \bigg)
\nonumber \\ & + \frac{G m_1 m_2}{r_{12}} \bigg( - \frac{5}{32}
(n_{12}v_1)^3 (n_{12}v_2)^3 + \frac{1}{8} (n_{12}v_1) (n_{12}v_2)^3
v_1^2 + \frac{5}{8} (n_{12}v_2)^4 v_1^2 \nonumber \\ & - \frac{11}{16}
(n_{12}v_1) (n_{12}v_2) v_1^4 + \frac{1}{4} (n_{12}v_2)^2 v_1^4 +
\frac{11}{16} v_1^6 \nonumber \\ & - \frac{15}{32} (n_{12}v_1)^2
(n_{12}v_2)^2 (v_1v_2) + (n_{12}v_1) (n_{12}v_2) v_1^2 (v_1v_2)
\nonumber \\ & + \frac{3}{8} (n_{12}v_2)^2 v_1^2 (v_1v_2) -
\frac{13}{16} v_1^4 (v_1v_2) + \frac{5}{16} (n_{12}v_1) (n_{12}v_2)
(v_1v_2)^2 \nonumber \\ & + \frac{1}{16} (v_1v_2)^3 - \frac{5}{8}
(n_{12}v_1)^2 v_1^2 v_2^2 - \frac{23}{32} (n_{12}v_1) (n_{12}v_2)
v_1^2 v_2^2 + \frac{1}{16} v_1^4 v_2^2 \nonumber \\ & - \frac{1}{32}
v_1^2 (v_1v_2) v_2^2 \bigg) \nonumber \\ & - \frac{3 G^4 m_1^4 m_2}{8
  r_{12}^4} + \frac{G^4 m_1^3 m_2^2}{r_{12}^4} \left( -
\frac{9707}{420} + \frac{22}{3} \ln \left(\frac{r_{12}}{r'_1} \right)
\right) \nonumber \\ & + \frac{G^3 m_1^2 m_2^2}{r_{12}^3} \bigg(
\frac{383}{24} (n_{12}v_1)^2 - \frac{889}{48} (n_{12}v_1) (n_{12}v_2)
\nonumber \\ & - \frac{123}{64} (n_{12}v_1)(n_{12}v_{12}) \pi^2 -
\frac{305}{72} v_1^2 + \frac{41}{64} \pi^2 (v_1v_{12}) +
\frac{439}{144} (v_1v_2) \bigg) \nonumber \\ & + \frac{G^3 m_1^3
  m_2}{r_{12}^3} \bigg( - \frac{8243}{210} (n_{12}v_1)^2 +
\frac{15541}{420} (n_{12}v_1) (n_{12}v_2) + \frac{3}{2} (n_{12}v_2)^2
+ \frac{15611}{1260} v_1^2 \nonumber \\ & - \frac{17501}{1260}
(v_1v_2) + \frac{5}{4} v_2^2 + 22 (n_{12}v_1)(n_{12}v_{12}) \ln \left(
\frac{r_{12}}{r'_1} \right) \nonumber \\ & - \frac{22}{3} (v_1v_{12})
\ln \left( \frac{r_{12}}{r'_1} \right) \bigg) + 1 \leftrightarrow 2\,.
\end{align}
\end{subequations}}\noindent
with $v_{12}^i=v_1^i-v_2^i$.
 
Next we present the 4PN term. As we have discussed this term is the
sum of an instantaneous contribution and a non-local tail piece, say
\begin{equation}\label{L4PNdecomp}
L_\text{4PN} = L^\text{inst}_\text{4PN} + L^\text{tail}_\text{4PN}\,.
\end{equation}
The tail piece has been found in
Eqs.~\eqref{Stailsym}--\eqref{StailHad}, but here we have to replace
the Hadamard partie-finie scale $s_0$ therein with the particle
separation $r_{12}$. Specifically, we insert $r_{12}$ into the
form~\eqref{StailHad} of the action (after integrations by parts), so
the Lagrangian reads
\begin{eqnarray}
L^\text{tail}_\text{4PN} &=& \frac{G^2M}{5}
\,I_{ij}^{(3)}(t)\!\mathop{\text{Pf}}_{2r_{12}/c}
\int_{-\infty}^{+\infty} \frac{\ud t'}{\vert t-t'\vert}
I_{ij}^{(3)}(t') \nonumber\\ &=& \frac{G^2M}{5} \,I_{ij}^{(3)}(t)
\int_0^{+\infty} \ud\tau
\ln\left(\frac{c\tau}{2r_{12}}\right)\left[I_{ij}^{(4)}(t-\tau) -
  I_{ij}^{(4)}(t+\tau)\right]\,.\label{Ltail4PN}
\end{eqnarray}
Again, note that when varying the Lagrangian we shall have to take
into account the variation of the ``constant''
$r_{12}=|\bm{y}_1(t)-\bm{y}_2(t)|$ in~\eqref{Ltail4PN}. The Lagrangian
is defined up to a total time derivative, and with the choice made
in~\eqref{Ltail4PN}, the tail term is a functional of
$\bm{y}_A$, $\bm{v}_A$, accelerations $\bm{a}_A$ and also derivatives
of accelerations $\bm{b}_A=\ud\bm{a}_A/\ud t$. Splitting for
convenience the very long instantaneous contribution according to
powers of $G$ as
\begin{equation}\label{split4PN}
L^\text{inst}_\mathrm{4PN} = L^{(0)}_\mathrm{4PN} +
G\,L^{(1)}_\mathrm{4PN} + G^2\,L^{(2)}_\mathrm{4PN} +
G^3\,L^{(3)}_\mathrm{4PN} + G^4\,L^{(4)}_\mathrm{4PN} +
G^5\,L^{(5)}_\mathrm{4PN}\,,
\end{equation}
we find
{\allowdisplaybreaks
\begin{subequations}\label{result4PN}
\begin{align}
L_\text{4PN}^{(0)}&={}\frac{7}{256} m_{1} v_1^{10} + 1 \leftrightarrow
2\,,\\[1ex]
L_\text{4PN}^{(1)}&={}m_{1} m_{2} \Biggl[\frac{5}{128} (a_2 n_{12})
  (n_{12} v_1)^6 + (n_{12} v_1)^5 \biggl\{- \frac{13}{64} (a_2 v_2) +
  \frac{5}{64} (a_2 n_{12}) (n_{12} v_2)\biggr\}\nonumber\\ & +
  \frac{33}{16} (a_1 v_1) (n_{12} v_2) (v_1 v_2)^2 + \frac{9}{16} (a_1
  n_{12}) (v_1 v_2)^3 + (a_2 n_{12}) (n_{12} v_1)^4
  \biggl\{\frac{11}{64} (v_1 v_2)\nonumber\\ & - \frac{27}{128}
  v_1^{2}\biggr\} + (a_2 n_{12}) \biggl\{(v_1 v_2)^2 v_1^{2} +
  \frac{49}{64} (v_1 v_2) v_1^{4} - \frac{75}{128}
  v_1^{6}\biggr\}\nonumber\\ & + (n_{12} v_1)^3 \biggl\{\frac{3}{32}
  (a_1 v_2) (n_{12} v_2)^2 - \frac{5}{32} (a_1 n_{12}) (n_{12} v_2)^3
  - \frac{2}{3} (a_2 v_2) (v_1 v_2)\nonumber\\ & + \frac{77}{96} (a_2
  v_2) v_1^{2} + (a_2 n_{12}) \Bigl[\frac{1}{2} (n_{12} v_2) (v_1 v_2)
    - \frac{11}{32} (n_{12} v_2) v_1^{2}\Bigr] - \frac{5}{32} (a_1
  v_2) v_2^{2}\nonumber\\ & + \frac{15}{32} (a_1 n_{12}) (n_{12} v_2)
  v_2^{2}\biggr\} + (n_{12} v_1)^2 \biggl\{\frac{9}{32} (a_1 v_1)
  (n_{12} v_2)^3\nonumber\\ & + \frac{9}{32} (a_1 n_{12}) (n_{12}
  v_2)^2 (v_1 v_2) + (a_2 n_{12}) \Bigl[- \frac{1}{2} (v_1 v_2)^2 -
    \frac{13}{32} (v_1 v_2) v_1^{2} + \frac{53}{128}
    v_1^{4}\Bigr]\nonumber\\ & - \frac{15}{32} (a_1 n_{12}) (v_1 v_2)
  v_2^{2} + (n_{12} v_2) \Bigl[- \frac{11}{16} (a_1 v_2) (v_1 v_2) -
    \frac{23}{32} (a_1 v_1) v_2^{2}\Bigr]\biggr\}\nonumber\\ & +
  v_1^{2} \biggl\{- \frac{19}{32} (a_1 v_1) (n_{12} v_2)^3 -
  \frac{9}{32} (a_1 n_{12}) (n_{12} v_2)^2 (v_1 v_2) + \frac{23}{32}
  (a_1 n_{12}) (v_1 v_2) v_2^{2}\nonumber\\ & + (n_{12} v_2)
  \Bigl[\frac{33}{16} (a_1 v_2) (v_1 v_2) + \frac{93}{32} (a_1 v_1)
    v_2^{2}\Bigr]\biggr\}\nonumber\\ & + (n_{12} v_1) \biggl\{-
  \frac{9}{16} (a_1 v_1) (n_{12} v_2)^2 (v_1 v_2) + \frac{27}{16} (a_1
  v_2) (v_1 v_2)^2\nonumber\\ & - \frac{11}{16} (a_1 n_{12}) (n_{12}
  v_2) (v_1 v_2)^2 - \frac{123}{64} (a_2 v_2) v_1^{4} + (a_2 n_{12})
  \Bigl[- (n_{12} v_2) (v_1 v_2) v_1^{2}\nonumber\\ & + \frac{31}{64}
    (n_{12} v_2) v_1^{4}\Bigr] + \frac{23}{16} (a_1 v_1) (v_1 v_2)
  v_2^{2} + v_1^{2} \Bigl[- \frac{9}{32} (a_1 v_2) (n_{12}
    v_2)^2\nonumber\\ & + \frac{9}{32} (a_1 n_{12}) (n_{12} v_2)^3 +
    \frac{7}{2} (a_2 v_2) (v_1 v_2) + \frac{23}{32} (a_1 v_2) v_2^{2}
    - \frac{23}{32} (a_1 n_{12}) (n_{12} v_2)
    v_2^{2}\Bigr]\biggr\}\nonumber\\ & + (a_2 v_1)
  \biggl\{\frac{13}{64} (n_{12} v_1)^5 + \frac{11}{64} (n_{12} v_1)^4
  (n_{12} v_2) + 2 (n_{12} v_2) (v_1 v_2) v_1^{2}\nonumber\\ & +
  (n_{12} v_1)^2 \Bigl[- (n_{12} v_2) (v_1 v_2) - \frac{13}{32}
    (n_{12} v_2) v_1^{2}\Bigr] + \frac{49}{64} (n_{12} v_2)
  v_1^{4}\nonumber\\ & + (n_{12} v_1)^3 \Bigl[\frac{1}{4} (n_{12}
    v_2)^2 - \frac{1}{16} (v_1 v_2) - \frac{77}{96} v_1^{2} -
    \frac{1}{3} v_2^{2}\Bigr] + (n_{12} v_1) \Bigl[\frac{3}{2} (v_1
    v_2)^2 + \frac{123}{64} v_1^{4}\nonumber\\ & + v_1^{2} \bigl(-
    \frac{1}{2} (n_{12} v_2)^2 - \frac{3}{16} (v_1 v_2) + \frac{7}{4}
    v_2^{2}\bigr)\Bigr]\biggr\} + \frac{1}{r_{12}} \biggl\{-
  \frac{17}{32} (v_1 v_2)^4 + \frac{75}{128} v_1^{8}\nonumber\\ & +
  v_1^{6} \Bigl[- \frac{11}{32} (n_{12} v_2)^2 - \frac{5}{4} (v_1 v_2)
    + \frac{5}{8} v_2^{2}\Bigr] + (n_{12} v_1)^5 \Bigl[\frac{35}{128}
    (n_{12} v_2)^3 - \frac{75}{128} (n_{12} v_2)
    v_2^{2}\Bigr]\nonumber\\ & + (n_{12} v_1)^4 \Bigl[- \frac{35}{256}
    (n_{12} v_2)^4 - \frac{75}{128} (n_{12} v_2)^2 (v_1 v_2) +
    \frac{75}{128} (v_1 v_2) v_2^{2}\Bigr]\nonumber\\ & + v_1^{2}
  \Bigl[\frac{37}{16} (n_{12} v_2)^2 (v_1 v_2)^2 + \frac{51}{32} (v_1
    v_2)^3 - \frac{7}{4} (v_1 v_2)^2 v_2^{2}\Bigr]\nonumber\\ & +
  (n_{12} v_1)^3 \Bigl[\frac{15}{16} (n_{12} v_2)^3 (v_1 v_2) +
    \frac{39}{32} (n_{12} v_2) (v_1 v_2)^2 + v_1^{2} \bigl(-
    \frac{55}{64} (n_{12} v_2)^3\nonumber\\ & + \frac{99}{64} (n_{12}
    v_2) v_2^{2}\bigr)\Bigr] + (n_{12} v_1)^2 \Bigl[- \frac{57}{32}
    (n_{12} v_2)^2 (v_1 v_2)^2 - \frac{29}{32} (v_1
    v_2)^3\nonumber\\ & + v_1^{2} \bigl(\frac{45}{64} (n_{12} v_2)^4 +
    (n_{12} v_2)^2 (\frac{99}{64} (v_1 v_2) - \frac{3}{4} v_2^{2}) -
    \frac{99}{64} (v_1 v_2) v_2^{2}\bigr)\Bigr]\nonumber\\ & + (n_{12}
  v_1) \Bigl[\frac{11}{8} (n_{12} v_2) (v_1 v_2)^3 + v_1^{4}
    \bigl(\frac{93}{128} (n_{12} v_2)^3 - \frac{185}{128} (n_{12} v_2)
    v_2^{2}\bigr)\nonumber\\ & + v_1^{2} \bigl(- \frac{21}{8} (n_{12}
    v_2)^3 (v_1 v_2) + (n_{12} v_2) (- \frac{73}{32} (v_1 v_2)^2 +
    \frac{19}{8} (v_1 v_2) v_2^{2})\bigr)\Bigr]\nonumber\\ & + v_1^{4}
  \Bigl[- \frac{15}{128} (n_{12} v_2)^4 + \frac{3}{4} (v_1 v_2)^2 +
    (n_{12} v_2)^2 \bigl(\frac{3}{128} (v_1 v_2) + \frac{23}{64}
    v_2^{2}\bigr) - \frac{7}{128} (v_1 v_2) v_2^{2}\nonumber\\ & +
    \frac{3}{256} v_2^{4}\Bigr]\biggr\}\Biggr] + 1 \leftrightarrow
2\,,\\[1ex]
L_\text{4PN}^{(2)}&={}m_{1}^2 m_{2} \Biggl[\frac{1}{r_{12}}
  \biggl\{\Bigl[- \frac{4247}{960} (a_1 n_{12}) - \frac{2}{3} (a_2
    n_{12})\Bigr] (n_{12} v_1)^4 + (n_{12} v_1)^3 \Bigl[-
    \frac{29}{12} (a_2 v_1)\nonumber\\ & - \frac{4501}{480} (a_1 v_2)
    + \frac{51}{8} (a_2 v_2) + \frac{519}{80} (a_1 n_{12}) (n_{12}
    v_2) - \frac{25}{6} (a_2 n_{12}) (n_{12} v_2)\Bigr]\nonumber\\ & -
  \frac{367}{10} (a_2 v_2) (n_{12} v_2) (v_1 v_2) + \Bigl[-
    \frac{13129}{480} (a_1 v_2) (n_{12} v_2) + \frac{437}{30} (a_2
    v_2) (n_{12} v_2)\Bigr] v_1^{2}\nonumber\\ & + (a_1 v_2)
  \Bigl[\frac{8653}{480} (n_{12} v_2)^3 + (n_{12} v_2)
    \bigl(\frac{8291}{240} (v_1 v_2) - \frac{6669}{160}
    v_2^{2}\bigr)\Bigr] + (a_2 v_1) \Bigl[\frac{42}{5} (n_{12}
    v_2)^3\nonumber\\ & - \frac{107}{12} (n_{12} v_2) v_1^{2} +
    (n_{12} v_2) \bigl(\frac{112}{15} (v_1 v_2) - \frac{367}{20}
    v_2^{2}\bigr)\Bigr] + (a_2 n_{12}) \Bigl[\frac{126}{5} (n_{12}
    v_2)^2 (v_1 v_2)\nonumber\\ & + \frac{56}{15} (v_1 v_2)^2 +
    \frac{19}{12} v_1^{4} - \frac{367}{20} (v_1 v_2) v_2^{2} + v_1^{2}
    \bigl(- \frac{47}{5} (n_{12} v_2)^2 - \frac{107}{12} (v_1 v_2) +
    \frac{437}{60} v_2^{2}\bigr)\Bigr]\nonumber\\ & + (n_{12} v_1)^2
  \Bigl[\frac{1}{2} (a_2 v_1) (n_{12} v_2) + \frac{10463}{480} (a_1
    v_2) (n_{12} v_2) - \frac{77}{15} (a_2 v_2) (n_{12}
    v_2)\nonumber\\ & + (a_2 n_{12}) \bigl(\frac{28}{5} (n_{12} v_2)^2
    + \frac{1}{2} (v_1 v_2) - \frac{89}{12} v_1^{2} - \frac{77}{30}
    v_2^{2}\bigr) + (a_1 n_{12}) \bigl(- \frac{9661}{480} (n_{12}
    v_2)^2\nonumber\\ & - \frac{94}{15} (v_1 v_2) + \frac{3017}{240}
    v_1^{2} + \frac{2177}{240} v_2^{2}\bigr)\Bigr] + (a_1 v_1)
  \Bigl[\frac{2507}{160} (n_{12} v_1)^3 - \frac{16183}{480} (n_{12}
    v_1)^2 (n_{12} v_2)\nonumber\\ & - \frac{2543}{160} (n_{12} v_2)^3
    + \frac{16589}{480} (n_{12} v_2) v_1^{2} + (n_{12} v_1)
    \bigl(\frac{6261}{160} (n_{12} v_2)^2 + \frac{5213}{80} (v_1
    v_2)\nonumber\\ & - \frac{16429}{480} v_1^{2} - \frac{5113}{160}
    v_2^{2}\bigr) + (n_{12} v_2) \bigl(- \frac{13129}{240} (v_1 v_2) +
    \frac{18191}{480} v_2^{2}\bigr)\Bigr]\nonumber\\ & + (n_{12} v_1)
  \Bigl[\bigl(\frac{7063}{160} (a_1 v_2) - \frac{583}{24} (a_2
    v_2)\bigr) v_1^{2} + (a_1 n_{12}) \bigl(\frac{2233}{240} (n_{12}
    v_2)^3 - \frac{489}{40} (n_{12} v_2) v_1^{2}\nonumber\\ & +
    (n_{12} v_2) (\frac{37}{60} (v_1 v_2) - \frac{56}{15}
    v_2^{2})\bigr) + (a_2 v_2) \bigl(\frac{127}{10} (n_{12} v_2)^2 +
    \frac{329}{30} (v_1 v_2) - \frac{83}{5} v_2^{2}\bigr)\nonumber\\ &
    + (a_2 v_1) \bigl(- \frac{77}{15} (n_{12} v_2)^2 + \frac{1}{4}
    (v_1 v_2) + \frac{299}{12} v_1^{2} + \frac{329}{60} v_2^{2}\bigr)
    + (a_1 v_2) \bigl(- \frac{13807}{480} (n_{12} v_2)^2\nonumber\\ &
    - \frac{4469}{240} (v_1 v_2) + \frac{9511}{480} v_2^{2}\bigr) +
    (a_2 n_{12}) \bigl(- \frac{184}{15} (n_{12} v_2)^3 + \frac{55}{3}
    (n_{12} v_2) v_1^{2}\nonumber\\ & + (n_{12} v_2) (- \frac{154}{15}
    (v_1 v_2) + \frac{127}{10} v_2^{2})\bigr)\Bigr] + (a_1 n_{12})
  \Bigl[- \frac{1811}{960} (n_{12} v_2)^4 + \frac{1341}{80} (v_1
    v_2)^2\nonumber\\ & + \frac{9143}{960} v_1^{4} + (n_{12} v_2)^2
    \bigl(\frac{31}{10} (v_1 v_2) - \frac{3}{5} v_2^{2}\bigr) -
    \frac{4277}{240} (v_1 v_2) v_2^{2} + v_1^{2} \bigl(\frac{109}{30}
    (n_{12} v_2)^2\nonumber\\ & - \frac{1213}{240} (v_1 v_2) +
    \frac{1363}{480} v_2^{2}\bigr) + \frac{1603}{320}
    v_2^{4}\Bigr]\biggr\} + \frac{1}{r_{12}^{2}} \biggl\{\frac{11}{40}
  (n_{12} v_1)^6 + \frac{109}{40} (n_{12} v_1)^5 (n_{12}
  v_2)\nonumber\\ & + \frac{110}{3} (n_{12} v_2)^4 (v_1 v_2) -
  \frac{527}{120} (v_1 v_2)^3 + \frac{33}{16} v_1^{6} + (n_{12} v_1)^4
  \Bigl[10 (n_{12} v_2)^2 + \frac{105}{16} (v_1 v_2)\nonumber\\ & -
    \frac{727}{48} v_1^{2} - \frac{65}{6} v_2^{2}\Bigr] + v_1^{4}
  \Bigl[\frac{37}{3} (n_{12} v_2)^2 + \frac{175}{16} (v_1 v_2) -
    \frac{287}{48} v_2^{2}\Bigr] - \frac{649}{60} (v_1 v_2)^2
  v_2^{2}\nonumber\\ & + (n_{12} v_1)^3 \Bigl[- \frac{237}{10} (n_{12}
    v_2)^3 + \frac{541}{12} (n_{12} v_2) v_1^{2} + (n_{12} v_2)
    \bigl(- \frac{91}{6} (v_1 v_2) + \frac{1691}{60}
    v_2^{2}\bigr)\Bigr]\nonumber\\ & + (n_{12} v_1) \Bigl[-
    \frac{92}{5} (n_{12} v_2)^5 - \frac{207}{8} (n_{12} v_2) v_1^{4} +
    v_1^{2} \bigl(\frac{794}{15} (n_{12} v_2)^3 + (n_{12} v_2)
    (\frac{617}{6} (v_1 v_2)\nonumber\\ & - \frac{2513}{40}
    v_2^{2})\bigr) + (n_{12} v_2)^3 \bigl(- \frac{1052}{15} (v_1 v_2)
    + \frac{113}{3} v_2^{2}\bigr) + (n_{12} v_2) \bigl(-
    \frac{1109}{60} (v_1 v_2)^2\nonumber\\ & + \frac{1144}{15} (v_1
    v_2) v_2^{2} - \frac{171}{8} v_2^{4}\bigr)\Bigr] + v_1^{2} \Bigl[-
    \frac{78}{5} (n_{12} v_2)^4 - \frac{293}{24} (v_1 v_2)^2 +
    \frac{2959}{240} (v_1 v_2) v_2^{2}\nonumber\\ & + (n_{12} v_2)^2
    \bigl(- \frac{623}{15} (v_1 v_2) + \frac{1819}{60} v_2^{2}\bigr) -
    \frac{1169}{240} v_2^{4}\Bigr] + (n_{12} v_2)^2
  \Bigl[\frac{189}{5} (v_1 v_2)^2\nonumber\\ & - \frac{273}{4} (v_1
    v_2) v_2^{2} + \frac{3}{16} v_2^{4}\Bigr] + \frac{75}{8} (v_1 v_2)
  v_2^{4} + (n_{12} v_1)^2 \Bigl[\frac{148}{5} (n_{12} v_2)^4 -
    \frac{3}{2} (v_1 v_2)^2\nonumber\\ & + \frac{231}{16} v_1^{4} +
    (n_{12} v_2)^2 \bigl(\frac{2063}{60} (v_1 v_2) - \frac{1253}{30}
    v_2^{2}\bigr) - \frac{3503}{240} (v_1 v_2) v_2^{2} + v_1^{2}
    \bigl(- \frac{883}{12} (n_{12} v_2)^2\nonumber\\ & -
    \frac{1009}{16} (v_1 v_2) + \frac{1693}{48} v_2^{2}\bigr) +
    \frac{989}{120} v_2^{4}\Bigr] + \frac{115}{32}
  v_2^{6}\biggr\}\Biggr] + 1 \leftrightarrow 2\,,\\[1ex]
L_\text{4PN}^{(3)}&={}\frac{m_1^3 m_2}{r_{12}^{2}}
\Biggl[\biggl\{\frac{2582267}{16800} (a_1 n_{12}) - \frac{89763}{1400}
  (a_2 n_{12})\biggr\} (n_{12} v_1)^2 +
  \biggl\{\frac{1111}{25200}\nonumber\\ & + \frac{110}{3}
  \ln\Bigl[\frac{r_{12}}{r'_{1}}\Bigr]\biggr\} (a_2 v_1) (n_{12} v_2)
  + \frac{487591}{25200} (a_1 v_2) (n_{12} v_2) - 6 (a_2 v_2) (n_{12}
  v_2)\nonumber\\ & + (a_1 v_1) \biggl\{- \frac{163037}{1200} (n_{12}
  v_1) + \Bigl[\frac{15929}{1400} - \frac{110}{3}
    \ln\bigl(\frac{r_{12}}{r'_{1}}\bigr)\Bigr] (n_{12}
  v_2)\biggr\}\nonumber\\ & + (n_{12} v_1) \biggl\{\frac{435011}{5040}
  (a_2 v_1) + \Bigl[\frac{31309}{560} + 44
    \ln\bigl(\frac{r_{12}}{r'_{1}}\bigr)\Bigr] (a_1 v_2) + \Bigl[-
    \frac{212641}{6300}\nonumber\\ & - 44
    \ln\bigl(\frac{r_{12}}{r'_{1}}\bigr)\Bigr] (a_2 v_2) -
  \frac{268169}{2100} (a_1 n_{12}) (n_{12} v_2) +
  \Bigl[\frac{5421}{700}\nonumber\\ & + 22
    \ln\bigl(\frac{r_{12}}{r'_{1}}\bigr)\Bigr] (a_2 n_{12}) (n_{12}
  v_2)\biggr\} + (a_1 n_{12}) \biggl\{\frac{27203}{1200} (n_{12}
  v_2)^2 + \Bigl[\frac{888179}{6300}\nonumber\\ & - 44
    \ln\bigl(\frac{r_{12}}{r'_{1}}\bigr)\Bigr] (v_1 v_2) + \Bigl[-
    \frac{1391897}{12600} + 44
    \ln\bigl(\frac{r_{12}}{r'_{1}}\bigr)\Bigr] v_1^{2} -
  \frac{89129}{2016} v_2^{2}\biggr\}\nonumber\\ & + (a_2 n_{12})
  \biggl\{\frac{51}{2} (n_{12} v_2)^2 + \Bigl[\frac{1111}{25200} +
    \frac{110}{3} \ln\bigl(\frac{r_{12}}{r'_{1}}\bigr)\Bigr] (v_1 v_2)
  + \Bigl[\frac{128867}{8400}\nonumber\\ & - \frac{110}{3}
    \ln\bigl(\frac{r_{12}}{r'_{1}}\bigr)\Bigr] v_1^{2} - 3
  v_2^{2}\biggr\}\Biggr] + \frac{m_1^3 m_2}{r_{12}^{3}} \Biggl[-
  \frac{906349}{3360} (n_{12} v_1)^4 + \frac{399851}{672} (n_{12}
  v_1)^3 (n_{12} v_2)\nonumber\\ & + \frac{85}{2} (n_{12} v_2)^4 +
  \biggl\{- \frac{34003}{525} + \frac{110}{3}
  \ln\Bigl[\frac{r_{12}}{r'_{1}}\Bigr]\biggr\} (v_1 v_2)^2 + \biggl\{-
  \frac{1195969}{16800}\nonumber\\ & + \frac{55}{3}
  \ln\Bigl[\frac{r_{12}}{r'_{1}}\Bigr]\biggr\} v_1^{4} + (n_{12}
  v_2)^2 \biggl\{\Bigl[- \frac{28403}{1680} + 99
    \ln\bigl(\frac{r_{12}}{r'_{1}}\bigr)\Bigr] (v_1 v_2) -
  \frac{131}{4} v_2^{2}\biggr\}\nonumber\\ & + \biggl\{-
  \frac{193229}{25200} - \frac{44}{3}
  \ln\Bigl[\frac{r_{12}}{r'_{1}}\Bigr]\biggr\} (v_1 v_2) v_2^{2} +
  v_1^{2} \biggl\{\Bigl[\frac{735527}{8400} - 99
    \ln\bigl(\frac{r_{12}}{r'_{1}}\bigr)\Bigr] (n_{12}
  v_2)^2\nonumber\\ & + \Bigl[\frac{7879619}{50400} - 55
    \ln\bigl(\frac{r_{12}}{r'_{1}}\bigr)\Bigr] (v_1 v_2) + \Bigl[-
    \frac{540983}{25200} + \frac{44}{3}
    \ln\bigl(\frac{r_{12}}{r'_{1}}\bigr)\Bigr]
  v_2^{2}\biggr\}\nonumber\\ & + (n_{12} v_1)^2 \biggl\{\Bigl[-
    \frac{46577}{140} - 55 \ln\bigl(\frac{r_{12}}{r'_{1}}\bigr)\Bigr]
  (n_{12} v_2)^2 - \frac{1160909}{2400} (v_1 v_2) +
  \Bigl[\frac{1732751}{4200}\nonumber\\ & - 55
    \ln\bigl(\frac{r_{12}}{r'_{1}}\bigr)\Bigr] v_1^{2} +
  \Bigl[\frac{77801}{840} + 66
    \ln\bigl(\frac{r_{12}}{r'_{1}}\bigr)\Bigr] v_2^{2}\biggr\} +
  (n_{12} v_1) \biggl\{\Bigl[- \frac{9559}{280}\nonumber\\ & + 55
    \ln\bigl(\frac{r_{12}}{r'_{1}}\bigr)\Bigr] (n_{12} v_2)^3 +
  \Bigl[- \frac{2617007}{5600} + 165
    \ln\bigl(\frac{r_{12}}{r'_{1}}\bigr)\Bigr] (n_{12} v_2) v_1^{2} +
  (n_{12} v_2) \Bigl[\bigl(\frac{65767}{150}\nonumber\\ & - 88
    \ln(\frac{r_{12}}{r'_{1}})\bigr) (v_1 v_2) + \bigl(-
    \frac{129667}{4200} - 88 \ln(\frac{r_{12}}{r'_{1}})\bigr)
    v_2^{2}\Bigr]\biggr\} + \frac{139}{16} v_2^{4}\Biggr]\nonumber\\ &
+ \frac{m_1^2 m_2^2}{r_{12}^{2}} \Bigl[\bigl((\frac{17811527}{33600} -
  \frac{8769}{512} \pi^2) (a_1 n_{12}) + (- \frac{12448339}{33600} +
  \frac{1017}{64} \pi^2) (a_2 n_{12})\bigr) (n_{12}
  v_1)^2\nonumber\\ & + (a_1 v_1) \bigl((- \frac{3168457}{10080} -
  \frac{2095}{256} \pi^2) (n_{12} v_1) + (\frac{11535007}{50400} +
  \frac{177}{64} \pi^2) (n_{12} v_2)\bigr)\nonumber\\ & + (n_{12} v_1)
  \bigl((\frac{12111653}{50400} + \frac{133}{8} \pi^2) (a_2 v_1) +
  (\frac{12496303}{50400} + \frac{1023}{64} \pi^2) (a_1 v_2) + (-
  \frac{4383363}{5600}\nonumber\\ & + \frac{2157}{64} \pi^2) (a_1
  n_{12}) (n_{12} v_2)\bigr) + \bigl(\frac{3213347}{100800} +
  \frac{55}{32} \pi^2\bigr) (a_2 n_{12}) v_1^{2} + (a_1 n_{12})
  \bigl((\frac{1263331}{16800}\nonumber\\ & + \frac{1107}{64} \pi^2)
  (v_1 v_2) - \frac{11}{4608} (29656 + 1989 \pi^2) v_1^{2}\bigr)\Bigr]
+ \frac{m_1^2 m_2^2}{r_{12}^{3}} \Bigl[\bigl(-
  \frac{465431}{480}\nonumber\\ & + \frac{27075}{1024} \pi^2\bigr)
  (n_{12} v_1)^4 + \bigl(\frac{10701209}{3360} - \frac{53445}{512}
  \pi^2\bigr) (n_{12} v_1)^3 (n_{12} v_2) + \bigl(-
  \frac{8248733}{50400}\nonumber\\ & - \frac{8379}{512} \pi^2\bigr)
  (v_1 v_2)^2 + (n_{12} v_1)^2 \bigl((- \frac{14873539}{6720} +
  \frac{79815}{1024} \pi^2) (n_{12} v_2)^2 + (-
  \frac{27374071}{16800}\nonumber\\ & - \frac{9033}{512} \pi^2) (v_1
  v_2) + (\frac{2079017}{2100} - \frac{2037}{512} \pi^2) v_1^{2}\bigr)
  + (n_{12} v_1) \bigl((\frac{1040673}{700}\nonumber\\ & +
  \frac{4587}{256} \pi^2) (n_{12} v_2) (v_1 v_2) + (-
  \frac{5303279}{3360} + \frac{7953}{512} \pi^2) (n_{12} v_2)
  v_1^{2}\bigr) + \bigl(- \frac{1177829}{10080}\nonumber\\ & -
  \frac{4057}{1024} \pi^2\bigr) v_1^{4} + v_1^{2}
  \bigl((\frac{12260653}{16800} - \frac{6057}{512} \pi^2) (n_{12}
  v_2)^2 + (\frac{17958959}{50400} + \frac{11049}{512} \pi^2) (v_1
  v_2)\nonumber\\ & + (- \frac{7672087}{100800} - \frac{1283}{1024}
  \pi^2) v_2^{2}\bigr)\Bigr] + 1 \leftrightarrow 2\,,\\[1ex]
L_\text{4PN}^{(4)}&={}\frac{1}{r_{12}^{3}} \biggl\{m_{1}^4 m_{2}
\Bigl[\frac{1691807}{25200} (a_1 n_{12}) - \frac{149}{6} (a_2
  n_{12})\Bigr] + m_{1}^3 m_{2}^2 \Bigl[\bigl(-
  \frac{2470667}{16800}\nonumber\\ & + \frac{1099}{96} \pi^2\bigr)
  (a_1 n_{12}) + \bigl(\frac{9246557}{50400} - \frac{555}{64}
  \pi^2\bigr) (a_2 n_{12})\Bigr]\biggr\} + \frac{1}{r_{12}^{4}}
\biggl\{m_{1}^4 m_{2} \Bigl[\bigl(\frac{2146}{75}\nonumber\\ & -
  \frac{880}{3} \ln(\frac{r_{12}}{r'_{1}})\bigr) (n_{12} v_1)^2 +
  \bigl(\frac{3461}{50} + \frac{880}{3}
  \ln(\frac{r_{12}}{r'_{1}})\bigr) (n_{12} v_1) (n_{12} v_2) -
  \frac{1165}{12} (n_{12} v_2)^2\nonumber\\ & + \bigl(-
  \frac{11479}{300} - \frac{220}{3} \ln(\frac{r_{12}}{r'_{1}})\bigr)
  (v_1 v_2) + \bigl(\frac{317}{25} + \frac{220}{3}
  \ln(\frac{r_{12}}{r'_{1}})\bigr) v_1^{2} + \frac{1237}{48}
  v_2^{2}\Bigr]\nonumber\\ & + m_{1}^3 m_{2}^2
\Bigl[\bigl(\frac{9102109}{16800} - \frac{3737}{96} \pi^2 -
  \frac{286}{3} \ln(\frac{r_{12}}{r'_{1}})\bigr) (n_{12} v_1)^2 +
  \bigl(- \frac{1409257}{1680} + \frac{179}{4} \pi^2\nonumber\\ & + 44
  \ln(\frac{r_{12}}{r'_{1}}) + 64 \ln(\frac{r_{12}}{r'_{2}})\bigr)
  (n_{12} v_1) (n_{12} v_2) + \bigl(\frac{5553521}{16800} -
  \frac{559}{96} \pi^2 + \frac{110}{3}
  \ln(\frac{r_{12}}{r'_{1}})\nonumber\\ & - 64
  \ln(\frac{r_{12}}{r'_{2}})\bigr) (n_{12} v_2)^2 +
  \bigl(\frac{1637809}{6300} - \frac{2627}{192} \pi^2 - \frac{154}{3}
  \ln(\frac{r_{12}}{r'_{1}}) - 16 \ln(\frac{r_{12}}{r'_{2}})\bigr)
  (v_1 v_2)\nonumber\\ & + \bigl(- \frac{1887121}{12600} +
  \frac{527}{48} \pi^2 + \frac{121}{3}
  \ln(\frac{r_{12}}{r'_{1}})\bigr) v_1^{2} + \bigl(- \frac{44389}{450}
  + \frac{173}{64} \pi^2 + \frac{22}{3}
  \ln(\frac{r_{12}}{r'_{1}})\nonumber\\ & + 16
  \ln(\frac{r_{12}}{r'_{2}})\bigr) v_2^{2}\Bigr]\biggr\} + 1
\leftrightarrow 2\,,\\[1ex]
L_\text{4PN}^{(5)}&={}\frac{3}{8} \frac{m_{1}^5 m_{2}}{r_{12}^5} +
\frac{m_{1}^3 m_{2}^3}{r_{12}^5} \bigl(\frac{587963}{5600} -
\frac{71}{32} \pi^2 - \frac{110}{3} \ln(\frac{r_{12}}{r'_{1}})\bigr) +
\frac{m_{1}^4 m_{2}^2}{r_{12}^5} \bigl(\frac{1690841}{25200} +
\frac{105}{32} \pi^2\nonumber\\ & - \frac{242}{3}
\ln(\frac{r_{12}}{r'_{1}}) - 16 \ln(\frac{r_{12}}{r'_{2}})\bigr) + 1
\leftrightarrow 2\,.
\end{align}
\end{subequations}}\noindent
These expressions depend linearly on accelerations $\bm{a}_A$ and do
not contain derivatives of accelerations. The only remaining constants
are the two UV scales $r'_1$ and $r'_2$ which are gauge constants and
will disappear from physical invariant results. The correct value
  of $\alpha$ given by~\eqref{valeuralpha} has been inserted.

We have checked that the full 4PN Lagrangian is invariant under global
Lorentz-Poincar\'e transformations. Indeed, the tail part of the
Lagrangian is separately Lorentz invariant. We have transformed the
variables $\bm{y}_A$, $\bm{v}_A$ and $\bm{a}_A$ in Eqs.~\eqref{L3PN}
and~\eqref{result4PN} according to a Lorentz boost (with constant
boost velocity), and verified that the Lagrangian is merely changed at
linear order by a total time derivative irrelevant for the dynamics.

Finally, we have verified that our 4PN Lagrangian, when restricted to terms
up to quadratic order in Newton's constant $G$, \textit{i.e.}, for
$L_\text{4PN}^{(0)}$, $L_\text{4PN}^{(1)}$ and
$L_\text{4PN}^{(2)}$, is equivalent to the Lagrangian obtained using
the effective field theory by Foffa \& Sturani~\cite{FS4PN}.

\subsection{Removal of accelerations from the Lagrangian}
\label{sec:ordlagrangian}
 
We shall now perform a shift of the particle's dynamical variables (or
``contact'' transformation) to a new Lagrangian whose instantaneous
part will be ordinary, in the sense that it depends only on positions
and velocities. Furthermore, the shift will be such that the
logarithms $\ln(r_{12}/r'_1)$ and $\ln(r_{12}/r'_2)$ are
canceled. This directly shows that the scales $r'_A$ are pure gauge
constants. Here we simply report the resulting ordinary Lagrangian,
which is comparatively much simpler than the harmonic one (the shift
is too lengthy to be presented here). Our choice for this ordinary
Lagrangian is that it is the closest possible one from the ADM
Lagrangian (see the discussion in Sec.~\ref{sec:hamiltonian}). We have
\begin{equation}\label{Lstructtilde}
\tilde{L} = \tilde{L}_\text{N} + \frac{1}{c^2}\tilde{L}_\text{1PN} +
\frac{1}{c^4}\tilde{L}_\text{2PN} + \frac{1}{c^6}\tilde{L}_\text{3PN}
+ \frac{1}{c^8}\tilde{L}_\text{4PN} + \mathcal{O}\left(10\right)\,,
\end{equation}
where $\tilde{L}_\text{N}$ and $\tilde{L}_\text{1PN}$ are actually
unchanged since the shift starts only at the 2PN order, and
{\allowdisplaybreaks
\begin{subequations}\label{L3PNtilde}
\begin{align}
\tilde{L}_\text{N} &= \frac{G m_1 m_2}{2 r_{12}} + \frac{m_1 v_1^2}{2} + 1
\leftrightarrow 2\,,\\
\tilde{L}_\text{1PN} &= - \frac{G^2 m_1^2 m_2}{2 r_{12}^2} + \frac{m_1
  v_1^4}{8} + \frac{G m_1 m_2}{r_{12}} \left( - \frac{1}{4}
(n_{12}v_1) (n_{12}v_2) + \frac{3}{2} v_1^2 - \frac{7}{4} (v_1v_2)
\right) + 1 \leftrightarrow 2\,,\\[1ex]
\tilde{L}_\text{2PN} &= \frac{1}{16} m_{1} v_1^{6} + \frac{G m_{1}
  m_{2}}{r_{12}} \bigl(\frac{3}{16} (n_{12} v_1)^2 (n_{12} v_2)^2 +
\frac{1}{8} (v_1 v_2)^2 + (n_{12} v_1) (\frac{3}{4} (n_{12} v_2) (v_1
v_2)\nonumber\\ & - \frac{1}{4} (n_{12} v_2) v_1^{2}) + \frac{7}{8}
v_1^{4} + v_1^{2} (- \frac{5}{8} (n_{12} v_2)^2 - \frac{7}{4} (v_1
v_2) + \frac{11}{16} v_2^{2})\bigr)\nonumber\\ & + \frac{G^2 m_{1}^2
  m_{2}}{r_{12}^2} \bigl(\frac{15}{8} (n_{12} v_1)^2 - \frac{15}{4}
(v_1 v_2) + \frac{11}{8} v_1^{2} + 2 v_2^{2}\bigr) + \frac{1}{4}
\frac{G^3 m_{1}^3 m_{2}}{r_{12}^3} + \frac{5}{8} \frac{G^3 m_{1}^2
  m_{2}^2}{r_{12}^3} \nonumber \\ & + 1 \leftrightarrow 2, \\[1ex]
\tilde{L}_\text{3PN}&=\frac{5}{128} m_{1} v_1^{8} + \frac{G m_{1}
  m_{2}}{r_{12}} \biggl\{- \frac{5}{32} (n_{12} v_1)^3 (n_{12} v_2)^3
+ \frac{1}{16} (v_1 v_2)^3 + \frac{11}{16} v_1^{6}\nonumber\\ & +
(n_{12} v_1) \Bigl[\frac{5}{16} (n_{12} v_2) (v_1 v_2)^2 -
  \frac{3}{16} (n_{12} v_2) v_1^{4} + v_1^{2} \bigl(\frac{9}{16}
  (n_{12} v_2)^3 + (n_{12} v_2) (\frac{3}{4} (v_1 v_2)\nonumber\\ & -
  \frac{9}{32} v_2^{2})\bigr)\Bigr] + (n_{12} v_1)^2 \Bigl[-
  \frac{15}{32} (n_{12} v_2)^2 (v_1 v_2) + v_1^{2} \bigl(\frac{3}{16}
  (n_{12} v_2)^2 - \frac{5}{16} v_2^{2}\bigr)\Bigr]\nonumber\\ & +
v_1^{4} \Bigl[- \frac{5}{16} (n_{12} v_2)^2 - \frac{21}{16} (v_1 v_2)
  + \frac{7}{8} v_2^{2}\Bigr] + v_1^{2} \Bigl[- \frac{1}{16} (n_{12}
  v_2)^2 (v_1 v_2) + \frac{1}{8} (v_1 v_2)^2\nonumber\\ & -
  \frac{15}{32} (v_1 v_2) v_2^{2}\Bigr]\biggr\} + \frac{G^2 m_{1}^2
  m_{2}}{r_{12}^2} \biggl\{- \frac{5}{12} (n_{12} v_1)^4 -
\frac{13}{8} (n_{12} v_1)^3 (n_{12} v_2) + \frac{341}{48} (v_1
v_2)^2\nonumber\\ & + \frac{21}{16} v_1^{4} + (n_{12} v_1)
\Bigl[\frac{1}{4} (n_{12} v_2) v_1^{2} + (n_{12} v_2)
  \bigl(\frac{1}{3} (v_1 v_2) - v_2^{2}\bigr)\Bigr] - \frac{71}{8}
(v_1 v_2) v_2^{2}\nonumber\\ & + (n_{12} v_1)^2 \Bigl[- \frac{23}{24}
  (n_{12} v_2)^2 - \frac{1}{2} (v_1 v_2) + \frac{13}{16} v_1^{2} +
  \frac{29}{24} v_2^{2}\Bigr] + v_1^{2} \Bigl[\frac{5}{6} (n_{12}
  v_2)^2 - \frac{97}{16} (v_1 v_2)\nonumber\\ & + \frac{43}{12}
  v_2^{2}\Bigr] + \frac{47}{16} v_2^{4}\biggr\} + \frac{G^3 m_{1}^2
  m_{2}^2}{r_{12}^3} \biggl\{\frac{1}{64} \Bigl[292 + 3 \pi^2\Bigr]
(n_{12} v_1)^2 + \Bigl[-11\nonumber\\ & - \frac{3}{64} \pi^2\Bigr]
(n_{12} v_1) (n_{12} v_2) + \frac{1}{64} \Bigl[472 + \pi^2\Bigr] (v_1
v_2) + \Bigl[- \frac{265}{48} - \frac{1}{64} \pi^2\Bigr]
v_1^{2}\biggr\}\nonumber\\ & + \frac{G^3 m_{1}^3 m_{2}}{r_{12}^3}
\biggl\{-5 (n_{12} v_1)^2 - \frac{1}{8} (n_{12} v_1) (n_{12} v_2) -
\frac{27}{8} (v_1 v_2) + \frac{173}{48} v_1^{2} + \frac{13}{8}
v_2^{2}\biggr\}\nonumber\\ & - \frac{1}{8} \frac{G^4 m_{1}^4
  m_{2}}{r_{12}^4} + \frac{1}{96} \biggl\{-908 + 63 \pi^2\biggr\}
\frac{G^4 m_{1}^3 m_{2}^2}{r_{12}^4} + 1 \leftrightarrow 2\,.
\end{align}
\end{subequations}}\noindent
Next the 4PN term is of the form
\begin{subequations}\label{split4PNtilde}
\begin{align}
\tilde{L}_\text{4PN} &= \tilde{L}^\text{inst}_\text{4PN} +
L^\text{tail}_\text{4PN}\,,\\ \tilde{L}^\text{inst}_\mathrm{4PN} &=
\tilde{L}^{(0)}_\mathrm{4PN} + G\,\tilde{L}^{(1)}_\mathrm{4PN} +
G^2\,\tilde{L}^{(2)}_\mathrm{4PN} + G^3\,\tilde{L}^{(3)}_\mathrm{4PN}
+ G^4\,\tilde{L}^{(4)}_\mathrm{4PN} +
G^5\,\tilde{L}^{(5)}_\mathrm{4PN}\,,
\end{align}\end{subequations}
where the tail piece $L^\text{tail}_\text{4PN}$ is exactly the same as
in Eqs.~\eqref{Ltail4PN} and where
{\allowdisplaybreaks
\begin{subequations}\label{result4PNtilde}
\begin{align}
\tilde{L}_\text{4PN}^{(0)}&={}\frac{7}{256} m_{1} v_1^{10} + 1
\leftrightarrow 2\,, \\[1ex]
\tilde{L}_\text{4PN}^{(1)}&={}\frac{m_{1} m_{2}}{r_{12}} \biggl\{-
\frac{25}{64} (n_{12} v_1)^3 (n_{12} v_2) (v_1 v_2)^2 + \frac{3}{64}
(n_{12} v_1)^2 (v_1 v_2)^3 + \frac{75}{128} v_1^{8}\nonumber\\ & +
v_1^{6} \Bigl[- \frac{5}{32} (n_{12} v_1) (n_{12} v_2) - \frac{15}{64}
  (n_{12} v_2)^2 - \frac{35}{32} (v_1 v_2) + \frac{45}{64}
  v_2^{2}\Bigr]\nonumber\\ & + (n_{12} v_1)^5 \Bigl[\frac{35}{256}
  (n_{12} v_2)^3 - \frac{55}{256} (n_{12} v_2) v_2^{2}\Bigr] + (n_{12}
v_1)^4 \Bigl[\frac{85}{256} (n_{12} v_2)^2 (v_1 v_2)\nonumber\\ & +
  \frac{23}{256} (v_1 v_2) v_2^{2}\Bigr] + v_1^{2} \Bigl[- \frac{1}{8}
  (n_{12} v_2)^2 (v_1 v_2)^2 + \frac{9}{64} (v_1 v_2)^3 + \frac{1}{32}
  (v_1 v_2)^2 v_2^{2}\nonumber\\ & + (n_{12} v_1)^3 \bigl(-
  \frac{85}{128} (n_{12} v_2)^3 + \frac{115}{128} (n_{12} v_2)
  v_2^{2}\bigr) + (n_{12} v_1)^2 \bigl(\frac{5}{32} (n_{12}
  v_2)^4\nonumber\\ & + (n_{12} v_2)^2 (- \frac{135}{128} (v_1 v_2) -
  \frac{21}{64} v_2^{2}) - \frac{19}{128} (v_1 v_2) v_2^{2}\bigr) +
  (n_{12} v_1) \bigl(\frac{1}{2} (n_{12} v_2)^3 (v_1 v_2)\nonumber\\ &
  + (n_{12} v_2) (\frac{53}{64} (v_1 v_2)^2 - \frac{1}{16} (v_1 v_2)
  v_2^{2})\bigr)\Bigr] + v_1^{4} \Bigl[- \frac{7}{32} (n_{12} v_2)^4 +
  \frac{3}{32} (v_1 v_2)^2\nonumber\\ & + (n_{12} v_1)
  \bigl(\frac{183}{256} (n_{12} v_2)^3 + (n_{12} v_2) (\frac{9}{16}
  (v_1 v_2) - \frac{167}{256} v_2^{2})\bigr) + (n_{12} v_1)^2
  \bigl(\frac{9}{64} (n_{12} v_2)^2\nonumber\\ & - \frac{15}{64}
  v_2^{2}\bigr) + (n_{12} v_2)^2 \bigl(- \frac{23}{256} (v_1 v_2) +
  \frac{3}{16} v_2^{2}\bigr) - \frac{185}{256} (v_1 v_2) v_2^{2} +
  \frac{31}{128} v_2^{4}\Bigr]\biggr\} + 1 \leftrightarrow 2\,, \\[1ex]
\tilde{L}_\text{4PN}^{(2)}&={}\frac{m_{1}^2 m_{2}}{r_{12}^2} \biggl\{-
\frac{369}{160} (n_{12} v_1)^6 + \frac{549}{128} (n_{12} v_1)^5
(n_{12} v_2) - \frac{21}{16} (n_{12} v_2)^2 (v_1 v_2)^2 -
\frac{53}{96} (v_1 v_2)^3\nonumber\\ & + \frac{143}{64} v_1^{6} +
(n_{12} v_1)^4 \Bigl[\frac{2017}{1280} (n_{12} v_2)^2 -
  \frac{1547}{256} (v_1 v_2) + \frac{243}{64} v_1^{2} -
  \frac{4433}{1920} v_2^{2}\Bigr]\nonumber\\ & + \frac{335}{32} (v_1
v_2)^2 v_2^{2} + v_1^{4} \Bigl[\frac{1869}{1280} (n_{12} v_2)^2 -
  \frac{1947}{256} (v_1 v_2) + \frac{5173}{1280}
  v_2^{2}\Bigr]\nonumber\\ & + (n_{12} v_1)^3 \Bigl[- \frac{11}{8}
  (n_{12} v_2)^3 - \frac{81}{16} (n_{12} v_2) v_1^{2} + (n_{12} v_2)
  \bigl(\frac{4531}{320} (v_1 v_2) + \frac{205}{96}
  v_2^{2}\bigr)\Bigr]\nonumber\\ & + (n_{12} v_1) \Bigl[\frac{7}{2}
  (n_{12} v_2)^3 (v_1 v_2) + \frac{295}{128} (n_{12} v_2) v_1^{4} +
  v_1^{2} \bigl(\frac{841}{192} (n_{12} v_2)^3\nonumber\\ & + (n_{12}
  v_2) (- \frac{771}{160} (v_1 v_2) - \frac{125}{32} v_2^{2})\bigr) +
  (n_{12} v_2) \bigl(\frac{37}{192} (v_1 v_2)^2 + \frac{15}{4} (v_1
  v_2) v_2^{2} - \frac{3}{2} v_2^{4}\bigr)\Bigr]\nonumber\\ & +
(n_{12} v_1)^2 \Bigl[\frac{7}{4} (n_{12} v_2)^4 - \frac{5629}{1280}
  (v_1 v_2)^2 - \frac{53}{16} v_1^{4} + (n_{12} v_2)^2 \bigl(-
  \frac{4013}{384} (v_1 v_2) - \frac{45}{16}
  v_2^{2}\bigr)\nonumber\\ & + \frac{527}{384} (v_1 v_2) v_2^{2} +
  v_1^{2} \bigl(- \frac{859}{160} (n_{12} v_2)^2 + \frac{875}{128}
  (v_1 v_2) + \frac{2773}{1280} v_2^{2}\bigr) + \frac{11}{64}
  v_2^{4}\Bigr]\nonumber\\ & - \frac{381}{32} (v_1 v_2) v_2^{4} +
v_1^{2} \Bigl[- \frac{7}{4} (n_{12} v_2)^4 + \frac{10087}{1280} (v_1
  v_2)^2 - \frac{5395}{384} (v_1 v_2) v_2^{2}\nonumber\\ & + (n_{12}
  v_2)^2 \bigl(\frac{629}{384} (v_1 v_2) + \frac{17}{16} v_2^{2}\bigr)
  + \frac{379}{64} v_2^{4}\Bigr] + \frac{59}{16} v_2^{6}\biggr\} + 1
\leftrightarrow 2\,, \\[1ex]
\tilde{L}_\text{4PN}^{(3)}&={}\frac{m_1^3 m_2}{r_{12}^{3}} \Bigl[-
  \frac{5015}{384} (n_{12} v_1)^4 + \frac{46493}{1920} (n_{12} v_1)^3
  (n_{12} v_2) + \frac{7359}{400} (v_1 v_2)^2 + \frac{4799}{1152}
  v_1^{4}\nonumber\\ & + (n_{12} v_1) \bigl(- \frac{6827}{640} (n_{12}
  v_2) v_1^{2} + (n_{12} v_2) (\frac{23857}{2400} (v_1 v_2) -
  \frac{31}{16} v_2^{2})\bigr)\nonumber\\ & + (n_{12} v_1)^2
  \bigl(\frac{3521}{960} (n_{12} v_2)^2 - \frac{6841}{384} (v_1 v_2) +
  \frac{11923}{960} v_1^{2} - \frac{2027}{1600} v_2^{2}\bigr) -
  \frac{357}{16} (v_1 v_2) v_2^{2}\nonumber\\ & + v_1^{2} \bigl(-
  \frac{13433}{4800} (n_{12} v_2)^2 - \frac{468569}{28800} (v_1 v_2) +
  \frac{54061}{4800} v_2^{2}\bigr) + \frac{203}{32} v_2^{4}\Bigr]
\nonumber\\ & + \frac{m_1^2 m_2^2}{r_{12}^{3}} \Bigl[\frac{3}{40960}
  \bigl(182752 - 625 \pi^2\bigr) (n_{12} v_1)^4 + \bigl(- \frac{72}{5}
  - \frac{35655}{16384} \pi^2\bigr) (n_{12} v_1)^3 (n_{12}
  v_2)\nonumber\\ & + \bigl(\frac{2051549}{57600} - \frac{10631}{8192}
  \pi^2\bigr) (v_1 v_2)^2 + (n_{12} v_1)^2 \bigl((\frac{16523}{960} +
  \frac{36405}{16384} \pi^2) (n_{12} v_2)^2 \nonumber\\ & +
  (\frac{578461}{6400} - \frac{56955}{16384} \pi^2) (v_1 v_2) + (-
  \frac{64447}{1600} + \frac{1107}{1024} \pi^2) v_1^{2}\bigr)
  \nonumber\\ & + (n_{12} v_1) \bigl(\frac{7}{51200} (-668104 + 21975
  \pi^2) (n_{12} v_2) (v_1 v_2) \nonumber\\ & + (\frac{1295533}{19200}
  - \frac{43869}{16384} \pi^2) (n_{12} v_2) v_1^{2}\bigr) +
  \bigl(\frac{65463}{6400} - \frac{2877}{8192} \pi^2\bigr) v_1^{4}
  \nonumber\\ & + v_1^{2} \bigl(\frac{1}{38400} (-1487258 + 79425
  \pi^2) (n_{12} v_2)^2 + (- \frac{836017}{14400} +
  \frac{40739}{16384} \pi^2) (v_1 v_2)\nonumber\\ & +
  (\frac{787817}{57600} - \frac{13723}{16384} \pi^2)
  v_2^{2}\bigr)\Bigr] + 1 \leftrightarrow 2\,, \\[1ex]
\tilde{L}_\text{4PN}^{(4)}&={}\frac{m_{1}^4 m_{2}}{r_{12}^{4}}
\Bigl[\frac{19341}{1600} (n_{12} v_1)^2 - \frac{15}{8} (v_1 v_2) -
  \frac{16411}{4800} v_1^{2} + \frac{31}{32} v_2^{2}\Bigl]
\nonumber\\ & + \frac{m_{1}^3 m_{2}^2}{r_{12}^{4}} \Bigl[(-
  \frac{3461303}{403200} - \frac{15857}{16384} \pi^2) (n_{12} v_1)^2
  \nonumber\\ & + (\frac{46994113}{403200} - \frac{79385}{24576}
  \pi^2) (n_{12} v_1) (n_{12} v_2) \nonumber\\ & + (-
  \frac{5615591}{134400}+ \frac{35603}{24576} \pi^2) (n_{12} v_2)^2 +
  (\frac{2827397}{57600} - \frac{171041}{24576} \pi^2) (v_1 v_2)
  \nonumber\\ & + (- \frac{1830673}{57600} + \frac{193801}{49152}
  \pi^2) v_1^{2}+ (- \frac{1158323}{57600} + \frac{21069}{8192} \pi^2)
  v_2^{2}\Bigr] + 1 \leftrightarrow 2\,, \\[1ex]
\tilde{L}_\text{4PN}^{(5)}&={}\frac{1}{16} \frac{m_{1}^5
  m_{2}}{r_{12}^5} + (\frac{3421459}{50400} - \frac{6237}{1024} \pi^2)
\frac{m_{1}^4 m_{2}^2}{r_{12}^5} + (\frac{4121669}{50400} -
\frac{44825}{6144} \pi^2) \frac{m_{1}^3 m_{2}^3}{r_{12}^5} \nonumber
\\ & + 1 \leftrightarrow 2\,.
\end{align}
\end{subequations}}\noindent

\subsection{Comparison with the Hamiltonian formalism}
\label{sec:hamiltonian}

In principle, by properly adjusting the contact transformation or
shift from harmonic coordinates, the ordinary Lagrangian obtained in
the previous section,
Eqs.~\eqref{Lstructtilde}--\eqref{result4PNtilde}, should correspond
to ADM like coordinates, and by an ordinary Legendre transformation we
should obtain the (instantaneous part of the) ADM
Hamiltonian. Concerning the tails we also need to find a shift (which
will be non-local~\cite{DJS15}) that removes the accelerations and
derivatives of accelerations from the tail part of the
Lagrangian~\eqref{Ltail4PN}, or, rather, from the corresponding
action. Once the tail part of the Lagrangian becomes ordinary, we can
obtain the corresponding tail part in the Hamiltonian.

The tail part of the action is
\begin{equation}\label{tailpart}
S_\text{F}^\text{tail} = \frac{G^2M}{5c^8}
\mathop{\text{Pf}}_{2r_{12}/c}\int\!\!\!\int \frac{\ud t\ud t'}{\vert
  t-t'\vert} \,I_{ij}^{(3)}(t) \,I_{ij}^{(3)}(t')\,,
\end{equation}
where the Hadamard scale $s_0$ in Eq.~\eqref{StailHad} has been
replaced by $r_{12}=r_{12}(t)$; the time derivatives of the
Newtonian quadrupole moment $I_{ij}=\sum_A m_A \,y_A^{\langle
  i}y_A^{j\rangle}$ are evaluated without replacement of
accelerations, \textit{i.e.},
\begin{equation}\label{Iij3}
I_{ij}^{(3)} = \sum_A 2 m_A \Bigl(3 v_A^{\langle i}a_A^{j\rangle} +
y_A^{\langle i}b_A^{j\rangle}\Bigr)\,.
\end{equation}
Here we look for a shift that transforms the action into the same
expression but with the derivatives of the quadrupole evaluated using
the Newtonian equations of motion, \textit{i.e.},
\begin{equation}\label{Iij3couchemasse}
\hat{I}_{ij}^{(3)} = \frac{2G m_1 m_2}{r_{12}^2}\left(-4
n_{12}^{\langle i}v_{12}^{j\rangle} + 3(n_{12}v_{12})n_{12}^{\langle
  i}n_{12}^{j\rangle}\right)\,.
\end{equation}
Note that here $\hat{I}_{ij}^{(3)}$ is not the third time derivative
of the quadrupole moment unless the equations of motion are
satisfied. The requested shift is easy to find and we get, after
removal of some double-zero terms which do not contribute to the
dynamics,
\begin{equation}\label{tailpartcouchemasse}
S_\text{F}^\text{tail} = \hat{S}_\text{F}^\text{tail} + \sum_A m_A
\int_{-\infty}^{+\infty} \ud t \,\Bigl[ a_A^i - (\partial_i U)_A\Bigl]
\xi_A^i\,,
\end{equation}
where $\hat{S}_\text{F}^\text{tail}$ is given by the same expression
as~\eqref{tailpart} but with the derivatives of the quadrupole moment
computed on-shell, Eq.~\eqref{Iij3couchemasse}, while the second term
takes the form of a shift explicitly given by\footnote{Note that in
  the shift vector itself, it does not matter whether we replace the
  accelerations with the equations of motion or not.}
\begin{equation}\label{shiftxi}
\xi_A^i = \frac{4G^2M}{5c^8} \left[2v_A^j
  \mathop{\text{Pf}}_{2r_{12}/c}\int \frac{\ud t'}{\vert t-t'\vert}
  \,\hat{I}_{ij}^{(3)}(t') - y_A^j \mathop{\text{Pf}}_{2r_{12}/c}\int
  \frac{\ud t'}{\vert t-t'\vert} \,\hat{I}_{ij}^{(4)}(t') +
  2\frac{(n_{12}v_{12})}{r_{12}}y_A^j \hat{I}_{ij}^{(3)}\right]\,.
\end{equation}
Once the total action $\hat{S}_\text{F} = S_\text{F}^\text{inst} +
\hat{S}_\text{F}^\text{tail}$ is ordinary, the (Fokker) Hamiltonian is
defined by the usual Legendre transformation as
\begin{equation}\label{Hamiltoniandef}
\hat{S}_\text{F} = \int_{-\infty}^{+\infty}\ud t\biggl[\sum_A
  p_A^iv_A^i - H\biggr]\,.
\end{equation}
The Hamiltonian is a functional of positions $\bm{y}_A$ and momenta
$\bm{p}_A$, and reads then ${H = H^\text{inst} + \hat{H}^\text{tail}}$,
where the tail part is just the opposite of the tail part of the
Lagrangian, as also found in Eq.(4.5) of~\cite{DJS14},
\begin{equation}\label{Htail4PN}
\hat{H}^\text{tail} = - \frac{G^2M}{5c^8}
\,\hat{I}_{ij}^{(3)}(t)\!\mathop{\text{Pf}}_{2r_{12}/c}
\int_{-\infty}^{+\infty} \frac{\ud t'}{\vert t-t'\vert}
\hat{I}_{ij}^{(3)}(t')\,.
\end{equation}
To prove this we notice that the tail term is a small 4PN quantity,
and that its contribution in the velocity expressed as a function of
the momentum cancels out in the Legendre transformation at leading
order. On the right-hand side of~\eqref{Htail4PN}, the velocities
present in the quadrupole moment~\eqref{Iij3couchemasse} are to be
replaced with this approximation by $\bm{v}_A\rightarrow\bm{p}_A/m_A$.

An important point is that since the action is non-local in time the
Hamiltonian is only defined in an ``integrated'' sense by
Eq.~\eqref{Hamiltoniandef} but not in a local sense~\cite{Llosa,Ferialdi}.
Thus, the Hamiltonian equations of motion will be valid in a sense of
\textit{functional derivatives}, and the value of the Hamiltonian ``on-shell''
does not yield in general a strictly conserved energy. Indeed, we shall find
in the companion paper~\cite{BBBFM15b} that in order to obtain an energy $E$
that consistently includes the non-local tails at the 4PN order and is
strictly conserved, \textit{i.e.}, $\ud E/\ud t=0$ at any time, we must take
into account an extra contribution with respect to the Hamiltonian computed
on-shell. The latter extra contribution is however zero for circular orbits.
We shall show in Sec.~\ref{sec:Ehamiltonian} how to compute, in that case, the
energy from the Hamiltonian.

We have compared our 4PN dynamics with the 4PN Hamiltonian published
in Refs.~\cite{JaraS12,JaraS13,DJS14,JaraS15}, but unfortunately we
have not been able to match our results with these works. Moreover,
we fundamentally disagree with Ref.~\cite{DJS14} regarding the
contribution of tails to the energy for circular orbits (see the
details in Sec.~\ref{sec:Ehamiltonian}), but taking into account that
disagreement is not sufficient to explain the full discrepancy.

We did two comparisons. One at the level of the equations of motion,
looking for a shift of the trajectories such that the equations of
motion derived from the 4PN harmonic Lagrangian in Sec.~\ref{sec:res}
are transformed into the equations of motion derived from the 4PN
Hamiltonian published in Eqs.~(A3)--(A4) of~\cite{DJS14}. Our second
comparison was directly at the level of the Lagrangian, constructing
from the harmonic Lagrangian the ordinary Lagrangian (see the result
in Sec.~\ref{sec:ordlagrangian}), then shifting the tail part
according to Eq.~\eqref{tailpartcouchemasse}, and constructing the 4PN
Hamiltonian following~\eqref{Hamiltoniandef}.

However these comparisons failed. The best we could do was to match
all the terms with powers $G^0$, $G^1$, $G^2$ (the
terms $G^0$, $G^1$ and $G^2$ in our Lagrangian also match with those
of Ref.~\cite{FS4PN}), $G^3$ and $G^5$, as well as
many terms with powers $G^4$ in the acceleration, but there are
residual terms with powers $G^4$ that are impossible to
reconcile. When looking for the ADM Lagrangian, the closest one we
could find is given by~\eqref{L3PNtilde}--\eqref{result4PNtilde} in
Sec.~\ref{sec:ordlagrangian}, but its Legendre transform disagrees
with the published ADM Hamiltonian by $G^4$ and $G^5$ terms.

Finally the contact transformation which minimizes the number of
irreconcilable terms in both formalisms gives the difference between
our harmonic-transformed acceleration $a_1^i$ and their ADM
acceleration $(a_1^i)_\text{DJS}$ as
\begin{equation}\label{discrepacc}
a_1^i - (a_1^i)_\text{DJS} = \frac{2}{15}\frac{G^4 m \,m_{1}
  m_{2}^2}{c^8 r_{12}^5} \biggl[- \frac{472}{3} v_{12}^{i} (n_{12}
  v_{12}) + n_{12}^{i} \Big(- \frac{1429}{7} (n_{12} v_{12})^2 +
  \frac{1027}{7} v_{12}^{2}\Big)\biggr]\,,
\end{equation}
where we denote $m=m_1+m_2$ and $v_{12}^{i} =
v_{1}^{i}-v_{2}^{i}$. Such a difference of accelerations cannot be
eliminated by a further contact transformation. It corresponds to the
following difference between Hamiltonians,
\begin{align}\label{discrepH}
H - (H)_\text{DJS} &= \frac{G^4m}{315\,c^8\,r_{12}^{4}}\biggl[
  1429\bigl(m_2^2(n_{12}p_{1})^2 -2\,m_1
  m_2(n_{12}p_{1})(n_{12}p_{2})+m_1^2(n_{12}p_{2})^2 \bigr)
  \nonumber\\ & \qquad\quad\quad ~+ \,826 \bigl( m_{2}^{2}\,p_{1}^{2}
  - 2\,m_{1}m_{2}(p_{1}p_{2})+m_{1}^{2}\,p_{2}^{2} \bigr) +
  902\frac{G m \,m_1^2m_2^2}{r_{12}}\biggr] \,.
\end{align}
Our Hamiltonian $H$ is defined by the sum of the tail
part~\eqref{Htail4PN} and of the Legendre transformation of the
ordinary Lagrangian given
by~\eqref{L3PNtilde}--\eqref{result4PNtilde}. In conclusion, from
Eqs.~\eqref{discrepacc}--\eqref{discrepH} we face a true
discrepancy. Note, however, that this discrepancy concerns only a few
terms; for many terms our Hamiltonian agrees with the
Hamiltonian of~\cite{DJS14}.

Furthermore, we observe the paradoxical fact that the difference of
accelerations~\eqref{discrepacc} does not yield a zero contribution to
the energy in the case of circular orbits. Similarly, the difference
of Hamiltonians~\eqref{discrepH} does not vanish for circular
orbits. This is inconsistent with the fact that the two groups agree
on the conserved energy in that case. Recall that we have adjusted our
ambiguity parameter $\alpha$ to the value $\alpha=\frac{811}{672}$ so
that the 4PN energy for circular orbits (computed directly from the
4PN equations of motion in harmonic coordinates~\cite{BBBFM15b})
agrees with self-force calculations [see~\eqref{valeuralpha} and the
  preceding discussion]. On the other hand, the ambiguity parameter in
Ref.~\cite{DJS14}, which is denoted by $C$, has been adjusted (to the
value $C=-\frac{1681}{1536}$) using the same self-force results. This
contradiction leads us to investigate the validity of the derivation of
the conserved energy for circular orbits using the Hamiltonian
formalism as presented in Ref.~\cite{DJS14}. We address this point in
the next section.

\subsection{Energy for circular orbits computed with the Hamiltonian}
\label{sec:Ehamiltonian}

As discussed in the previous section we can consider the non-local but
ordinary Hamiltonian $H[\bm{y}_A, \bm{p}_A] = H^\text{inst}(\bm{y}_A,
\bm{p}_A) + \hat{H}^\text{tail}[\bm{y}_A, \bm{p}_A]$, where the tail
term given by~\eqref{Htail4PN} functionally depends on the canonical
positions $\bm{y}_A$ and momenta $\bm{p}_A$. In the frame of the
center of mass the Hamiltonian is a functional of
$\bm{y}=r\bm{n}\equiv\bm{y}_1-\bm{y}_2$ and
$\bm{p}\equiv\bm{p}_1=-\bm{p}_2$. Next, introducing polar coordinates
$(r, \varphi)$ in the binary's orbital plane and their conjugate
momenta $(p_r=\bm{n}\cdot\bm{p},p_\varphi)$, we make the substitution
$\bm{p}^2=p_r^2+p_\varphi^2/r^2$ to obtain the reduced Hamiltonian
$H_\text{red}$ which is a (non-local) functional of the canonical
variables $r$, $p_r$ and $p_\varphi$.\footnote{Because of the
  non-local tail term, the Hamiltonian depends also on $\varphi$, so
  that $p_\varphi$ is not strictly conserved. However, we can neglect
  this dependence on $\varphi$ and the variation of $p_\varphi$ in the
  present calculation, since in particular $p_\varphi$ is constant in
  the case of circular orbits.}  For circular orbits we have $r=r_0$
(a constant) and $p^0_r=0$.  The angular momentum $p^0_\varphi$ is
then obtained as a function of the radius $r_0$ by solving the radial
equation
\begin{equation}\label{radeq}
\frac{\delta H_\text{red}}{\delta r}\bigl[r_0, p^0_r=0,
  p^0_\varphi\bigr] = 0\,,
\end{equation}
while the orbital frequency $\Omega$ of the circular motion is given
by
\begin{equation}\label{omegaeq}
\frac{\delta H_\text{red}}{\delta p_\varphi}\bigl[r_0, p^0_r=0,
  p^0_\varphi\bigr] = \Omega\,.
\end{equation}
The circular energy is then $E=H_\text{red}[r_0, 0,
  p^0_\varphi(r_0)]$, the function $p^0_\varphi(r_0)$ representing
here the solution of Eq.~\eqref{radeq}. Finally, by inverting
Eq.~\eqref{omegaeq}, we can express the radius $r_0$ as a function of
the frequency $\Omega$, or rather, of the PN parameter $x=(G
m\Omega/c^3)^{2/3}$. This leads to the invariant circular energy
$E(x)$.

The only tricky calculation is that of the contribution of the tail
part of the Hamiltonian. Because of the non-locality, the
differentiation occurring in Eqs.~\eqref{radeq}--\eqref{omegaeq}
should be performed in the sense of functional derivatives. As such,
the functional variation of the tail term with respect to $r(t)$
(where $t$ is the coordinate time with respect to which the binary's
dynamics is measured) yields
\begin{equation}\label{variationHtaila}
\frac{\delta\hat{H}^\text{tail}}{\delta r(t)} = - \frac{2G^2M}{5c^8}
\left[\frac{\partial\hat{I}_{ij}^{(3)}(t)}{\partial
    r(t)}\mathop{\text{Pf}}_{2r(t)/c} \int_{-\infty}^{+\infty}
  \frac{\ud t'}{\vert t-t'\vert} \hat{I}_{ij}^{(3)}(t') -
  \frac{1}{r(t)}\left(\hat{I}_{ij}^{(3)}(t)\right)^2\right]\,.
\end{equation}
The second term in the square brackets comes from the variation of the
Hadamard partie finie scale $r(t)\equiv r_{12}(t)$ present in
Eq.~\eqref{Htail4PN}. Similarly, the functional variation with respect
to $p_\varphi(t)$ reads
\begin{equation}\label{variationHtailb}
\frac{\delta\hat{H}^\text{tail}}{\delta p_\varphi(t)} = -
\frac{2G^2M}{5c^8} \frac{\partial\hat{I}_{ij}^{(3)}(t)}{\partial
  p_\varphi(t)}\mathop{\text{Pf}}_{2r(t)/c} \int_{-\infty}^{+\infty}
\frac{\ud t'}{\vert t-t'\vert} \hat{I}_{ij}^{(3)}(t') \,.
\end{equation}

We substitute into the radial equation~\eqref{radeq} all the
instantaneous contributions --- this poses no problem since the
partial derivatives are ordinary --- and add to that the tail
piece~\eqref{variationHtaila}. Solving iteratively for $p_\varphi^0$
as a function of $r_0$, we find the standard Newtonian result
\begin{equation}\label{pphi0N}
p_\varphi^0(r_0) = m\nu \sqrt{G m r_0} + \mathcal{O}(2)\,,
\end{equation}
which we can insert back into the tail integral
entering~\eqref{variationHtaila}, because the tail term is a small 4PN
quantity. At this stage, and \textit{only at this stage}, we are
allowed to reduce the tail integral and compute it in the case of
circular orbits, \textit{i.e.}, for $r=r_0$, $p_r^0=0$,
$p_\varphi^0(r_0)$ being given by~\eqref{pphi0N}, and for
$\dot{\varphi}=\omega_0 + \mathcal{O}(2)$, with $\omega_0=\sqrt{G
  m/r_0^3}$. A straightforward calculation~\cite{BS93} leads to the
formula (modulo higher-order PN radiation-reaction corrections)
\begin{equation}\label{evaluatetail}
\left(\mathop{\text{Pf}}_{2r(t)/c} \int_{-\infty}^{+\infty} \frac{\ud
  t'}{\vert t-t'\vert}
\hat{I}_{ij}^{(3)}(t')\right)\!\!{\bigg|}_{[r_0,p_r^0=0,p_\varphi^0(r_0)]}
\hspace{-0.5cm} = -2
\left(\hat{I}_{ij}^{(3)}(t)\right)\!\!{\bigg|}_{[r_0,p_r^0=0,p_\varphi^0(r_0)]}
\biggl[\ln\left(\frac{4\omega_0 r_0}{c}\right) +
  \gamma_\text{E}\biggr]\,,
\end{equation}
with $\gamma_\text{E}$ denoting the Euler constant. This result is
used to determine the contribution of the tail term in the link
between $p_\varphi^0$ and $r_0$ at the 4PN order. Denoting such
contribution by $\Delta p_\varphi^0(r_0)$ we explicitly find
\begin{align}
\Delta p_\varphi^0(r_0) = \frac{G^2M}{5c^8\omega_0}\left\{
\left(r\frac{\partial\bigl(\hat{I}_{ij}^{(3)}\bigr)^2}{\partial r}
\right)\!\!{\bigg|}_{[r_0,p_r^0=0,p_\varphi^0(r_0)]}
\biggl[\ln\left(\frac{4\omega_0 r_0}{c}\right) +
  \gamma_\text{E}\biggr] +
\left(\bigl(\hat{I}_{ij}^{(3)}\bigr)^2\right)
\!\!{\bigg|}_{[r_0,p_r^0=0,p_\varphi^0(r_0)]}\right\}\,.\label{correctprescr}
\end{align}
This is easily reduced by employing the Newtonian expression of the
quadrupole moment, valid for a general orbit (\textit{i.e.}, for any
$r$, $p_r$ and $p_\varphi$), hence
\begin{equation}\label{I3carre}
\bigl(\hat{I}_{ij}^{(3)}\bigr)^2 =
\frac{G^2m^2}{r^4}\left(\frac{8}{3}p_r^2 + 32
\frac{p_\varphi^2}{r^2}\right) + \mathcal{O}(2)\,.
\end{equation}
Treating $(r,\varphi,p_r,p_\varphi)$ as independent variables, we
differentiate~\eqref{I3carre} partially with respect to $r$ and take
$r=r_0$, $p_r=0$, $p_\varphi=p_\varphi^0(r_0)$ afterwards. We get
[using also $M=m+\mathcal{O}(2)$]
\begin{equation}\label{Deltapphi}
\Delta p_\varphi^0(r_0) = \frac{G^{9/2}
  m^{11/2}\nu^2}{5c^8r_0^{7/2}}\left(-192
\biggl[\ln\left(\frac{4\omega_0 r_0}{c}\right) +
  \gamma_\text{E}\biggr] + 32\right)\,.
\end{equation}
Next, we consider the orbital frequency $\Omega\equiv \ud \varphi/\ud
t$ given by Eq.~\eqref{omegaeq}. The tail contribution therein has
been displayed in~\eqref{variationHtailb}. Using Eq.~\eqref{pphi0N} to
Newtonian order we simply find
$\Omega(r_0)=\omega_0+\mathcal{O}(2)$. The tail term is consistently
evaluated thanks to~\eqref{evaluatetail} as before. The tail induced
modification of the frequency, say $\Delta \Omega(r_0)$, is then the
sum of the direct effect of the tail term in~\eqref{variationHtailb}
and of a contribution due to the tail modification of the angular
momentum~\eqref{Deltapphi},
\begin{align}
\Delta \Omega(r_0) &= \frac{\Delta p_\varphi^0(r_0)}{m \nu r_0^2} +
\frac{128}{5}\frac{G^{9/2}
  m^{9/2}\nu}{c^8r_0^{11/2}}\biggl[\ln\left(\frac{4\omega_0
    r_0}{c}\right) + \gamma_\text{E}\biggr]\nonumber\\&= \frac{G^{9/2}
  m^{9/2}\nu}{5c^8r_0^{11/2}}\left(-64 \biggl[\ln\left(\frac{4\omega_0
    r_0}{c}\right) + \gamma_\text{E}\biggr] +
32\right)\,. \label{Deltaomega}
\end{align}
With the two results~\eqref{Deltapphi}--\eqref{Deltaomega} in hand
the contribution of tails in the invariant energy for circular orbits
expressed as a function of $x=(G m\Omega/c^3)^{2/3}$ is readily found
to be
\begin{equation}\label{invEx} 
\Delta E(x) = - \frac{224}{15} m c^2 \nu^2 x^5\biggl[\ln\left(16
  x\right) + 2\gamma_\text{E} - \frac{4}{7}\biggr] \,. \end{equation}
This result fully agrees with our alternative derivation based on the
direct construction of the conserved circular energy from the
equations of motion in harmonic coordinates (see the companion
paper~\cite{BBBFM15b}). Thus, by following the above Hamiltonian
procedure, \textit{i.e.}, by carefully taking into account the
non-local character of the tail term during the variation of the
Hamiltonian [see notably~\eqref{variationHtaila}], we have shown that
our Hamiltonian $H$ defined by the Legendre transformation of the
ordinary Lagrangian~\eqref{L3PNtilde}--\eqref{result4PNtilde} plus the
tail part~\eqref{Htail4PN} leads to the correct conserved invariant
energy for circular orbits. This calculation confirms our value
$\alpha=\frac{811}{672}$ for the ambiguity parameter.

However, we find that, applying the same Hamiltonian procedure to the
Hamiltonian $(H)_\text{DJS}$, we do not recover the part of the
invariant energy for circular orbits that is known from self-force
calculations, unless the ambiguity parameter $C$ is adjusted to a
different value, which would then in turn change several coefficients
in the Hamiltonian for general orbits~\cite{DJS14}. We obtain that the
value for which that Hamiltonian gives the correct circular energy is
\begin{equation}\label{Ccorr}
C^\text{new} = C + \frac{3}{7} = - \frac{7159}{10752}\,.
\end{equation}

One possible explanation for the discrepancy could reside in the
treatment of the non-local part of the Hamiltonian when reducing to
circular orbits. Recall from Eq.~\eqref{variationHtaila} that
one must evaluate the tail integral for circular orbits \textit{after}
the differentiation with respect to $r$. We think that the treatment
of Ref.~\cite{DJS14} effectively amounts to doing the reverse,
\textit{i.e.}, to computing first the tail integral for circular
orbits by means of~\eqref{evaluatetail}, and only then performing the
differentiation with respect to $r$. Indeed, we have been told by
G. Sch\"afer (private communication) that Ref.~\cite{DJS14} uses a
local version of the Hamiltonian computed with
Eq.~\eqref{evaluatetail}, and then differentiates it with respect to
the independent canonical variables $r$, $p_r$ and $p_\varphi$, using
$\omega=p_\varphi/(m\nu r^2)$ for the circular orbit frequency,
therefore arriving at
\begin{equation}\label{varDJS}
\left(\Delta p_\varphi^0\right)_\text{DJS} =
\frac{G^2M}{5c^8\omega_0}\Biggl(r\frac{\partial}{\partial
  r}\left(\bigl(\hat{I}_{ij}^{(3)}\bigr)^2\biggl[\ln\left(\frac{4p_\varphi}{m\nu
    r c}\right) + \gamma_\text{E}\biggr]
\right)\Biggr)\!\,{\bigg|}_{[r_0,p_r^0=0,p_\varphi^0(r_0)]}\,,
\end{equation}
instead of our prescription~\eqref{correctprescr}. If one now applies
the derivative with respect to $r$, one finds that the tail induced
contribution to the angular momentum as a function of $r_0$
in~\cite{DJS14} differs from ours by the amount
\begin{equation}\label{diffnousDJS}
\Delta p_\varphi^0 - \left(\Delta p_\varphi^0\right)_\text{DJS} =
\frac{2}{5}\frac{G^2M}{c^8\omega_0}\left(\bigl(\hat{I}_{ij}^{(3)}\bigr)^2\right)
\!\!{\bigg|}_{[r_0,p_r^0=0,p_\varphi^0(r_0)]}
\hspace{-0.3cm} = \frac{64}{5}\frac{G^{9/2}
  m^{11/2}\nu^2}{c^8r_0^{7/2}}\,.
\end{equation}
Furthermore we find that the tail contribution to the orbital
frequency $\Omega(r_0)$ as a function of the radius agrees with us, so
that, in the end, the tail contribution in the invariant circular energy
differs from ours by
\begin{equation}\label{diffnousDJS}
\Delta E - \left(\Delta E\right)_\text{DJS} = \omega_0 \left[\Delta
  p_\varphi^0 - \left(\Delta p_\varphi^0\right)_\text{DJS}\right] =
\frac{64}{5} m c^2 \nu^2 x^5\,.
\end{equation}
Finally the prescription~\eqref{varDJS}, with which we disagree,
leads to an incorrect invariant energy $E(x)$ for circular orbits when
starting from our Hamiltonian or from the one in~\cite{DJS14} but with $C$
given by~\eqref{Ccorr}. On the other hand, if one applies this prescription,
different from ours, for the circular orbit reduction of the Hamiltonian
in~\cite{DJS14}, without modifying the constant $C$, one ends up with the
correct $E(x)$.

Similarly, we disagree with the computation of the effective-one-body (EOB)
potentials at the 4PN order in Ref.~\cite{DJS15}. Indeed, a local ansatz has
been made for the EOB Hamiltonian, since it has been obtained by evaluating
the tail term on shell for an explicit solution of the motion (see
Eqs.~(4.10)--(4.11) in~\cite{DJS15}), which means effectively using
Eq.~\eqref{evaluatetail} in the case of circular orbits, and results in a
local Hamiltonian. Note that the comparison to the self-force results of the
work~\cite{DJS14,DJS15} have been recently complemented by deriving and
confirming with another method the EOB potential $D(u)$~\cite{BiniDG15}.
However, the problem in the treatment of the non-locality described above
might affect this comparison as well.\footnote{After this work was submitted
  for publication, the authors of~\cite{DJS14, DJS15} described in more
  details their method for reducing the dynamics to a local-in-time
  Hamiltonian in~\cite{djs16}.}

Still, if we now make the comparison with the Hamiltonian~\cite{DJS14}
but with the new value of the ambiguity parameter~\eqref{Ccorr}, we
cannot reduce the difference to zero. Indeed, the right-hand sides of
Eqs.~\eqref{discrepacc}--\eqref{discrepH} do not correspond to a mere
rescaling of the ambiguity parameter. We get instead
\begin{subequations}\label{discrepacc_corr}
\begin{align}
{a'}_1^i - (a_1^i)^\text{new}_\text{DJS} &= \frac{2}{15}\frac{G^4 m
  \,m_{1} m_{2}^2}{c^8 r_{12}^5} \biggl[ \frac{680}{3} v_{12}^{i}
  (n_{12} v_{12}) + n_{12}^{i} \Big(- 595 (n_{12} v_{12})^2 + 85
  v_{12}^{2}\Big)\biggr]\,,\\ H' - (H)^\text{new}_\text{DJS} &=
\frac{1}{315}\frac{G^4m}{c^8\,r_{12}^{4}}\biggl[
  4165\bigl(m_2^2(n_{12}p_{1})^2 -2\,m_1
  m_2(n_{12}p_{1})(n_{12}p_{2})+m_1^2(n_{12}p_{2})^2 \bigr)
  \nonumber\\ & \qquad\quad - \,1190 \bigl( m_{2}^{2}\,p_{1}^{2} -
  2\,m_{1}m_{2}(p_{1}p_{2})+m_{1}^{2}\,p_{2}^{2} \bigr) + 1190\frac{G
    m \,m_1^2m_2^2}{r_{12}}\biggr] \,.
\end{align}
\end{subequations}
With respect to~\eqref{discrepacc}--\eqref{discrepH} we have performed
for convenience an additional shift, hence we denote our new
acceleration and Hamiltonian with a prime. When using the
value~\eqref{Ccorr} of the ambiguity parameter, the
differences~\eqref{discrepacc_corr} can now be seen not to contribute
to the conserved invariant energy for circular orbits, which resolves
our paradox. Unfortunately, we have no explanation for the remaining
discrepancy in Eqs.~\eqref{discrepacc_corr}. 

\acknowledgments We are grateful to Thibault Damour for bringing to
our attention, at an early stage of this work, useful properties of
the Fokker action, and for informative comments regarding the DJS
paper~\cite{DJS14}. We also thank Gilles Esposito-Far\`ese and Gerhard
Sch\"afer for interesting discussions. S.M. was supported by the NASA
grant 11-ATP-046, as well as the NASA grant NNX12AN10G at the
University of Maryland College Park.

\appendix

\section{Complement about the method ``$n+2$''}
\label{app:np2}

We compute the Fokker action $S_\text{F}$ for a full-fledge solution
$h$ of the field equations reducing to the PN expansion $\overline{h}$
in the near zone and to the multipole expansion $\mathcal{M}(h)$ in
the far zone. These two expansions obey the matching equation
$\overline{\mathcal{M}(h)} = \mathcal{M}\bigl(\overline{h}\bigr)$. We
suppose that this solution is of the type
\begin{equation}
h = h_n + \delta h\,,
\end{equation}
where $h_n$ is some known \textit{approximate} solution and $\delta h$
is a remainder (or error) term defined everywhere. We denote the
approximate solution $h_n$ with the label $n$ because we assume that
in the near zone the PN expansion of this solution agrees with the
known solution considered in Sec.~\ref{sec:np2}, \textit{i.e.},
$\overline{h}_n$. However we extend here that solution to the far zone
as well, where it agrees with the multipole expansion
$\mathcal{M}(h_n)$. Similarly the error in that solution is defined
both in the near zone, \textit{i.e.},  $\overline{\delta h}$, and in
the far zone, $\mathcal{M}(\delta h)$.

Since we are considering here the true solution $h$ (and not merely
its PN expansion $\overline{h}$) there is no regulator $r^B$ in the
first place, and we can freely integrate by parts the action and write
the usual expansion
\begin{equation}\label{SFexp}
S_\text{F}[h] = S_\text{F}[h_n] + \int\!\ud t\!\int
\ud^3\mathbf{x}\,\frac{\delta S_\text{F}}{\delta h}[h_n]\,\delta h +
\mathcal{O}\left(\delta h^2\right)\,.
\end{equation}
At this stage we apply the lemma~\eqref{lemma1} to the second term
in~\eqref{SFexp}. We introduce the regulator $r^B$ and transform it
into an expression that integrates over the formal PN expansion, plus
a contribution that integrates over the multipole expansion (with,
say, $r_0=1$):
\begin{equation}\label{decompPNmult}
\int \ud^3\mathbf{x}\,\frac{\delta S_\text{F}}{\delta h}\,\delta h =
\mathop{\text{{\rm FP}}}_{B=0}\int \ud^{3}\mathbf{x}
\,r^B\,\overline{\frac{\delta S_\text{F}}{\delta h}}\,\overline{\delta
  h} + \mathop{\text{{\rm FP}}}_{B=0}\int \ud^{3}\mathbf{x} \,r^B
\mathcal{M}\left(\frac{\delta S_\text{F}}{\delta h}\right)
\mathcal{M}\left(\delta h\right)\,.
\end{equation}
The first term of~\eqref{decompPNmult} corresponds exactly to the PN
remainder that is investigated in Sec.~\ref{sec:np2} and yields to our
method $n+2$ [see Eq.~\eqref{SFexpand}]. Here we worry about the
second, multipolar contribution in~\eqref{decompPNmult} that was not
considered in the arguments of Sec.~\ref{sec:np2}. We shall argue that
its contribution is completely negligible when re-expanded in the near
zone as compared with the 4PN order.

For the method $n+2$ we proved in Sec.~\ref{sec:np2} that if the PN
solution $\overline{h_n}$ is known to order $\mathcal{O}(n+2,n+1,n+2)$
when $n$ is even and $\mathcal{O}(n+1,n+2,n+1)$ when $n$ is odd, then
the contribution of the first, PN term in~\eqref{decompPNmult} is very
small, and the action is finally controlled up to $n$PN order [see
Eq.~\eqref{resultnp2}]. We evaluate the contribution due to the second,
multipolar term in~\eqref{decompPNmult}, namely
\begin{equation}\label{TF}
T_\text{F} = \mathop{\text{{\rm FP}}}_{B=0}\int\!\ud t\!\int
\ud^{3}\mathbf{x} \,r^B \mathcal{M}\left(\frac{\delta
  S_\text{F}}{\delta h}\right) \mathcal{M}\left(\delta h\right)\,.
\end{equation}
We must impose that the error made in the multipole expansion in the
far zone, \textit{i.e.}, $\mathcal{M}(\delta h)$, becomes equal when
re-expanded into the near zone to the error assumed in the PN
expansion, \textit{i.e.}, $\overline{\delta h}$. Note that
$\overline{\mathcal{M}(\delta h)}$ is not equal to
$\mathcal{M}(\overline{\delta h})$ as the matching equation is only
correct for the true, complete solution $h$ [see
Eq.~\eqref{matching}]. Since the multipole expansion is constructed
from a post-Minkowskian (PM) expansion (see~\cite{Bliving14})
\begin{equation}\label{multPM}
\mathcal{M}(h) = \sum_{m=1}^{+\infty} G^m h_{(m)}\,,
\end{equation}
we shall assume that the error $\mathcal{M}(\delta h)$ in the far zone
corresponds to some high PM order $m_0$, \textit{i.e.}, is of order
$\mathcal{O}(G^{m_0})$. Thus we have
\begin{subequations}\label{deltah}\begin{align}
      \mathcal{M}(h_n) &= \sum_{m=1}^{m_0-1} G^m
      h_{(m)}\,,\\\mathcal{M}(\delta h) &= \sum_{m=m_0}^{+\infty} G^m
      h_{(m)}\,.
\end{align}\end{subequations}
To determine what $m_0$ is we recall that the leading PN order of the
near zone re-expansion of the PM coefficients is $\overline{h}_{(m)} =
\mathcal{O}(2m,2m+1,2m)$.\footnote{Such statement can be proved by
  induction over the PM order $m$.} Imposing then that this
PN-expanded PM error is equal to the previous one assumed for
$\overline{\delta h}$, we find the minimal PM order of the error in
the far zone to be $2m_0 = n+2$ when $n$ is even and $2m_0 = n+3$ when
$n$ is odd, thus (with $[\,]$ being the integer part)
\begin{equation}\label{m0}
m_0 = \left[\hbox{$\frac{n+1}{2}$}\right] + 1\,.
\end{equation}

To obtain the magnitude of the term~\eqref{TF} we first notice that
any term in the integrand which is instantaneous in the sense of
having the structure~\eqref{inst} will yield zero contribution thanks
to our lemma~\eqref{lemma2}. Thus it remains only the hereditary
contributions which have the more complicated structure given
by~\eqref{hered}. Next, we remark that since $\mathcal{M}(\delta h)$
is a small error PM term of order $\mathcal{O}(G^{m_0})$, the
variation of the Fokker action $\mathcal{M}(\delta S_\text{F}/\delta
h)$ evaluated for the approximate solution $h_n$ must necessarily also
be a small PM term, because the Fokker action is stationary for the
exact solution. More precisely we find [because of the extra factor
  $\sim c^4/G$ in front of~\eqref{deltaSFokker}] that it must be of
order $\mathcal{O}(G^{m_0-1})$, hence
\begin{equation}\label{deltaSF}
\mathcal{M}\left(\frac{\delta S_\text{F}}{\delta h}\right) =
\sum_{m=m_0-1}^{+\infty} G^m k_{(m)}\,,
\end{equation}
with some PM coefficients $k_{(m)}$. Thus we conclude that only
hereditary terms that are at least of order $\mathcal{O}(G^{2m_0-1})$
can contribute to the PM expansion of~\eqref{TF}. For our 4PN
computation $n=4$ thus $m_0=3$ from~\eqref{m0}, thus such hereditary
terms must be $\mathcal{O}(G^{5})$.

We had argued at the end of Sec.~\ref{sec:fokkerPN} that cubic
$\sim\mathcal{O}(G^{3})$ hereditary terms will correspond for the
leading multipole interactions to ``tail-of-tails'' and are dominantly
of order 5.5PN when re-expanded in the near zone. Here the hereditary
terms $\sim\mathcal{O}(G^{5})$ should correspond minimally to say
``tail-of-tail-of-tail-of-tails'' and give an even smaller
contribution in the near zone, presumably starting at the order
8.5PN. In conclusion we can neglect the term~\eqref{TF} and our use of
Eq.~\eqref{SFexpand} for the method ``$n+2$'' is justified.

\section{Local expansion of the function $g^{(d)}$ in $d$ 
dimensions}
\label{app:g}

In this Appendix we work out the local expansion near the
singularities, say when $r_1\to 0$, of the function $g^{(d)}$, defined
in $d$ dimensions by
\begin{equation}\label{g}
g^{(d)} = \Delta^{-1}\left(r_1^{2-d}r_2^{2-d}\right) \,,
\end{equation}
where $\Delta^{-1}$ is the usual inverse Laplace operator in $d$
dimensions.  Such solution plays a crucial role when integrating the
non compact support source terms in the elementary
potentials~\eqref{defpotentials} for $d$ dimensions. The explicit form
of this function is known and has been displayed in the Appendix C
of~\cite{BDE04}. Here we shall complete the latter work by providing
the explicit expansion of $g^{(d)}$ when $r_1\to 0$. This expansion is
all that we need when computing the difference between DR and HR ---
since that difference can precisely be obtained solely from the local
expansions $r_1\to 0$ or $r_2\to 0$ near the singularities [see
notably Eq.~\eqref{Dres}].

\subsection{Derivation based on distribution theory}

Following Ref.~\cite{BDE04} we first obtain a local solution in an
expanded form near the particle 1, denoted as $g^{(d)}_\text{loc1}$,
by expanding near $r_1= 0$ the source of the Poisson equation for
$g^{(d)}$ in~\eqref{g} and integrating that source term by term. For
this purpose, we insert the well-known expansion when $r_1\to 0$,
\begin{equation}\label{exp}
r_2^{2-d} = r_{12}^{2-d}\sum_{\ell=0}^{+\infty}
\left(\frac{r_1}{r_{12}}\right)^\ell P_\ell^{(d)}(c_1)\,,
\end{equation}
where we have posed $c_1=-\bm{n}_1\cdot\bm{n}_{12}=\cos\theta_1$,
following exactly the notation of Appendix C of~\cite{BDE04}, and
denoted $P_\ell^{(d)}(c_1)=C_\ell^{(d/2-1)}(c_1)$ the Gegenbauer
polynomial representing the appropriate generalization of the
$\ell$th-degree Legendre polynomial in $d$ dimensions
\begin{equation}\label{Gegenbauer}
P_\ell^{(d)}(c_1) =
\frac{(-2)^\ell\Gamma\left(\frac{d}{2}+\ell-1\right)}
     {\ell!\,\Gamma\left(\frac{d}{2}-1\right)}
     \,\hat{n}_1^L\hat{n}_{12}^L\,.
\end{equation}
After replacing Eq.~\eqref{exp} into the right-hand side of~\eqref{g},
we integrate term by term, using the fact that $P_\ell^{(d)}(c_1)
\propto \hat{n}_1^L$, by means of the elementary formula
\begin{equation}\label{formula}
\Delta^{-1}\left(\hat{n}_1^L r_1^\alpha\right) = \frac{\hat{n}_1^L
  r_1^{\alpha+2}}{(\alpha-\ell+2)(\alpha+\ell+d)}\,.
\end{equation}
In this way, we arrive at the formal \textit{local} expansion when
$r_1\to 0$,
\begin{equation}\label{gloc1exp}
g^{(d)}_\text{loc1} =
\frac{r_{12}^{2-d}r_{1}^{4-d}}{2(4-d)}\sum_{\ell=0}^{+\infty}
\frac{1}{\ell+1}\left(\frac{r_1}{r_{12}}\right)^\ell
P_\ell^{(d)}(c_1)\,.
\end{equation}
The trick now is to rewrite~\eqref{gloc1exp} as an expression
formally valid ``everywhere'', \textit{i.e.}, not only in the vicinity
of the singular point 1, namely the following integral extending along
the segment of line joining the source point $\bm{y}_1$ to the field
point $\bm{x}$,
\begin{equation}\label{gloc1line}
g^{(d)}_\text{loc1} = \frac{r_{1}^{4-d}}{2(4-d)}\int_0^1 \ud \lambda \big|
\bm{y}_{12} + \lambda \,\bm{r}_1 \big|^{2-d}\,.
\end{equation}
See the Appendix C in Ref.~\cite{BDE04} for more details about this
procedure (we recall that here $\bm{y}_{12}=\bm{y}_{1}-\bm{y}_{2}$ and
$\bm{r}_1=\bm{x}-\bm{y}_1$).

Let us next add to $g^{(d)}_\text{loc1}$ given in the form of the line
integral~\eqref{gloc1line} the appropriate homogeneous solution in
such a way that the requested equation~\eqref{g} be satisfied
\textit{in the sense of distributions}. Computing $\Delta
g^{(d)}_\text{loc1}$ in the sense of distributions we readily
obtain~\cite{BDE04}
\begin{equation}\label{Deltagloc1}
\Delta g^{(d)}_\text{loc1} = r_1^{2-d}r_2^{2-d} +
\frac{r_{12}^{4-d}}{2(4-d)}\int_0^1 \frac{\ud\lambda}{\lambda^2}
\Delta\big| \bm{r}_{1} + \textstyle{\frac{1}{\lambda}} \,\bm{y}_{12}
\big|^{2-d}\,,
\end{equation}
showing that the true solution, valid in the sense of distributions,
actually reads
\begin{equation}\label{gtot}
g^{(d)} = g^{(d)}_\text{loc1} + g^{(d)}_\text{hom1}\,,
\end{equation}
where $g^{(d)}_\text{hom1}$ is obtained from the second term
in~\eqref{Deltagloc1}. Changing $\lambda$ into $1/\lambda$ we can
arrange this term as a semi infinite line integral extending from
$\bm{x}$ up to infinity in the direction $\bm{n}_{12}$, 
\begin{equation}\label{ghom1line}
g^{(d)}_\text{hom1} = - \frac{r_{12}^{4-d}}{2(4-d)}\int_1^{+\infty}
\ud\lambda \,\big| \bm{r}_{1} + \lambda \,\bm{y}_{12} \big|^{2-d}\,.
\end{equation}
This is an homogeneous solution in the sense that $\Delta
g^{(d)}_\text{hom1}=0$ in the sense of functions. One can prove that
the sum $g^{(d)} = g^{(d)}_\text{loc1} + g^{(d)}_\text{hom1}$ is
indeed symmetric in the exchange of $\bm{y}_1$ and $\bm{y}_2$ although
the two separate pieces are not.

Here we shall only need the expansion when $r_1\to 0$. An easy
calculation, inserting the expansion~\eqref{exp}
into~\eqref{ghom1line} and performing the integral over $\lambda$
using analytic continuation in $d$ (which is the essence of
dimensional regularization) readily yields (with $\varepsilon=d-3$)
\begin{equation}\label{ghom1exp}
g^{(d)}_\text{hom1} =
-\frac{r_{12}^{-2\varepsilon}}{2(1-\varepsilon)}\sum_{\ell=0}^{+\infty}
\frac{1}{\ell+\varepsilon}\left(\frac{r_1}{r_{12}}\right)^\ell
P_\ell^{(d)}(c_1)\,,
\end{equation}
where $\varepsilon=d-3$. Finally the complete expansion of $g^{(d)}$
when $r_1\to 0$ is obtained by adding the corresponding piece given
by~\eqref{gloc1exp}, as
\begin{equation}\label{gtotexp}
g^{(d)} =
\frac{r_{12}^{-2\varepsilon}}{2(1-\varepsilon)}\sum_{\ell=0}^{+\infty}
\biggl[\frac{1}{\ell+1}\left(\frac{r_1}{r_{12}}\right)^{1-\varepsilon}
  \!\!\!\!-\frac{1}{\ell+\varepsilon}\biggr]
\left(\frac{r_1}{r_{12}}\right)^{\ell} P_\ell^{(d)}(c_1)\,.
\end{equation}

\subsection{Derivation based on asymptotic matching}

It is instructive to present an alternative proof of
Eq.~\eqref{gtotexp} based on the same asymptotic matching techniques
as in the demonstration of Lemma~\ref{th1} exposed in
Sec.~\ref{sec:fokkerPN}. We start with the definition of $g^{(d)}$ in
the form of the $d$-dimensional Poisson integral, which we choose to
be centered on the particle 1,
\begin{equation}\label{gbis}
g^{(d)} = - \frac{\tilde{k}_d}{4\pi} \int \frac{\ud^d
  \boldsymbol{r}'_1}{|\boldsymbol{r}_1- \boldsymbol{r}'_1|^{d-2}}
\,r'^{2-d}_1 r'^{2-d}_2 \,,
\end{equation}
where $\tilde{k}_d$ stands for the constant factor
$\Gamma(d/2-1)/\pi^{d/2-1}$ and $\boldsymbol{r}_1 =
\mathbf{x}-\boldsymbol{y}_1$. In this definition, we decompose the
source into two terms. The first one is taken to be the Taylor
expansion of $S=r_1^{2-d} r_2^{2-d}$ near $r_1=0$, denoted as
$\mathcal{T}_1(S)$ henceforth. The second term is thus the difference
$\delta S(\mathbf{x},t)= S - \mathcal{T}_1(S)$. The key point consists
in noticing that $\delta S(\mathbf{x},t)$ vanishes in some open ball
of radius $R_1$ centered at the ``origin'' $\boldsymbol{r}_1=0$. This
means that one can restrict the integration domain of the Poisson
operator acting on $\delta S$ to the set of points verifying $r'_1 >
R_1$. Therefore, for $r_1 < R_1$, the Poisson kernel
$|\boldsymbol{r}_1-\boldsymbol{r}'_1|^{2-d}$ may be replaced by its
multipole expansion
\begin{equation}
\mathcal{M}\left(|\boldsymbol{r}_1-\boldsymbol{r}'_1|^{2-d} \right) =
\sum_{\ell = 0}^{+\infty}\frac{(-)^\ell}{\ell!} r^L_1 \partial_L
r'^{2-d}_1 \,,
\end{equation}
with the short notation $r^L_1=r^{i_1}_1r^{i_2}_1\cdots
r^{i_\ell}_1$. After this operation, the integration domain may be
extended again to the whole space, since the source is still zero for
$r'_1 < R_1$. It is implicitly understood here that all Taylor and
multipole expansions are actually performed at some finite but
arbitrary high orders. The formal use of infinite series in the
present discussion just allows us to elude technicalities related to
the control of remainders. However, we have checked that truncations
at finite orders do not change the backbone of our argument. In
particular, we are formally allowed to commute the sum and integral
symbols.

At this stage, we have shown that
\begin{align} \label{gdmatched}
g^{(d)} &= \Delta^{-1}\mathcal{T}_1 \left(r_1^{2-d}r_2^{2-d}\right)
\nonumber \\ &- \frac{\tilde{k}_d}{4\pi}
\sum_{\ell=0}^{+\infty}\frac{(-)^\ell}{\ell!}  r^L_1 \int \ud^d
\boldsymbol{r}'_1 \,\partial_L \left(\frac{1}{r'^{d-2}_1} \right)
\Bigl[ r'^{2-d}_1 r'^{2-d}_2 - \mathcal{T}_1\left(r'^{2-d}_1
  r'^{2-d}_2 \right) \Bigr] \,.
\end{align}
Because $\Delta^{-1}\mathcal{T}_1 (r_1^{2-d}r_2^{2-d})$ is precisely
what we have defined to be $g^{(d)}_\text{loc1}$ in Eq.~\eqref{gloc1exp},
the expression in the second line is identified with the homogeneous
solution $g^{(d)}_\text{hom1}$. Now, the second term within the square
brackets is made of pieces of the form
$\sim(\hat{n}'^L_1/r'^{d-2+\ell}_1) (r'^{2-d}_1)(r'^k_1\hat{n}'^K_1)$
for $\ell$, $k$ integers. Its radial integration leads to integrals
$\int_0^{+\infty} \ud r'_1 \, r'^{-1-\varepsilon}_1$. The latter are
just zero by analytic continuation on the parameter $\varepsilon=d-3$,
as explained in the proof of Lemma~\ref{th1} (with $\varepsilon$
playing the role of $-B$ there). Hence the homogeneous solution reads
\begin{equation}\label{ghom1int}
g^{(d)}_\text{hom1} = - \frac{\tilde{k}_d}{4\pi} \sum_{\ell =
  0}^{+\infty} \frac{(-)^\ell}{\ell!} r^L_1 \int \ud^d
\boldsymbol{r}'_1 \partial_L \left(\frac{1}{r'^{d-2}_1} \right)
r'^{2-d}_1 r'^{2-d}_2 \,.
\end{equation}
Next we complete the proof by evaluating explicitly the integral
entering the above formula with the help of the relation
$\hat{\partial}_L r'^\alpha_1 = (-2)^\ell \Gamma(\ell -
\alpha/2)/\Gamma(-\alpha/2) r'^{\alpha-\ell}_1 \hat{n}'^L_1$. We first
obtain\footnote{Recall that the spatial multi-derivative $\partial_L
  (r^{2-d})$ is trace-free. The traces are actually made of
  derivatives of $d$-dimensional Dirac functions but one can check
  that, when inserted into the integral of~\eqref{ghom1int}, they
  vanish by analytic continuation on $\varepsilon$.}
\begin{equation} \label{ghom1intbis}
g^{(d)}_\text{hom1} = - \frac{\tilde{k}_d}{4\pi} \sum_{\ell =
  0}^{+\infty} \frac{\Gamma(\ell + d/2 -1)\Gamma(d-2)}{\Gamma
  (d/2-1)\Gamma(\ell+d-2)} \frac{r^L_1}{\ell!}\frac{\partial}{\partial
  \hat{y}^L_1} \int \ud^d \boldsymbol{r}'_1 r'^{4-2d}_1 r'^{2-d}_2\,,
\end{equation}
and, in the last step, we compute $\int \ud^d \boldsymbol{r}'_1
r'^{4-d}_1 r'^{2-d}_2 $ by means of the Riesz formula, given
\textit{e.g.} by Eq.~(B.19) of Ref.~\cite{BDE04}. Using the
relation~\eqref{Gegenbauer} we find that the ensuing expression for
$g^{(d)}_\text{hom1}$ is in full agreement with Eq.~\eqref{ghom1exp}.

Notice finally that the function $g^{(d)}$ in $d$ dimensions contains
a pole in the dimension coming from the monopole part of the
expansion~\eqref{ghom1exp} or~\eqref{gtotexp}, namely
${g^{(d)}=-\frac{1}{2}\varepsilon^{-1}+\mathcal{O}(\varepsilon^0)}$.
However, since in practical computations $g^{(d)}$ will always be
differentiated, this pole is always cancelled out. Furthermore it was
proved in Ref.~\cite{BDE04} that the finite part of $g^{(d)}$ when
$\varepsilon\to 0$ recovers the 3-dimensional result
$\ln(r_1+r_2+r_{12})$~\cite{Fock} up to some irrelevant additive
constant, namely
\begin{equation}\label{gpole}
g^{(d)} = -\frac{1}{2\varepsilon} - \frac{1}{2} +
\ln\left(\frac{r_1+r_2+r_{12}}{2}\right) + \mathcal{O}(\varepsilon)\,.
\end{equation}
But here we only need the local expansion provided by~\eqref{gtotexp}
up to order $\varepsilon$ included.

\section{The complete 4PN shift}
\label{app:shift}

In this Appendix we show the complete shift at 4PN order that removes,
in particular, all the poles $\propto 1/\varepsilon$ and all the IR
constants $r_0$.\footnote{Another shift has been used in
  Secs.~\ref{sec:ordlagrangian} and~\ref{sec:hamiltonian} to remove
  the accelerations in the harmonic Lagrangian and compute the
  Hamiltonian. This shift, however, is too long to be presented.}
Furthermore, this shift cancels the dependence on the individual
positions $\bm{y}_A$ of the particles and is such that the shifted
equations of motion are manifestly Poincar\'e invariant (including
spatial translations and boosts). It reads
\begin{equation}\label{xi1}
    \bm{\xi}_1=\frac{11}{3}\frac{G^2\,m_1^2}{c^6} \left[
      \frac{1}{\varepsilon}-2\ln\left(
      \frac{\overline{q}^{1/2}r'_1}{\ell_0}\right)
      -\frac{327}{1540}\right] \bm{a}^{(d)}_{1,\,\mathrm{N}} +
    \frac{1}{c^8}\bm{\xi}_{1,\,\mathrm{4PN}}\,,
\end{equation}
where $G=G_\text{N}$ in this Appendix, $\bm{a}^{(d)}_{1,\,\mathrm{N}}$
represents the Newtonian acceleration of 1 in $d$ dimensions and we
recall that $\overline{q} = 4\pi e^{\gamma_\text{E}}$. For convenience
we divide the 4PN piece of the shift in several pieces,
\begin{equation}
\xi^i_{1,\, \text{4PN}} = \frac{1}{\varepsilon}\xi^{i\, (-1)}_{1,\,
  \text{4PN}} + \xi^{(0,n_{12})}_{1,\, \text{4PN}} n^i_{12} +
\xi^{(0,v_{1})}_{1,\,\text{4PN}} v^i_1 +
\xi^{(0,v_{12})}_{1,\,\text{4PN}} v^i_{12} \,,
\end{equation}
with $v^i_{12}=v^i_{1}-v^i_{2}$ and
\begin{subequations}\label{shift4PN}
\begin{align}
\xi^{i\, (-1)}_{1,\, \text{4PN}}&={}\frac{G^3 m_{1}^2 m_{2} v_{12}^{i}}{r_{12}^2} 
\bigl(11 (n_{12} v_{12})
 + \frac{11}{3} (n_{12} v_1)\bigr)
 + n_{12}^{i} \bigl( \frac{G^4}{r_{12}^3}(\frac{55}{3} m_{1}^3 m_{2}
 + \frac{22}{3} m_{1}^2 m_{2}^2
 + 4 m_{1} m_{2}^3)\nonumber\\
& + \frac{G^3 m_{1}^2 m_{2}}{r_{12}^2} (\frac{11}{2} (n_{12} v_{12})^2
 - 11 (n_{12} v_{12}) (n_{12} v_1)
 + \frac{11}{2} (n_{12} v_1)^2
 -  \frac{22}{3} v_{12}^{2})\bigr)\,,\\[1ex]
\xi^{(0,n_{12})}_{1,\, \text{4PN}}&={}G^3 m_{1} m_{2}^2 \biggl\{\frac{1}{r_{12}^{2}} 
\Bigl[\bigl(- \frac{2539}{560}
 + \frac{72}{5} \ln(\frac{r_{12}}{r_{0}})\bigr) (n_{12} v_{12})^2
 \nonumber\\
& + \bigl(\frac{18759}{280}+ \frac{96}{5} \ln(\frac{r_{12}}{r_{0}})\bigr) 
(n_{12} v_{12}) (n_{12} v_1)
 + \bigl(- \frac{6253}{140}
 -  \frac{64}{5} \ln(\frac{r_{12}}{r_{0}})\bigr) (v_{12} v_1)
 \nonumber\\
& + \bigl(\frac{5783}{840} 
 -  \frac{32}{5} \ln(\frac{r_{12}}{r_{0}})\bigr) v_{12}^{2}\Bigr]
 + \frac{1}{r_{12}^{3}} \Bigl[\bigl(\frac{519}{35}
 + 48 \ln(\frac{r_{12}}{r_{0}})\bigr) (n_{12} v_{12})^2 (n_{12} y_1)
\nonumber\\
& + \bigl(- \frac{289}{35} -  
\frac{144}{5} \ln(\frac{r_{12}}{r_{0}})\bigr) (n_{12} v_{12}) (v_{12} y_1)
 + \bigl(- \frac{171}{35}
 -  \frac{48}{5} \ln(\frac{r_{12}}{r_{0}})\bigr) 
(n_{12} y_1) v_{12}^{2}\Bigr]\biggr\} \nonumber\\
& + G^3 m_{1}^2 m_{2} \biggl\{\frac{1}{r_{12}^{2}} 
 \Bigl[\bigl(- \frac{2761}{168}
 -  \frac{33}{2} \ln(\frac{\bar{q}^{1/2} r'_{1}}{\ell_{0}})
 -  \frac{11}{2} \ln(\frac{r_{12}}{r'_{1}})\bigr) (n_{12} v_{12})^2 \nonumber\\
& + \bigl(\frac{41299}{840} + 33 \ln(\frac{\bar{q}^{1/2} r'_{1}}{\ell_{0}})
 + \frac{96}{5} \ln(\frac{r_{12}}{r_{0}})
 + 11 \ln(\frac{r_{12}}{r'_{1}})\bigr) (n_{12} v_{12}) (n_{12} v_1) \nonumber\\
& + \bigl(\frac{7489}{840} 
 -  \frac{33}{2} \ln(\frac{\bar{q}^{1/2} r'_{1}}{\ell_{0}})
 -  \frac{11}{2} \ln(\frac{r_{12}}{r'_{1}})\bigr) (n_{12} v_1)^2
 + \bigl(- \frac{6253}{140}
 -  \frac{64}{5} \ln(\frac{r_{12}}{r_{0}})\bigr) (v_{12} v_1) \nonumber\\
& + \bigl(\frac{3103}{168} + 22 \ln(\frac{\bar{q}^{1/2} r'_{1}}{\ell_{0}})
 + \frac{16}{5} \ln(\frac{r_{12}}{r_{0}})\bigr) v_{12}^{2}\Bigr]
 + \frac{1}{r_{12}^{3}} \Bigl[\bigl(\frac{519}{35}
 + 48 \ln(\frac{r_{12}}{r_{0}})\bigr) (n_{12} v_{12})^2 (n_{12} y_1)\nonumber\\
& + \bigl(- \frac{289}{35}
 - \frac{144}{5} \ln(\frac{r_{12}}{r_{0}})\bigr) (n_{12} v_{12}) (v_{12} y_1) + 
\bigl(- \frac{171}{35}
 -  \frac{48}{5} \ln(\frac{r_{12}}{r_{0}})\bigr) (n_{12} y_1)
 v_{12}^{2}\Bigr]\biggr\} \nonumber\\
&+ G^4
m_{1}^2 m_{2}^2 \Bigl[\frac{1}{r_{12}^{3}} \bigl(- \frac{811}{210}
 -  \frac{88}{3} \ln(\frac{\bar{q}^{1/2} r'_{1}}{\ell_{0}})
 -  \frac{16}{5} \ln(\frac{r_{12}}{r_{0}})\bigr)
\nonumber\\
& + \frac{1}{r_{12}^{4}} \bigl(- \frac{114}{35} - 
\frac{32}{5} \ln(\frac{r_{12}}{r_{0}})\bigr) (n_{12} y_1)\Bigr]
\nonumber\\
& + G^4 m_{1}^3 m_{2} \Bigl[\frac{1}{r_{12}^{3}} \bigl(- \frac{811}{105}
 -  \frac{220}{3} \ln(\frac{\bar{q}^{1/2} r'_{1}}{\ell_{0}})
 -  \frac{32}{5} \ln(\frac{r_{12}}{r_{0}})\bigr)\nonumber\\
& + \frac{1}{r_{12}^{4}} \bigl(- \frac{57}{35}
 -  \frac{16}{5} \ln(\frac{r_{12}}{r_{0}})\bigr) (n_{12} y_1)\Bigr]
 + G^4 m_{1} m_{2}^3 \Bigl[\frac{1}{r_{12}^{3}} \bigl(\frac{811}{210}
 - 16 \ln(\frac{\bar{q}^{1/2} r'_{1}}{\ell_{0}})\nonumber\\
& + \frac{16}{5} \ln(\frac{r_{12}}{r_{0}})\bigr)
 + \frac{1}{r_{12}^{4}} \bigl(- \frac{57}{35}
 -  \frac{16}{5} \ln(\frac{r_{12}}{r_{0}})\bigr) (n_{12} y_1)\Bigr]
 \,,\\[1ex]
\xi^{(0,v_{1})}_{1,\, \text{4PN}}&={}G^3 \Bigl[\frac{m_{1}^2 m_{2}}{r_{12}^2}
\bigl(\frac{839}{2520}
 + \frac{32}{15} \ln(\frac{r_{12}}{r_{0}})\bigr) (n_{12} v_{12})
 + \frac{m_{1} m_{2}^2}{r_{12}^2} \bigl(\frac{839}{2520}
 + \frac{32}{15} \ln(\frac{r_{12}}{r_{0}})\bigr) (n_{12} v_{12})\Bigr]\,,\\[1ex]
\xi^{(0,v_{12})}_{1,\, \text{4PN}}&={}G^3 m_{1} m_{2}^2
\biggl\{\frac{1}{r_{12}^{2}} \Bigl[\bigl(\frac{2041}{336}
 - 8 \ln(\frac{r_{12}}{r_{0}})\bigr) (n_{12} v_{12})
 + \bigl(- \frac{6253}{140}
 -  \frac{64}{5} \ln(\frac{r_{12}}{r_{0}})\bigr) (n_{12} v_1)\Bigr]\nonumber\\
& + \frac{1}{r_{12}^{3}} \Bigl[\bigl(- \frac{289}{35}
 -  \frac{144}{5} \ln(\frac{r_{12}}{r_{0}})\bigr) (n_{12} v_{12}) (n_{12} y_1)
 + \bigl(\frac{228}{35}
 + \frac{64}{5} \ln(\frac{r_{12}}{r_{0}})\bigr) 
 (v_{12} y_1)\Bigr]\biggr\}\nonumber\\
& + G^3 m_{1}^2 m_{2} 
 \biggl\{\frac{1}{r_{12}^{2}} \Bigl[\bigl(\frac{23113}{1680}
 - 33 \ln(\frac{\bar{q}^{1/2} r'_{1}}{\ell_{0}})
 + \frac{11}{3} \ln(\frac{r_{12}}{r'_{1}})\bigr) (n_{12} v_{12})
\nonumber\\
& + \bigl(- \frac{1398}{35} - 11 \ln(\frac{\bar{q}^{1/2} r'_{1}}{\ell_{0}})
 -  \frac{64}{5} \ln(\frac{r_{12}}{r_{0}})
 -  \frac{11}{3} \ln(\frac{r_{12}}{r'_{1}})\bigr) (n_{12} v_1)\Bigr]
\nonumber\\
& + \frac{1}{r_{12}^{3}} \Bigl[\bigl(- \frac{289}{35} 
-  \frac{144}{5} \ln(\frac{r_{12}}{r_{0}})\bigr) (n_{12} v_{12}) (n_{12} y_1)
 + \bigl(\frac{228}{35}
 + \frac{64}{5} \ln(\frac{r_{12}}{r_{0}})\bigr) (v_{12} y_1)\Bigr]\biggr\}\,.
\end{align}
\end{subequations}

\bibliography{BBBFM15a_resub_apr16_grqc.bib}

\end{document}